\documentclass[11pt,british,english]{article}
\pdfoutput=1
\usepackage[T1]{fontenc}
\usepackage[latin9]{inputenc}
\usepackage{geometry}
\usepackage{float}
\usepackage{amsmath}
\usepackage{graphicx}
\usepackage{esint}

\makeatletter

\newcommand*\LyXThinSpace{\,\hspace{0pt}}
\providecommand{\tabularnewline}{\\}

\numberwithin{equation}{section}
\numberwithin{figure}{section}
\numberwithin{table}{section}

\makeatother

\usepackage{babel}
\begin{document}

\title{Initial states in integrable quantum field theory quenches from an
integral equation hierarchy}

\author{D. X. Horváth%
\thanks{esoxluciuslinne@gmail.com%
}$\;{}^{1,2}$, S. Sotiriadis%
\thanks{sotiriad@sissa.it%
}$\;{}^{3}$ and G. Takács%
\thanks{takacsg@eik.bme.hu%
}$\;{}^{1,2}$\\
~\\
$^{1}${\small{}MTA-BME \textquotedbl{}Momentum\textquotedbl{} Statistical
Field Theory Research Group}\\
{\small{}1111 Budapest, Budafoki út 8, Hungary}\\
~\\
 $^{2}${\small{}Department of Theoretical Physics, }\\
{\small{} Budapest University of Technology and Economics}\\
{\small{}1111 Budapest, Budafoki út 8, Hungary}\\
~\\
$^{3}${\small{}SISSA and INFN,}\\
{\small{}Via Bonomea 265, 34136 Trieste, Italy}}

\date{23rd November 2015}
\maketitle
\begin{abstract}
We consider the problem of determining the initial state of integrable
quantum field theory quenches in terms of the post-quench eigenstates.
The corresponding overlaps are a fundamental input to most exact methods
to treat integrable quantum quenches. We construct and examine an
infinite integral equation hierarchy based on the form factor bootstrap,
proposed earlier as a set of conditions determining the overlaps.
Using quenches of the mass and interaction in Sinh-Gordon theory as
a concrete example, we present theoretical arguments that the state
has the squeezed coherent form expected for integrable quenches, and
supporting an Ansatz for the solution of the hierarchy. Moreover we
also develop an iterative method to solve numerically the lowest equation
of the hierarchy. The iterative solution along with extensive numerical
checks performed using the next equation of the hierarchy provide
a strong numerical evidence that the proposed Ansatz gives a very
good approximation for the solution.
\end{abstract}

\section{Introduction}

The study of quantum dynamics in one-dimensional integrable systems
has led to intriguing discoveries, like the experimental observation
of lack of thermalization \cite{exp1,exp2,exp3}, the theoretical
prediction \cite{Rigol} and experimental observation \cite{exp4}
of an unconventional statistical ensemble known as the \emph{Generalized
Gibbs Ensemble} (GGE), the discovery of unexpected effects on quantum
transport \cite{transport-1,transport-2} and of novel quasi-local
conserved charges \cite{Prosen,PPSA,completeGGE}. A typical protocol
employed for the preparation of a closed quantum system in an out-of-equilibrium
initial state is the instantaneous change of some parameter of its
Hamiltonian, a process called \emph{quantum quench} \cite{CC1,CC2}.
After a quantum quench, the initial state of the system, which is
typically the ground state of the pre-quench Hamiltonian, evolves
unitarily under a different post-quench Hamiltonian. 

Obviously in order to derive the time evolution of the system we generally
need to know the post-quench excitation amplitudes in the initial
state, i.e. the overlaps of the initial state with any post-quench
energy eigenstates. Determining the excitation content of the initial
state is typically easy for quantum quenches in which both the pre-
and the post-quench Hamiltonians are quadratic in terms of some suitable
physical fields. In this class of problems, which correspond to non-interacting
models or interacting models that can be mapped into non-interacting
ones, the relation between pre-quench and post-quench excitations
is typically described by a so-called \textit{Bogoliubov transformation}
and the initial state is then a squeezed coherent state\textit{,}
more precisely a \emph{squeezed vacuum} state, of the post-quench
Hamiltonian \cite{RSMSS,CEF,SE12}. Such states consist of pairs of
excitations with opposite momenta and have the characteristic exponential
form

\begin{equation}
|S\rangle=\exp\left(\int_{0}^{\infty}dk\; K(k)A^{\dagger}(-k)A^{\dagger}(k)\right)|0\rangle\:,\label{sv-1}
\end{equation}
where $A^{\dagger}(k)$ are operators that create particles of momentum
$k$ and $|0\rangle$ is the corresponding vacuum state. However,
the task of determining the excitation content of the initial state
is far more difficult in the context of genuinely interacting integrable
systems (except in special cases \cite{RS14}). The reason is that
the pre-quench and post-quench excitations are no longer related through
a simple linear transformation, but instead a nonlinear one that in
general corresponds to an infinite series \cite{SFM}. Even though
the initial state has been derived exactly in a number of special
cases \cite{KozPozs,LLQQ,Pozsi_overlap,overlaps,overlaps2,overlaps3},
a general and systematic method for its determination remains so far
elusive. On the other hand, while in those earlier studies the initial
state was derived by means of finite volume calculations based on
the Bethe Ansatz, in fact we are mostly interested in the thermodynamic
limit where all finite size effects vanish and integrable quantum
field theoretic methods come into play \cite{FM,SFM,STM,Delfino14,Schuricht15}.
In particular, it has been argued \cite{QuenchAction} that in the
thermodynamic limit, i.e. in the limit of large system size $L$ and
particle number $N$ with a fixed density of particles, it is sufficient
to know only the extensive part of the logarithmic overlaps between
the initial state $|\Psi\rangle$ and the post-quench eigenstates
$|\Phi\rangle$, i.e. the quantity

\begin{equation}
\mathcal{E}_{\Psi}[\Phi]=\lim_{N,L\to\infty}N^{-1}\log\langle\Phi|\Psi\rangle\:.\label{eq:QA_overlap}
\end{equation}
Indeed it was shown in \cite{QuenchAction} that a single post-quench
eigenstate that is representative of the initial state, is sufficient
for the description of the asymptotic values of local observables
at times $t\to\infty$. Such a representative state is completely
determined by the quantity $\mathcal{E}_{\Psi}[\Phi]$ above, moreover,
the full time evolution can also be derived from the same quantity.
The argument is based on the fundamental idea of statistical physics
that microstates can be classified according to their macroscopic
properties into macrostates, which in integrable models are fully
characterized by their Bethe Ansatz root density \cite{YangYang}.
This approach, also known as \textit{Quench Action method,} has successfully
predicted the stationary values of observables at large times \cite{fGGE1,fGGE2,fGGE4,fGGE5,LMV_QA,PCE}
as well as their full time evolution \cite{BSE14,QA_timeevol,QAevol1,QAevol2}.

The thermodynamic quantity (\ref{eq:QA_overlap}) for any specific
quench initial state is generally given by an expression much simpler
than the exact expression for the finite volume overlaps \cite{LLQQ},
suggesting that it may be easier to derive it \emph{directly} in infinite
volume by a field theoretic method. One of the earliest field theoretic
approaches was given in \cite{CC1,CC2} in which the quantum quench
problem was reformulated as a boundary problem defined in an infinite
strip geometry with boundary conditions given by the initial state.
This way the authors exploited results already known from Boundary
Renormalization Group (RG) theory to study a broad class of quantum
quenches in Conformal Field Theories (CFT). In particular it is known
that universal boundary behaviour at equilibrium can be classified
within a small number of different types, each corresponding to a
scale invariant boundary state. For example, the Dirichlet boundary
state $|D\rangle$ corresponds to a quantum quench with vanishing
initial field fluctuations, as in the case of an infinite pre-quench
particle mass $m_{0}\to\infty$.

Exploring the generalization of this approach to systems described
by an Integrable Field Theory (IFT) with no off-diagonal scattering,
the authors of \cite{FM,highly_excited} studied a special class of
initial states of the form
\begin{equation}
|B\rangle=\exp\left(\int_{0}^{\infty}d\theta\; K(\theta)Z^{\dagger}(-\theta)Z^{\dagger}(\theta)\right)|\Omega\rangle\:,\label{sv}
\end{equation}
where $Z^{\dagger}(\theta)$ is an operator that creates a post-quench
excitation of rapidity $\theta$, $|\Omega\rangle$ is the ground
state of the IFT, while $K(\theta)$ is the pair excitation amplitude.
States of this form, which is a generalization of the squeezed vacuum
state (\ref{sv-1}), include (but are not limited to) the so-called
\emph{boundary integrable states} \cite{GhoshalZamo}, i.e. boundary
states that respect the integrability of the bulk. This approach was
later extended to include a slightly more general form of initial
states \cite{Mussardo13}. For this type of states it was possible
to analytically show that the system equilibrates and is described
by a GGE. Furthermore the large time decay of observables can also
be analytically calculated for quenches starting from an analogous
class of initial states in the sine-Gordon model \cite{BSE14}, using
a form factor based approach. In addition, using the semiclassical
approach proposed in \cite{SachdevYoung} and extended to a quench
situation in \cite{igloi_semicl}, the time evolution of correlation
functions after a quench in the sine-Gordon model starting from a
state of the form (\ref{sv}) was determined in \cite{kormos_zarand}.
Moreover, in recent investigations of the Loschmidt echo and statistics
of work done during a quantum quench \cite{work1,work2}, an initial
state of the form (\ref{sv}) was considered as a starting point.

Therefore the knowledge of the amplitude $K$, which fully characterizes
the state (\ref{sv}), is sufficient to determine the expectation
values of local operators in the post-quench stationary state. In
addition, the amplitude $K$ serves as an input quantity to form factor
based and semiclassical approaches to correlators, and to determinations
of Loschmidt echo and the statistics of work. As a result, knowledge
of $K$ is of primary interest in the description of integrable quantum
field theory quenches. 

However, there is no a priori reason to believe that a quantum quench
in an IFT would lead to an initial state of this form. Indeed, even
though Dirichlet states are of this form and correspond to a quantum
quench with an infinite pre-quench particle mass $m_{0}$, these are
ill-defined states: they exhibit ultraviolet divergent physical observables
due to the fact that their excitation amplitude $K_{D}(\theta)$ does
not decay for large momenta. This is easily understood: initial state
excitations are cut-off at large momenta by the mass scale $m_{0}$
which in this case is taken to be infinite. To regularize the boundary
state in the CFT case, the authors of \cite{CC1,CC2} used the concept
of ``extrapolation length'' $\tau_{0}$, known from Boundary RG
theory, where it expresses the difference of the actual boundary state
from the idealized Dirichlet state. The parameter $\tau_{0}$ plays
the role of an exponential large momentum cut-off, dependent in general
on the initial parameters and typically proportional to $1/m_{0}$
in the CFT case. A generalization of this idea to IFT \cite{FM} (although
not justified by RG theory since massive IFT are non-critical) would
amount to replacing the Dirichlet state $|D\rangle$ by the regularized
state $e^{-H\tau_{0}}|D\rangle$ where $H$ is the post-quench Hamiltonian,
or equivalently replacing $K_{D}(\theta)$ by 
\begin{equation}
K_{D}(\theta)e^{-2E(\theta)\tau_{0}}\:,\label{eq:extrapol_length}
\end{equation}
where $E(\theta)$ is the energy of an excitation with rapidity $\theta$.
In the IFT case $\tau_{0}$ can be considered as a phenomenological
parameter. Although it may be expected to be of order $1/m_{0}$ as
in the CFT case, its precise dependence on the quench parameters is
not generally known. The relation of this phenomenological parameter
to the quench parameters is crucial, since the values of physical
observables depend explicitly on it. In \cite{STM} an estimation
of $\tau_{0}$ in the free massive case by comparison of the field
fluctuations in the actual and approximate state demonstrated that
this approach does not reproduce correctly the known exact results
for the large time values of observables, as far as numerical factors
are concerned. Moreover it was shown that the estimation of $\tau_{0}$
is not independent of the choice of observable used to make the comparison
between actual and approximate state. Different choices lead to different
scaling for the $m_{0}$ dependence of $\tau_{0}$ as $m_{0}\to\infty$.
This discrepancy is an indication that the effect of large momentum
excitations that are present in the initial state cannot be incorporated
in a suitable and unique definition of the momentum cut-off. 

On the other hand, there is no fundamental reason preventing us from
choosing $\tau_{0}$ to be momentum dependent itself. In this way
the actual $K(\theta)$ may not necessarily decay exponentially with
the momentum and such alternative choices would modify the predicted
values of physical observables. In fact, such a generalization is
justified from the point of view of the same boundary formulation:
RG theory teaches us that, in estimating the difference of the actual
boundary state from the idealized Dirichlet state $|D\rangle$, any
boundary irrelevant operator could be inserted as a boundary perturbation
\cite{extrapol_length_GGE}. In an IFT such boundary irrelevant operators
include (but are not limited to) all conserved charges of the bulk
theory. Indeed adding such perturbations does not critically change
the system's behaviour \cite{EKS}. This means that, in the same way
that the extrapolation length $\tau_{0}$ is introduced essentially
as a perturbative parameter associated to the Hamiltonian, one could
in principle introduce a different parameter $\tau_{s}$ for each
conserved charge $Q_{s}=\int d\theta q_{s}(\theta)Z^{\dagger}(\theta)Z(\theta)$.
This generalizes our Ansatz for the regularized initial state from
$e^{-H\tau_{0}}|D\rangle$ to the more general $e^{-\sum_{s}Q_{s}\tau_{s}}|D\rangle$,
which is clearly equivalent to introducing a momentum-dependent ``extrapolation
length'' $\tau(\theta)$. Such a generalized regularization $K_{D}(\theta)\to K_{D}(\theta)e^{-2E(\theta)\tau(\theta)}$
could be any function of $\theta$ that fulfills the ultraviolet convergence
condition. Other irrelevant perturbations that may be included as
perturbations are not simply quadratic in $Z(\theta)$ and $Z^{\dagger}(\theta)$
(as are the charges) but of higher order instead \cite{extrapol_length_GGE}.
These would lead to deviations from the form (\ref{sv}). Overall
this conclusion brings us back to our starting question of how to
determine correctly the pair excitation amplitude $K(\theta)$ of
(\ref{sv}), or more generally all (possibly independent) excitation
amplitudes of the initial state. 

Note that, as mentioned above, this is a physically important problem
in the context of quantum quenches, since the values of observables
depend significantly on the initial state and even its regularization.
In particular a modification of the amplitude $K(\theta)$ may crucially
affect the subsequent evolution and the final equilibrium. In the
CFT case, for example, considering only the Hamiltonian as perturbation
leads to the conclusion that the system tends to an effectively thermal
equilibrium with a temperature proportional to $\tau_{0}$ \cite{CC1,CC2}.
On the contrary, including more perturbations results in the emergence
of a generalized equilibrium (GGE) with as many different temperatures
as the perturbative parameters \cite{extrapol_length_GGE}. General
studies in interacting to free quenches show that in the case of massive
evolution the only information that survives at large times is that
of the initial two point correlation function and the equilibration
is described by the GGE \cite{SC14}. On the other hand, in the massless
case this is not true and much more information about the initial
state must be taken into account \cite{S15}, unless the initial state
is gaussian in which case the GGE is still valid \cite{LL1,LL2,LL3,LL4}.

In \cite{STM} a systematic approach was proposed to determine the
expansion of the initial state in terms of post-quench excitations
from first principles. The specific problem under investigation was
a quantum quench of the particle mass $m$ and the interaction $g$
in the Sinh-Gordon model, a prototypical integrable model with a single
type of particle, starting from a large but not infinite initial mass
and zero interaction. It was shown that the condition that the initial
state, being the ground state of the pre-quench Hamiltonian, is annihilated
by all pre-quench annihilation operators, imposes an infinite system
of equations that must be satisfied by the initial state excitation
amplitudes. These are integral equations that involve infinite series
of form factors of the physical field $\phi$. The simplest of these
equations for the pair excitation amplitude $K(\theta)$ was derived
explicitly, assuming that the form (\ref{sv}) is valid. By truncating
the form factor series and analyzing the integration kernels, a simple
factorized Ansatz for the solution $K(\theta)$ was proposed and numerically
verified. This justified the expression (\ref{sv}) for the initial
state as the leading order result in a systematic expansion, rather
than an ambiguous approximation. On the other hand it also showed
that a consistent regularization of $K_{D}(\theta)$ for large initial
mass is decaying not exponentially as suggested by (\ref{eq:extrapol_length}),
but rather as a power of the momentum.

The Ansatz of \cite{STM} itself rested on a number of assumptions,
for which only partial justification was given, and the numerical
checks only treated a single member of an eventually infinite integral
equation hierarchy, which the amplitude has to satisfy. In this paper
we first derive the whole hierarchy of equations in explicit form
at all orders of the form factor series. To achieve this we employ
a simpler derivation method that works directly in the infinite volume
limit and verify its agreement with our earlier method that is based
on finite volume regularization. Second, we eliminate a large part
of the assumptions present in \cite{STM} and give plausible arguments
for the rest. In addition, we perform a thorough numerical analysis
of the hierarchy. Unlike \cite{STM} where it was only checked that
the Ansatz satisfies approximately a truncated version of the lowest
order equation, here we compute a numerical solution of the equation
by means of a newly developed iterative method without bias (i.e.,
without assuming the Ansatz). Moreover we perform numerical consistency
checks that the derived solution also satisfies the next order equation
and finally compare it with our Ansatz concluding that they are in
perfect agreement. We also give independent numerical evidence for
the correctness of our theoretical arguments. While we treat the same
quantum quench in Sinh-Gordon model, as before, our theoretical considerations,
supported by the numerics, also make it plausible that the expression
(\ref{sv}) for the initial state, and the factorized form of the
Ansatz is valid for a much larger class of models.

The paper is organized as follows. After setting up some necessary
notions and notations in Section \ref{sec:The-Sinh-Gordon-model},
we proceed to a general overview of the hierarchy of integral equations
in Section \ref{sec:The-integral-equation}, and consider the problem
of existence and uniqueness of its solution. The properties of the
solution are explored in Section \ref{sec:General-properties-of},
using both general principles of field theory and reasoning connected
to integrability, leading to a strong plausibility argument for the
Ansatz used in \cite{STM}. Then we proceed to the numerical investigation
of the integral equation hierarchy in Section \ref{sec:Numerical-solution-of},
and draw our conclusions in Section \ref{sec:Conclusions}. The paper
also contains two appendices, Appendix \ref{sec:Finite-volume-regularization}
containing concepts and results of finite volume regularization that
are necessary for the main text, and Appendix \ref{sec:Three-particle-condition}
collecting some numerical data for illustration.

\section{The Sinh-Gordon model \label{sec:The-Sinh-Gordon-model}}

The Sinh-Gordon theory is defined by the Hamiltonian 
\begin{equation}
H=\int dx\left[\frac{1}{2}\pi^{2}+\frac{1}{2}(\partial_{x}\phi)^{2}+\frac{\mu^{2}}{g^{2}}\cosh g\phi(x)\right]\:,\label{eq:hamiltonian}
\end{equation}
where $\mu$ is the classical particle mass and $g$ the coupling
constant. It is the simplest example of Lagrangian integrable field
theories, and is invariant under the $Z_{2}$ symmetry $\phi\rightarrow-\phi$.
The spectrum of the model is made up by multi-particle states of a
single massive bosonic particle, whose exact mass at the quantum level
is denoted by $m$. Parameterizing the dispersion relations in terms
of the rapidity $\theta$, the energy and momentum of a single particle
excitation are $E(\theta)=m\cosh\theta$ and $p(\theta)=m\sinh\theta$
and the two-particle $S$-matrix is given by \cite{AFZ} 
\begin{equation}
S(\theta,B)\,=\,\frac{\tanh\frac{1}{2}(\theta-i\frac{\pi B}{2})}{\tanh\frac{1}{2}(\theta+i\frac{\pi B}{2})}\:,
\end{equation}
where $\theta=\theta_{1}-\theta_{2}$ is the relative rapidity of
the particles, and $B$ is the so-called renormalized coupling constant
\begin{equation}
B(g)\,=\,\frac{2g^{2}}{8\pi+g^{2}}\:.
\end{equation}
For real values of $g$ the $S$-matrix has no poles in the physical
sheet and hence there are no bound states.

A complete basis of eigenstates of this Quantum Field Theory is provided
by the $n$ particle asymptotic states 
\begin{equation}
|\theta_{1},\theta_{2},...,\theta_{n}\rangle_{in}=Z^{\dagger}(\theta_{1})Z^{\dagger}(\theta_{2})...Z^{\dagger}(\theta_{n})|\Omega\rangle,\qquad\theta_{1}>\theta_{2}>...>\theta_{n}\:,\label{eq:basis}
\end{equation}
where the operator $Z^{\dagger}(\theta)$ creates a particle excitation
with rapidity $\theta$ and $|\Omega\rangle$  is the vacuum state
of the theory. The creation and annihilation operators $Z^{\dagger}(\theta)$
and $Z(\theta)$ satisfy the Zamolodchikov-Faddeev algebra 
\begin{eqnarray}
Z^{\dagger}(\theta_{1})Z^{\dagger}(\theta_{2}) & = & S(\theta_{1}-\theta_{2})Z^{\dagger}(\theta_{2})Z^{\dagger}(\theta_{1})\:,\nonumber \\
Z(\theta_{1})Z(\theta_{2}) & = & S(\theta_{1}-\theta_{2})Z(\theta_{2})Z(\theta_{1})\:,\nonumber \\
Z(\theta_{1})Z^{\dagger}(\theta_{2}) & = & S(\theta_{2}-\theta_{1})Z^{\dagger}(\theta_{2})Z(\theta_{1})+\delta(\theta_{1}-\theta_{2})\boldsymbol{1}\:.\label{eq:ZF}
\end{eqnarray}
The form factors $F_{n}^{{\cal O}}$ of the Sinh-Gordon model are
matrix elements of a generic local operator ${\cal O}(0,0)$ between
the vacuum and a set of $n$ particle asymptotic states 
\begin{equation}
F_{n}^{{\cal O}}(\theta_{1},\theta_{2},\ldots,\theta_{n})\,=\,\langle\Omega\mid{\cal O}(0,0)\mid\theta_{1},\theta_{2},\ldots,\theta_{n}\rangle_{in}\,\,\,.\label{eq:formfactordef}
\end{equation}
Any other matrix element of the operator ${\cal O}(0,0)$ can be obtained
from the form factors above by exploiting the crossing symmetry of
quantum field theory. The latter imposes that matrix elements of ${\cal O}(0,0)$
between in- and out-states can be obtained by analytical continuation
of the out-state rapidities, as discussed in detail in Smirnov's book
\cite{Smirnov}:
\begin{eqnarray}
_{out}\langle\beta_{1},\beta_{2},...,\beta_{l}\mid & {\cal O}(0,0)\mid\theta_{1},\theta_{2},\ldots,\theta_{n}\rangle_{in}=F_{l+n}^{{\cal O}}(\theta_{1},\theta_{2},\ldots,\theta_{n},\beta_{1}-i\pi,\beta_{2}-i\pi,...,\beta_{l}-i\pi)\nonumber \\
 & +\sum_{{\Theta=\{\Theta_{1},\Theta_{2}\}\atop B=\{B_{1},B_{2}\}}}S(\Theta,\Theta_{1})S(B,B_{1})F_{l+n}^{{\cal O}}(B_{1},\Theta_{1}-i\pi+i\epsilon)\Delta(\Theta_{2},B_{2})\label{eq:general_matrix_element}
\end{eqnarray}
where the sum is over all possible splittings of the sets $\Theta=\{\theta_{1},...,\theta_{n}\}$
and $B=\{\beta_{1},...,\beta_{l}\}$ into non-overlapping subsets
$\Theta_{1},\Theta_{2}$ and $B_{1},B_{2}$ respectively (with $\Theta_{2}$
non-empty), $S(\Theta,\Theta_{1})$ and $S(B,B_{1})$ the S-matrix
products needed to reorder the rapidities and $\Delta(\Theta_{2},B_{2})=\langle\Theta_{2}\mid B_{2}\rangle$.

Note that using the translation operator $U=e^{-i(Ht-Px)}$ we can
always shift any operator ${\cal O}(x,t)$ to the origin ${\cal O}(x,t)=U^{\dagger}{\cal O}(0,0)U$.
Based on general properties of a Quantum Field Theory (as unitarity,
analyticity and locality) and on the relations (\ref{eq:ZF}), the
form factor bootstrap approach leads to a system of linear and recursive
equations for the matrix elements $F_{n}^{{\cal O}}$ \cite{Smirnov}
\begin{equation}
\begin{array}{cc}
 & F_{n}^{\mathcal{O}}(\theta_{1},\dots,\theta_{i},\theta_{i+1},\dots,\theta_{n})=F_{n}^{\mathcal{O}}(\theta_{1},\dots,\theta_{i+1},\theta_{i},\dots,\theta_{n})\, S(\theta_{i}-\theta_{i+1})\,\,,\\
 & F_{n}^{\mathcal{O}}(\theta_{1}+2\pi i,\dots,\theta_{n-1},\theta_{n})=F_{n}^{\mathcal{O}}(\theta_{2},\ldots,\theta_{n-1},\theta_{n},\theta_{1})\,\,,\\
 & -i\lim_{\tilde{\theta}\rightarrow\theta}(\tilde{\theta}-\theta)F_{n+2}^{\mathcal{O}}(\tilde{\theta}+i\pi,\theta,\ldots,\theta_{n})=\left(1-\prod_{i=1}^{n}S(\theta-\theta_{i})\right)\, F_{n}^{\mathcal{O}}(\theta_{1},\ldots,\theta_{n})\,\,\,.
\end{array}\label{axioms}
\end{equation}
In addition, for an operator ${\cal O}(x)$ of spin $s$, relativistic
invariance implies 
\begin{equation}
F_{n}^{{\cal O}}(\theta_{1}+\Lambda,\theta_{2}+\Lambda,\ldots,\theta_{n}+\Lambda)\,=\, e^{s\Lambda}\, F_{n}^{{\cal O}}(\theta_{1},\theta_{2},\ldots,\theta_{n})\,\,.
\end{equation}
Form factor solutions for the Sinh-Gordon model were constructed in
\cite{FMS,KM}; explicit expressions of the form factors $F_{n}^{\phi}(\theta_{1},\theta_{2},...,\theta_{n})$
for the elementary field $\phi$ can be found in the Supplementary
Material of \cite{STM}.

Observable operators can be written as an expansion in terms of the
Zamolodchikov-Faddeev operators \cite{Smirnov} (for a more recent
exposition in the framework of algebraic QFT cf. \cite{BC15}) 
\begin{equation}
\mathcal{O}=\sum_{l,n=0}^{\infty}\frac{1}{l!n!}\int\prod_{i=1}^{l}\frac{d\theta_{i}}{2\pi}\int\prod_{j=1}^{n}\frac{d\eta_{j}}{2\pi}f_{l,n}^{\mathcal{O}}(\theta_{1},\dots,\theta_{l}|\eta_{1},\dots\eta_{n})Z^{\dagger}(\theta_{1})\dots Z^{\dagger}(\theta_{l})Z(\eta_{1})\dots Z(\eta_{n})\:,\label{eq:genff_expansion}
\end{equation}
where the functions $f$ can be expressed in terms of the form factors
\begin{equation}
f_{l,n}^{\mathcal{O}}(\theta_{1},\dots,\theta_{l}|\eta_{1},\dots,\eta_{n})=F_{l+n}^{\mathcal{O}}(\theta_{l}+i\pi+i0,\dots,\theta_{1}+i\pi+i0,\eta_{n}-i0,\dots,\eta_{1}-i0)\label{eq:general_ff}
\end{equation}
and $Z^{\dagger}$, $Z$ are the Zamolodchikov-Faddeev creation and
annihilation operators, which satisfy the algebra (\ref{eq:ZF}).
It can be easily verified that the above expansion reproduces correctly
the form factors of the local field $\mathcal{O}$. 

The above formula is equivalent to (\ref{eq:general_matrix_element}).
To show that is sufficient to consider the general matrix element
of the operator $\mathcal{O}$, substitute (\ref{eq:genff_expansion}),
perform the contractions using the algebra (\ref{eq:ZF}) and compare
the result to (\ref{eq:general_matrix_element}). The simplest case
is the matrix elements with no particle in the out-state, as those
appearing in the definition of form factors (\ref{eq:formfactordef}).
We can easily see that only one term of the expansion (\ref{eq:genff_expansion})
contributes to such a matrix element and therefore we obtain

\begin{equation}
f_{0,n}^{\mathcal{O}}(|\theta_{1},\theta_{2},\dots,\theta_{n})=\langle\Omega\mid{\cal O}(0,0)\mid\theta_{n},\ldots,\theta_{2},\theta_{1}\rangle_{in}=F_{n}^{{\cal O}}(\theta_{n},\ldots,\theta_{2},\theta_{1})\,\,\,.\label{eq:extra1}
\end{equation}
Notice that the ordering of contractions imposes that the order of
rapidities gets inverted. Considering a general matrix element instead,
there are also disconnected contributions from all lower order terms
of the expansion. Therefore to determine the coefficients $f_{l,n}^{\mathcal{O}}$
one can start from the $l=0$ case obtained above and work recursively
in $l$, deriving each new one using all earlier derived coefficients.
It turns out that the disconnected contributions match one by one
the disconnected terms of Smirnov's formula, and we find that (\ref{eq:general_ff})
holds.

From (\ref{axioms}) the functions $f_{l,n}^{\mathcal{O}}$ satisfy
the permutation relations 
\begin{eqnarray}
f_{l,n}^{\mathcal{O}}(\ldots\theta_{i},\theta_{i+1}\ldots|\ldots) & = & S(\theta_{i+1}-\theta_{i})f_{l,n}^{\mathcal{O}}(\ldots\theta_{i+1},\theta_{i}\ldots|\ldots)\:,\nonumber \\
f_{l,n}^{\mathcal{O}}(\ldots|\ldots,\eta_{i},\eta_{i+1}\ldots) & = & S(\eta_{i+1}-\eta_{i})f_{l,n}^{\mathcal{O}}(\ldots|\ldots,\eta_{i+1},\eta_{i}\ldots)\:.\label{eq:S-exchange}
\end{eqnarray}

We also introduce the Dirichlet boundary states, defined as the eigenstates
of the field operator located at the temporal boundary $\phi(t=0)$.
Below we only need the Dirichlet state $|D\rangle$ that corresponds
to zero field value, i.e. defined by the condition

\begin{equation}
\phi(t=0)|D\rangle=0\:.\label{eq:D}
\end{equation}
This is an integrable boundary state, i.e. its exact expansion on
the eigenstate basis (\ref{eq:basis}) is of the squeezed vacuum form
\cite{GhoshalZamo} 
\begin{equation}
|D\rangle=\exp\left(\int_{0}^{\infty}d\theta\; K_{D}(\theta)Z^{\dagger}(-\theta)Z^{\dagger}(\theta)\right)|\Omega\rangle\:,\label{eq:boundary_state}
\end{equation}
with amplitude $K_{D}(\theta)$ given by 
\begin{equation}
K_{D}(\theta)=i\tanh(\theta/2)\frac{\cosh\left(\theta/2-i\pi B/8\right)\sinh(\theta/2+i\pi(B+2)/8)}{\sinh\left(\theta/2+i\pi B/8\right)\cosh(\theta/2-i\pi(B+2)/8)}\:.\label{eq:KD}
\end{equation}
This formula can be obtained by analytically continuing the result
for the first breather of the Sine-Gordon model, obtained in \cite{Ghoshal},
to imaginary coupling. Under such continuation the Sine-Gordon model
turns into the Sinh-Gordon model to all orders of perturbation theory,
and the first Sine-Gordon breather is mapped into the Sinh-Gordon
particle, both of which are physically identical to the elementary
particle created by the field.

\section{The integral equation hierarchy for the initial state \label{sec:The-integral-equation}}

\subsection{The infinite hierarchy}

We consider a quench in the Sinh-Gordon theory from an arbitrary initial
mass $m_{0}$ and zero initial interaction to arbitrary (renormalized)
mass $m$ and interaction coupling $g$. The initial state $|B\rangle$
is the ground state of the pre-quench Hamiltonian i.e. it is defined
through the condition that it is annihilated by the pre-quench annihilation
operators 
\begin{equation}
A(p)|B\rangle=0
\end{equation}
for all momenta $p$. It is convenient to express the annihilation
operators $A(p)$ in terms of the elementary field $\phi$ and the
canonical conjugate field $\pi$, which by continuity of the time
evolution through the quench satisfy the initial conditions 
\begin{equation}
\phi(x,t\to0^{-})=\phi(x,t\to0^{+})\quad\text{and}\quad\pi(x,t\to0^{-})=\pi(x,t\to0^{+})
\end{equation}
This can be done easily since the pre-quench annihilation operators
$A(p)$ correspond to a free field theory and are therefore given
in terms of $\phi$ and $\pi$ by the standard relation 
\begin{equation}
A(p)=\frac{1}{\sqrt{2E_{0}(p)}}\left(E_{0}(p)\hat{\phi}(p)+i\hat{\pi}(p)\right)\:,\label{cao-1}
\end{equation}
where 
\begin{equation}
\hat{\phi}(p)=\int dx\, e^{-ipx}\phi(x),\label{eq:Fourier}
\end{equation}
is the Fourier transform of $\phi(x)$ and
\begin{equation}
E_{0}(p)=\sqrt{p^{2}+m_{0}^{2}}\label{eq:E0}
\end{equation}
is the energy of a pre-quench excitation with momentum $p$. Moreover,
from the relation $\pi=\partial_{t}\phi=-i[\phi,H]$ where $H$ is
the post-quench Hamiltonian, we obtain the equation

\begin{equation}
\left\{ \hat{\phi}(p)+\frac{1}{E_{0}(p)}\left[\hat{\phi}(p),H\right]\right\} |B\rangle=0\:,\label{eq:hierarchy_opform}
\end{equation}
This equation was first derived in \cite{STM} and, as we will now
see, it has the advantage that all operators are straightforwardly
expressible in the framework of the post-quench field theory, i.e.
the Sinh-Gordon theory.

Expanding the initial state on a complete set of states we have
\begin{equation}
|B\rangle=\sum_{r=0}^{\infty}\frac{1}{r!}\int\prod_{j=1}^{r}\frac{d\beta_{j}}{2\pi}K_{r}(\beta_{1},\dots,\beta_{r})|\beta_{1},\dots,\beta_{r}\rangle\:,\label{eq:gen_state}
\end{equation}
where $|\beta_{1},\dots,\beta_{r}\rangle$ are the multi-particle
eigenstates of the post-quench Hamiltonian with eigenvalues 
\begin{equation}
\sum_{j=1}^{r}E(\beta_{j})=\sum_{j=1}^{r}m\cosh\beta_{j}\:.
\end{equation}
Without loss of generality, the functions $K_{r}(\beta_{1},\dots,\beta_{r})$
are chosen to satisfy the symmetry relation 
\begin{equation}
K_{r}(\beta_{1},\dots,\beta_{i},\beta_{i+1},\dots,\beta_{r})=K_{r}(\beta_{1},\dots,\beta_{i+1},\beta_{i},\dots,\beta_{r})\, S(\beta_{i+1}-\beta_{i})\:.\label{eq:K_permut}
\end{equation}
by virtue of the commutation rules of the Zamolodchikov-Faddeev algebra
(\ref{eq:ZF}). Projecting (\ref{eq:hierarchy_opform}) to all the
linearly independent states of the post-quench Hilbert space, we can
write an infinite hierarchy of integral equations for the functions
$K_{r}$ 
\begin{equation}
\langle\theta_{1},\dots,\theta_{N}|\left\{ \hat{\phi}(p)+\frac{1}{E_{0}(p)}\left[\hat{\phi}(p),H\right]\right\} |B\rangle=0\label{eq:hierarchy_general}
\end{equation}
by taking the inner product with all possible post-quench eigenstates
$|\theta_{1},\dots,\theta_{N}\rangle$. We expect that this set of
equations determines all the functions $K_{r}$ uniquely (up to overall
normalization).

Note that due to translational invariance of the global quench, the
initial state $|B\rangle$ is in the subspace of zero total momentum,
i.e all amplitudes $K_{r}$ contain a delta function $\delta\bigl(\sum_{j=1}^{r}m\sinh\beta_{j}\bigr)$.
As a result, the nontrivial equations are only obtained for 
\begin{equation}
p=-\sum_{j=1}^{N}m\sinh\theta_{j}\:,\label{eq:pvalue}
\end{equation}
i.e. when the momentum $p$ is opposite to the total momentum of the
test state. Also note that the number $N$ of particles in the test
state must be chosen odd, otherwise the equations are trivially satisfied.
This can be seen easily from (\ref{eq:hierarchy_general}) since the
operator $\hat{\phi}(p)+\bigl[\hat{\phi}(p),H\bigr]/E_{0}(p)$ is
odd i.e. antisymmetric under the transformation $\phi\to-\phi$, while
the state $|B\rangle$ is even since the pre-quench Hamiltonian is
symmetric under the same transformation. This means in particular
that it contains only excitations with even number of particles, i.e.
all odd amplitudes $K_{r}$ in (\ref{eq:gen_state}) must vanish.

\subsection{The hierarchy in terms of form factors}

To derive explicit expressions for equations of the hierarchy (\ref{eq:hierarchy_general})
we substitute the expansion (\ref{eq:genff_expansion}) for the operator
$\hat{\phi}(p)+\bigl[\hat{\phi}(p),H\bigr]/E_{0}(p)$ and the general
expansion (\ref{eq:gen_state}) of the state $|B\rangle$ into (\ref{eq:hierarchy_general}).
Then the general equation with an $N$ particle test state is the
sum of all possible contractions between the $N$ particles of the
test state with the operator $\hat{\phi}(p)+\bigl[\hat{\phi}(p),H\bigr]/E_{0}(p)$
and the state $|B\rangle$. 

From (\ref{eq:Fourier}) and using the translation operator $e^{iPx}$
(where $P$ is the momentum operator) to shift the field $\phi(x)$
to the origin, we can easily find that 
\begin{align}
\langle\theta_{1},\dots,\theta_{l}| & \hat{\phi}(p)|\eta_{1},\dots,\eta_{n}\rangle=\int dx\, e^{-ipx}\langle\theta_{1},\dots,\theta_{l}|\phi(x)|\eta_{1},\dots,\eta_{n}\rangle\nonumber \\
 & =\int dx\, e^{-ipx}\langle\theta_{1},\dots,\theta_{l}|e^{-iPx}\phi(0)e^{iPx}|\eta_{1},\dots,\eta_{n}\rangle\nonumber \\
 & =\int dx\,\exp\left(-ipx-ix\sum_{i=1}^{l}m\sinh\theta_{i}+ix\sum_{j=1}^{n}m\sinh\eta_{j}\right)\langle\theta_{1},\dots,\theta_{l}|\phi(0)|\eta_{1},\dots,\eta_{n}\rangle\nonumber \\
 & =2\pi\delta\bigl(p+\sum_{i=1}^{l}m\sinh\theta_{i}-\sum_{j=1}^{n}m\sinh\eta_{j}\bigr)\langle\theta_{1},\dots,\theta_{l}|\phi(0)|\eta_{1},\dots,\eta_{n}\rangle\label{eq:Ftff}
\end{align}
so that the expansion of $\hat{\phi}(p)$ is
\begin{align}
\hat{\phi}(p) & =\sum_{l,n=0}^{\infty}\frac{1}{l!n!}\int\prod_{i=1}^{l}\frac{d\theta_{i}}{2\pi}\int\prod_{j=1}^{n}\frac{d\eta_{j}}{2\pi}\,2\pi\delta\bigl(p+\sum_{i=1}^{l}m\sinh\theta_{i}-\sum_{j=1}^{n}m\sinh\eta_{j}\bigr)\nonumber \\
 & \quad\times\, f_{l,n}^{\mathcal{\phi}}(\theta_{1},\dots,\theta_{l}|\eta_{1},\dots\eta_{n})Z^{\dagger}(\theta_{1})\dots Z^{\dagger}(\theta_{l})Z(\eta_{1})\dots Z(\eta_{n})\:,\label{eq:FTexpansion}
\end{align}
with expansion coefficients $f_{l,n}^{\phi}(\theta_{1},\dots,\theta_{l}|\eta_{1},\dots\eta_{n})$
given by the form factors of the elementary Sinh-Gordon field $\phi$
according to (\ref{eq:general_ff}). We also need the relations

\begin{equation}
\langle\theta_{1},\dots,\theta_{N}|H=\langle\theta_{1},\dots,\theta_{N}|\left(\sum_{s=1}^{N}E(\theta_{s})\right)\:,
\end{equation}
and 
\begin{equation}
H|B\rangle=\sum_{r=0}^{\infty}\frac{1}{r!}\int\prod_{j=1}^{r}\frac{d\beta_{j}}{2\pi}\left(\sum_{k=1}^{r}E(\beta_{k})\right)K_{r}(\beta_{1},\dots,\beta_{r})|\beta_{1},\dots,\beta_{r}\rangle\:,
\end{equation}
that give the action of the Hamiltonian operator $H$ on the bra and
ket states. We can now substitute the above relations to (\ref{eq:hierarchy_general})
\begin{align}
 & \sum_{l,n,r=0}^{\infty}\frac{1}{l!n!r!}\int\prod_{i=1}^{l}\frac{d\zeta_{i}}{2\pi}\int\prod_{j=1}^{n}\frac{d\eta_{j}}{2\pi}\int\prod_{k=1}^{r}\frac{d\beta_{k}}{2\pi}\,2\pi\delta\bigl(p+\sum_{i=1}^{l}m\sinh\zeta_{i}-\sum_{j=1}^{n}m\sinh\eta_{j}\bigr)\nonumber \\
 & \quad\times f_{l,n}^{\phi}(\zeta_{1},\dots,\zeta_{l}|\eta_{1},\dots\eta_{n})\left(E_{0}(p)-\sum_{s=1}^{N}E(\theta_{s})+\sum_{k=1}^{r}E(\beta_{k})\right)K_{r}(\beta_{1},\dots,\beta_{r})\nonumber \\
 & \quad\times\langle\theta_{1},\dots,\theta_{N}|Z^{\dagger}(\zeta_{1})\dots Z^{\dagger}(\zeta_{l})Z(\eta_{1})\dots Z(\eta_{n})|\beta_{1},\dots,\beta_{r}\rangle=0\:.\label{eq:hierarchy_explct}
\end{align}
and use the Zamolodchikov-Faddeev algebra (\ref{eq:ZF}) to perform
the contractions. From the latter it is easy to see that the matrix
element 
\begin{equation}
\langle\theta_{1},\dots,\theta_{N}|Z^{\dagger}(\zeta_{1})\dots Z^{\dagger}(\zeta_{l})Z(\eta_{1})\dots Z(\eta_{n})|\beta_{1},\dots,\beta_{r}\rangle
\end{equation}
is only non-zero when all $\eta_{j}$ are contracted with some of
the $\beta_{k}$ and all $\zeta_{i}$ are contracted with some of
the $\theta_{s}$, while the remaining $\beta_{k}$ are contracted
with the remaining $\theta_{s}$. This means in particular that $l$
and $n$ are constrained to vary between the values $0\le l\le N$,
$0\le n\le r$ and to satisfy the equation $N-l+n-r=0$. These conditions
restrict the sums in the above equation. Moreover, according to earlier
comments, $l+n$ must be an odd positive integer, $r$ must be even
and $N$ odd. There are $r!/(r-n)!$ ways to contract $\eta_{j}$
and $\beta_{k}$, $N!/(N-l)!$ ways to contract $\zeta_{i}$ and $\theta_{s}$
and $(r-n)!$ ways to contract the remaining $\beta_{k}$ and $\theta_{s}$,
which are all equivalent up to S-matrix factors due to permutations
of the rapidities. Note that S-matrix factors due to permutations
of \emph{integrated} rapidities can always be absorbed by re-ordering
the rapidities to a fixed ordering using (\ref{eq:S-exchange}) and
(\ref{eq:K_permut}) and by renaming them since they are dummy integration
variables. However there still remain S-matrix factors corresponding
to the permutations of the \emph{test particle} rapidities $\theta_{s}$
that are necessary in order to choose those that are contracted with
$\zeta_{i}$. That is, there are $\binom{N}{l}$ distinct choices
that give terms multiplied by different S-matrix products. Overall
the number of combinations of contractions that give equal contributions
is 
\begin{equation}
\frac{r!}{(r-n)!}\times\frac{N!}{(N-l)!}\times(r-n)!/\binom{N}{l}=l!r!
\end{equation}

After all contractions have been performed, (\ref{eq:hierarchy_explct})
becomes

\begin{align}
 & \sum_{r=0}^{\infty}\sum_{n=0}^{r}\frac{1}{n!}\int\prod_{k=1}^{n}\frac{d\beta_{k}}{2\pi}\,2\pi\delta\bigl(p+\sum_{s=r-n+1}^{N}m\sinh\theta_{s}-\sum_{k=1}^{n}m\sinh\beta_{k}\bigr)\nonumber \\
 & \quad\times f_{N+n-r,n}^{\phi}(\theta_{N},\dots,\theta_{r-n+1}|\beta_{n},\dots,\beta_{1})\left(E_{0}(p)-\sum_{s=r-n+1}^{N}E(\theta_{s})+\sum_{k=1}^{n}E(\beta_{k})\right)\nonumber \\
 & \quad\times K_{r}(\beta_{1},\dots,\beta_{n},\theta_{r-n},\dots,\theta_{1})+\text{perm.s}=0\:,\label{eq:hierarchy_explct-1}
\end{align}
where ``perm.s'' denotes the other choices of splitting the test
particle rapidities $\theta_{s}$ into those contracted with $\zeta_{i}$
and those contracted with $\beta_{k}$, which contain S-matrix products
as explained above. Taking into account that the amplitudes $K_{r}$
contain a factor $\delta\bigl(\sum_{j=1}^{r}m\sinh\beta_{j}\bigr)$
expressing the translational invariance of $|B\rangle$, the $\delta$-function
in the above equation can be replaced by $\delta\bigl(p+\sum_{s=1}^{N}m\sinh\theta_{s}\bigr)$
which can be taken out of the integral and sum. This overall factor
means that, as mentioned earlier, nontrivial equations are only those
for which $p$ is opposite to the total momentum of the test state
(\ref{eq:pvalue}). Provided that this condition is fulfilled, the
final form of the $N$ test particle equation is 
\begin{align}
 & \sum_{r=0}^{\infty}\sum_{n=0}^{r}\frac{1}{n!}\int\prod_{k=1}^{n}\frac{d\beta_{k}}{2\pi}\,\nonumber \\
 & \quad\times f_{N+n-r,n}^{\phi}(\theta_{N},\dots,\theta_{r-n+1}|\beta_{n},\dots,\beta_{1})\left(E_{0}(p)-\sum_{s=r-n+1}^{N}E(\theta_{s})+\sum_{k=1}^{n}E(\beta_{k})\right)\nonumber \\
 & \quad\times K_{r}(\beta_{1},\dots,\beta_{n},\theta_{r-n},\dots,\theta_{1})+\text{perm.s}=0\:,\label{eq:hierarchy_explct-2}
\end{align}
In particular, for a single test particle, there are only two possible
values for $n$, either zero or one, and so $n=r-1$ or $n=r$ respectively.
Therefore the $N=1$ equation is 
\begin{align}
 & f_{1,0}^{\phi}(\theta|)\Big(E_{0}(p(\theta))-E(\theta)\Big)\nonumber \\
 & +\sum_{{r=2\atop \text{even}}}^{\infty}\frac{1}{(r-1)!}\int\prod_{k=1}^{r-1}\frac{d\beta_{k}}{2\pi}\Biggl[f_{0,r-1}^{\phi}(|\beta_{r-1},\dots,\beta_{1})\left(E_{0}(p(\theta))+\sum_{k=1}^{r-1}E(\beta_{k})\right)K_{r}(\beta_{1},\dots,\beta_{r-1},\theta)\nonumber \\
 & \quad+\frac{1}{r}\int\frac{d\beta_{r}}{2\pi}\, f_{1,r}^{\phi}(\theta|\beta_{r},\dots,\beta_{1})\left(E_{0}(p(\theta))-E(\theta)+\sum_{k=1}^{r}E(\beta_{k})\right)K_{r}(\beta_{1},\dots,\beta_{r})\Biggr]=0\:.\label{eq:one-test-particle}
\end{align}
with $p(\theta)=m\sinh\theta$ and therefore $E_{0}(p(\theta))=\sqrt{m^{2}\sinh^{2}\theta+m_{0}^{2}}$.
In the limit of large initial mass $m_{0}\to\infty$, the factors
in round brackets in the above equations are dominated by $E_{0}(p(\theta))\sim m_{0}$
and therefore they can be replaced by $m_{0}$, which, being an overall
factor multiplying the equation, can be omitted. As already explained,
in this limit the equations must be satisfied by the Dirichlet state
given by (\ref{eq:boundary_state}) and (\ref{eq:KD}).

\begin{figure}
\begin{centering}
\includegraphics[width=0.7\textwidth]{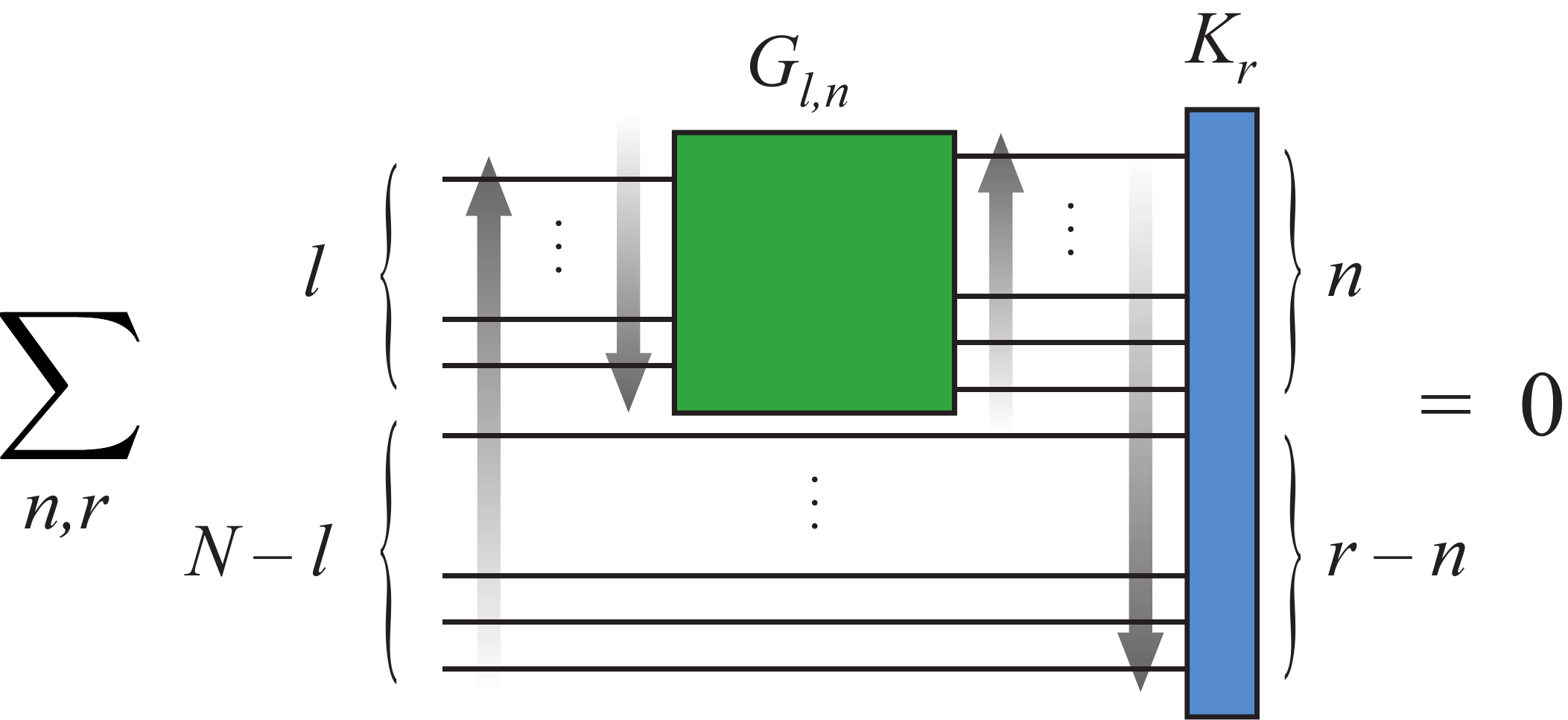}
\par\end{centering}

\protect\caption{Diagrammatic representation of the hierarchy of integral equations
(\ref{eq:hierarchy_explct-2}). The external lines on the left correspond
to the $N$ rapidities of the test state $\langle\theta_{1},\dots,\theta_{N}|$.
The green square represents the $(l,n)$-order term in the expansion
of the operator $\hat{\phi}(p)+\bigl[\hat{\phi}(p),H\bigr]/E_{0}(p)$,
while the blue rectangle on the right represents the $r$-order term
in the expansion of the initial state $|B\rangle$. The sum is over
$r$ and $n\leq r$, while $l$ is fixed to $N+n-r$. The arrows show
the ordering of contracted rapidities in equation (\ref{eq:hierarchy_explct-2}).}
\label{fig-1}
\end{figure}

The above equations can be represented diagrammatically as in Figure
\ref{fig-1}. The terms 
\begin{equation}
G_{l,n}(\theta_{N},\dots,\theta_{r-n+1}|\beta_{n},\dots,\beta_{1})\equiv f_{l,n}^{\phi}(\theta_{N},\dots,\theta_{r-n+1}|\beta_{n},\dots,\beta_{1})\left(E_{0}(p)-\sum_{s=r-n+1}^{N}E(\theta_{s})+\sum_{k=1}^{n}E(\beta_{k})\right)
\end{equation}
in the expansion of the operator $\hat{\phi}(p)+\bigl[\hat{\phi}(p),H\bigr]/E_{0}(p)$
are represented as square boxes with $l$ legs on the left and $n$
legs on the right, which should be contracted respectively with the
$N$ rapidities of the test state, represented as external legs on
the left side of the graph, and with the $r$ rapidities of $K_{r}$,
represented as legs emerging from the rectangular box on the right.
The remaining rapidities of $K_{r}$ should be contracted with the
remaining rapidities of the test state. The sum is over all possible
orders $n,r$ and all possible combinations of contractions. Note
that in principle the ordering of rapidities in such graphs matters
and that exchange of two consecutive rapidities results in multiplication
with the corresponding S-matrix factor.

\begin{figure}
\begin{centering}
\includegraphics[scale=0.5]{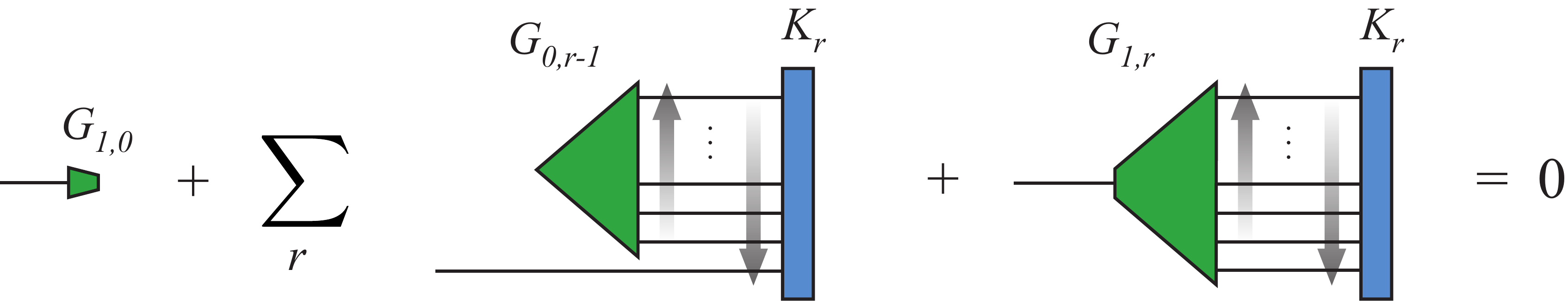}
\par\end{centering}

\protect\caption{Diagrammatic representation of one test particle equation (\ref{eq:one-test-particle}).}
\label{fig-1-1}
\end{figure}
The above derivation was based directly on the \emph{infinite volume}
formulation of Integrable Field Theories through the crucial use of
the Zamolodchikov-Faddeev operators and the related operator and state
expansions. In order to verify the validity of the equations from
the finite volume formulation of integrable models, in Appendix \ref{sec:Finite-volume-regularization}
we present an alternative derivation of the above equations (up to
a certain order) based on the Bethe-Yang equations. We compare the
two systems of equations and confirm that they are indeed identical.

\subsection{The hierarchy for free theory: Bogoliubov transformation}

To argue for the statement that the hierarchy uniquely determines
the initial state up to normalization, consider first the case when
the interaction is zero, corresponding to the free bosonic theory
with the Hamiltonian

\begin{equation}
H=\frac{1}{2}\int[\pi^{2}+(\nabla\phi)^{2}+m\phi^{2}]\, dx\:,\label{hamiltonian}
\end{equation}
where the canonical momentum is $\pi=\partial_{t}\phi.$ We can define
creation and annihilation operators as follows%
\footnote{For the free case it is convenient to parametrize the modes using
the momentum instead of the rapidity variable.%
}

\begin{equation}
A(p)=\frac{1}{\sqrt{2E(p)}}\left(E(p)\hat{\phi}(p)+i\hat{\pi}(p)\right)\:,\label{cao}
\end{equation}
where $E(p)=\sqrt{m^{2}+p^{2}}$ and the canonical commutation relations
are 
\begin{equation}
[A(p),A^{\dagger}(p')]=2\pi\delta(p-p')\:.
\end{equation}
Now consider a global quantum quench when at time $t=0$ the mass
parameter is abruptly changed in the theory $m_{0}\rightarrow m$.
As the time evolution of the fields $\phi$ and $\pi$ is continuous,
taking their spatial Fourier transform at time $t=0$ and defining
the creation and annihilation operators with them, the following equations
must hold 
\begin{equation}
\begin{aligned}\frac{1}{\sqrt{2E_{0}(p)}}\left(A_{0}(p)+A_{0}^{\dagger}(-p)\right) & =\frac{1}{\sqrt{2E(p)}}\left(A(p)+A^{\dagger}(-p)\right)\\
\frac{-i}{2}\sqrt{2E_{0}(p)}\left(A_{0}(p)-A_{0}^{\dagger}(-p)\right) & =\frac{-i}{2}\sqrt{2E(p)}\left(A(p)-A^{\dagger}(-p)\right)
\end{aligned}
\:,\label{caoeq}
\end{equation}
where $E_{0}(p)=m_{0}^{2}+p^{2}$, $E(p)=m^{2}+p^{2}$ and $A_{0}(p)$
and $A(p)$ are the pre- and post-quench mode operators. This is nothing
but the familiar Bogoliubov transformation, which allows one to express
$|B\rangle$ using the fact that it is the pre-quench ground state,
i.e. $A_{0}(p)|B\rangle=0$. We now demonstrate that the usual expression
for the initial state can also be obtained from the integral equation
hierarchy. For better transparency of the subsequent manipulations
we start directly with the operator form of the hierarchy (\ref{eq:hierarchy_opform})
and go through essentially the same steps that yielded (\ref{eq:hierarchy_explct-2})
from (\ref{eq:hierarchy_opform}).

Using (\ref{cao}), the operator form of the hierarchy can be written
as 
\begin{equation}
\left(\frac{1}{\sqrt{2E(p)}}(A(p)+A^{\dagger}(-p))+\frac{\sqrt{2E(p)}}{2E_{0}(p)}(A(p)-A^{\dagger}(-p))\right)|B\rangle=0\:.\label{wi}
\end{equation}
The fact that $A(p)$ and $A^{\dagger}(-p)$ occur together means
that the equations only link components which differ by pairs of particles
of opposite momenta. The lowest component is the post-quench vacuum,
so $|B\rangle$ can be expressed in terms of states composed entirely
of pairs of particles with opposite momenta. This allows us to write
the amplitude $K_{2n}^{free}(-k_{1},k_{1}\dots-k_{n},k_{n})$ as a
function of only $n$ variables $K_{2n}^{free}(k_{1}\dots k_{n}),$
hence 
\begin{equation}
|B\rangle=\mathcal{N}\sum_{n=0}^{\infty}\left(-\frac{1}{2}\right)^{n}\int_{-\infty}^{+\infty}\left(\prod_{i=1}^{n}\frac{\, dk_{i}}{2\pi}A^{\dagger}(k_{i})A^{\dagger}(-k_{i})\right)K_{2n}^{free}(k_{1}\dots k_{n})|0\rangle\:.\label{exp}
\end{equation}
Due to Bose symmetry, the functions $K_{2n}^{free}$ can be taken
to be invariant under permutations of their arguments and under $k_{i}\rightarrow-k_{i}$.

\begin{figure}
\begin{centering}
\includegraphics[width=0.5\textwidth]{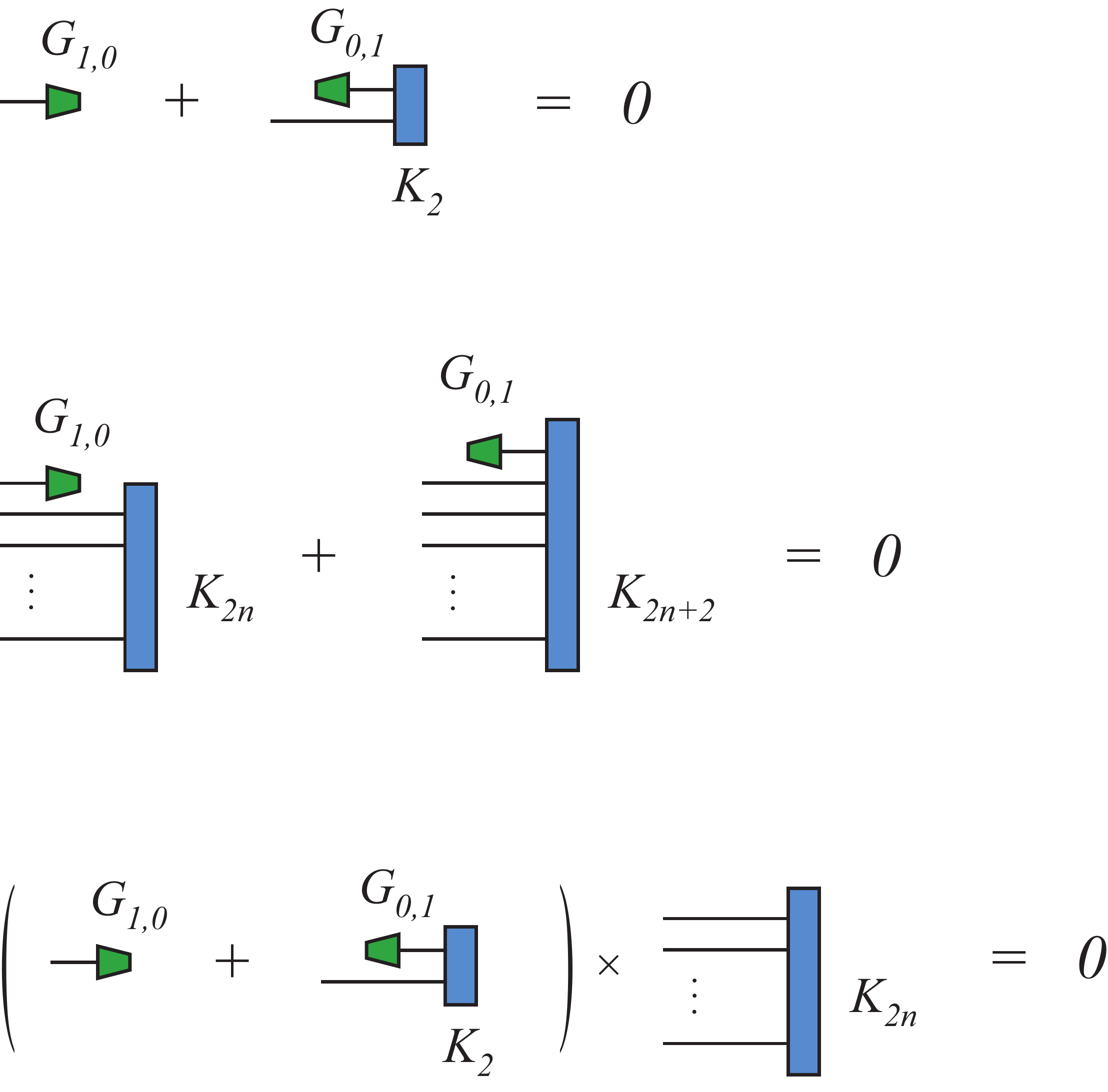}
\par\end{centering}

\protect\caption{\label{fig-2} Diagrammatic expansion of the integral equations for
a free theory. The first line represents the one-particle equation,
while the second line shows the general multi-particle equation. As
can be easily seen in the third line, a factorized solution $K_{2n+2}\sim K_{2n}\times K$
automatically satisfies the multi-particle equation which then reduces
to the one-particle equation for the function $K$. The latter is
therefore equal to the pair amplitude $K=K_{2}$. This explains the
exponential form of the solution (\ref{exp2}).}
\end{figure}

In order to find $K_{2n}$, eqn. (\ref{wi}) can be further specified
by applying a given test state on the left side of eqn. (\ref{wi}),
which gives the individual equations of the hierarchy. To find $K_{2}$,
we apply a one-particle test state $\langle p_{1}|=\langle0|A(p_{1})$
to eqn. (\ref{wi}) and substitute the expansion of (\ref{exp}) to
obtain 
\begin{eqnarray}
 &  & \left(\frac{1}{E(p)}-\frac{1}{E_{0}(p)}\right)\langle0|A(p_{1})A^{\dagger}(-p)|0\rangle\label{vmi2}\\
 &  & -\frac{1}{2}\left(\frac{1}{E(p)}+\frac{1}{E_{0}(p)}\right)\int_{-\infty}^{+\infty}\,\frac{dk}{2\pi}\langle0|A(p_{1})A(p)A^{\dagger}(k)A^{\dagger}(-k)|0\rangle K_{2}^{free}(k)=0\:,\nonumber 
\end{eqnarray}
where the matrix elements can be evaluated using the canonical commutation
relations (or, equivalently, Wick's theorem) which leads to
\begin{equation}
\left(\frac{1}{E(p)}-\frac{1}{E_{0}(p)}\right)-\left(\frac{1}{E(p)}+\frac{1}{E_{0}(p)}\right)K_{2}^{free}(p)=0\:,
\end{equation}
from which 
\begin{equation}
K_{2}^{free}(p)=\frac{E_{0}(p)-E(p)}{E_{0}(p)+E(p)}\label{K2}
\end{equation}
follows. 

To determine the functional form of $K_{2n}^{free}(k)$ for a general
$n$, test states with higher number of particles are applied on the
left side of (\ref{wi}); only test states with odd numbers of particles
give a nonzero result. Let there be $2n-1$ particles in the test
state; then only two terms from (\ref{exp}) give a nonzero contribution,
which can be attributed to $\hat{\phi}(p)$ containing only one creation
and one annihilation operator in its expansion (\ref{eq:FTexpansion})
in the free theory, or equivalently to $\hat{\phi}(p)$ having non-zero
form factors $f_{l,n}^{\mathcal{\phi}}$ only when $l=1$, $n=0$
or $l=0$, $n=1$. These considerations lead to
\begin{equation}
\left[nK_{2n}^{free}(k_{1},\dots k_{n-1},p)\left(\frac{1}{E(p)}+\frac{1}{E_{0}(p)}\right)-K_{2(n-1)}^{free}(k_{1},\dots k_{n-1})\left(\frac{1}{E(p)}-\frac{1}{E_{0}(p)}\right)\right]=0\:.\label{vmi8}
\end{equation}
Therefore 
\begin{equation}
\frac{K_{2n}^{free}(k_{1},\dots k_{n-1},p)}{K_{2(n-1)}^{free}(k_{1},\dots k_{n-1})}=\frac{1}{n}\frac{E_{0}(p)-E(p)}{E_{0}(p)+E(p)}\:.\label{vmi9}
\end{equation}
Due to the symmetry properties of $K_{2n}^{free}$, and using (\ref{K2})
\begin{equation}
K_{2n}^{free}(k_{1},\dots k_{n})=\frac{1}{n!}\prod_{i=1}^{n}K_{2}^{free}(k_{i})\:,\label{K2n}
\end{equation}
that is 
\begin{equation}
|B\rangle=\mathcal{N}\exp\left[-\frac{1}{2}\int_{-\infty}^{+\infty}\frac{\, dk}{2\pi}K^{free}(k)A{}^{\dagger}(k)A^{\dagger}(-k)\right]|0\rangle\label{exp2}
\end{equation}
where 
\begin{equation}
K^{free}(k)=K_{2}^{free}(k)=\frac{E_{0}(k)-E(k)}{E_{0}(k)+E(k)}\:.
\end{equation}
This is exactly the squeezed state form that can also be obtained
via a direct application of the Bogoliubov transformation (\ref{caoeq})
to the condition $A_{0}(p)|B\rangle=0$ (see e.g. \cite{SGS13} for
a derivation in the quantum quench context).

\subsection{Uniqueness of the solution for the interacting case\label{sub:Uniqueness-of-the}}

As we have seen, the hierarchy (\ref{eq:hierarchy_general}) has a
unique solution (up to normalization) for a mass quench in a free
field theory, which coincides with the well-known squeezed state resulting
from the Bogoliubov transformation. We now argue that for the interacting
case (Sinh-Gordon theory) we expect the same uniqueness property. 

The interacting systems is much more involved than the free one since
each of the equations (\ref{eq:hierarchy_explct}) involve all of
the $K_{r}$ excitation amplitudes and therefore it does not have
the chain-like organization of the equations of the free case, where
only two terms were present for each equation, allowing us to calculate
each amplitude of arbitrary order one-by-one from its predecessors.
However, since the zero-coupling limit gives back the free equation,
and at least for a small enough value of the coupling the dynamics
and form factors of Sinh-Gordon theory are known to be well-described
by perturbation theory, the introduction of a small coupling does
not spoil the uniqueness of the solution. If anything, we expect the
solution to be slightly deformed from the free solution, but to preserve
much of its properties. Therefore if a given Ansatz can give a solution,
or at least a very good approximation of a solution, we can argue
it is close to the unique solution of the hierarchy. 

We note that a remnant of the chain structure is still present in
the interacting case. Denoting the number of particles in the bra
state by $n_{L}$ and in the ket state by $n_{R}$, for a free field
theory the only nonzero matrix elements are the ones with $n_{L}-n_{R}=\pm1$.
From perturbation theory, it is clear that these terms dominate for
weak coupling. However, in a two-dimensional field theory this argument
goes even further: as a simple consideration of available phase space
shows, the Feynman graphs corresponding to form factors with larger
differences in the number of incoming and outgoing particles are suppressed.
The reason for this is that the difference in number of particles
can be interpreted as particle creation by the operator inserted;
however, in two space time dimensions the available phase space eventually
decreases with increasing particle multiplicity \cite{giuseppe_book},
therefore processes involving a change in particle number are suppressed;
the larger the change, the stronger the suppression. As a result,
we expect that the terms are hierarchically organized by the value
of $\Delta n=n_{L}-n_{R}$, the ones with $\Delta n=\pm1$ being the
largest, followed by the terms with $\Delta n=\pm3$ and so on%
\footnote{Note that $\Delta n$ must always be odd, as even form factors of
the elementary field $\phi$ vanish.%
}. This fact is important for our ability to treat the infinitely many
integral equations, each composed of an infinite number of terms,
making up the hierarchy, as it implies that they can be well-approximated
by equations that are truncated to a finite number of terms, and more
terms can be gradually included to improve the approximation.

\section{General properties of the initial state \label{sec:General-properties-of}}

\subsection{Pair structure and exponentiation of the initial state}

In the introduction we presented an argument for the exponentiation
of the initial state, based on RG theory. More specifically we argued
that the initial state after a quench in an IFT can be built by inclusion
of the IFT charges $Q_{s}$ as perturbations to the Dirichlet state:
$e^{-\sum_{s}Q_{s}\tau_{s}}|D\rangle$. Such perturbations preserve
the exponential form and merely modify the pair excitation amplitude
$K(\theta)$. Below we will show an independent argument based on
two general properties of the initial state: the extensivity of the
charges in the initial state and its pair structure. The first property
is due to the fact that the charges are local and the quench is a
change of a global parameter. The second property is expected to hold
from perturbation theory considerations. Combination of the two properties
leads to exponentiation as described by (\ref{sv}) and simplifies
our search for a solution to the hierarchy equations.

\subsubsection{Requirements from extensivity of local charges}

We consider a general state $|\Psi\rangle$ that satisfies translational
invariance and therefore contains only excitations with zero total
momentum. Without loss of generality we can write such a state in
the form of a cumulant expansion 

\begin{equation}
|\Psi\rangle=\exp\left(\sum_{r=1}^{\infty}\int K_{r}^{\Psi}(\theta_{1},\theta_{2},\dots,\theta_{r})\prod_{i=1}^{r}Z^{\dagger}(\theta_{i})d\theta_{i}\right)|0\rangle\:,\label{eq:psi}
\end{equation}
where the amplitudes $K_{r}^{\Psi}(\theta_{1},...,\theta_{r})$ contain
a factor $\delta(\sum_{i}p(\theta_{i}))$ which expresses the conservation
of total momentum due to the translation invariance. We further assume
that the state $|\Psi\rangle$ is $Z_{2}$ invariant, i.e. symmetric
under the transformation $\phi\to-\phi$, therefore it must contain
only excitations with even number of particles or, in other words,
all odd amplitudes $K_{2n+1}^{\Psi}(\theta_{1},...,\theta_{2n+1})$
must vanish.

We will show that the expectation values of the local charges 
\begin{equation}
Q_{s}=\int d\theta\, e^{s\theta}Z^{\dagger}(\theta)Z(\theta)
\end{equation}
in the above state are extensive quantities (i.e. they increase linearly
with the system size $L$) only if the amplitudes $K_{2n}$ do not
contain any other $\delta$-function factor except of the one that
accounts for the translation invariance, $\delta(\sum_{i}p(\theta_{i}))$.
The initial state after a quantum quench, by definition, must satisfy
the requirement of extensive local charges. Indeed, the charges $Q_{s}$,
being spatial integrals of local operators, are extensive thermodynamic
quantities for all states that satisfy the cluster decomposition principle,
as all ground states of physical Hamiltonians do.

\begin{figure}
\begin{centering}
\includegraphics[width=0.8\textwidth]{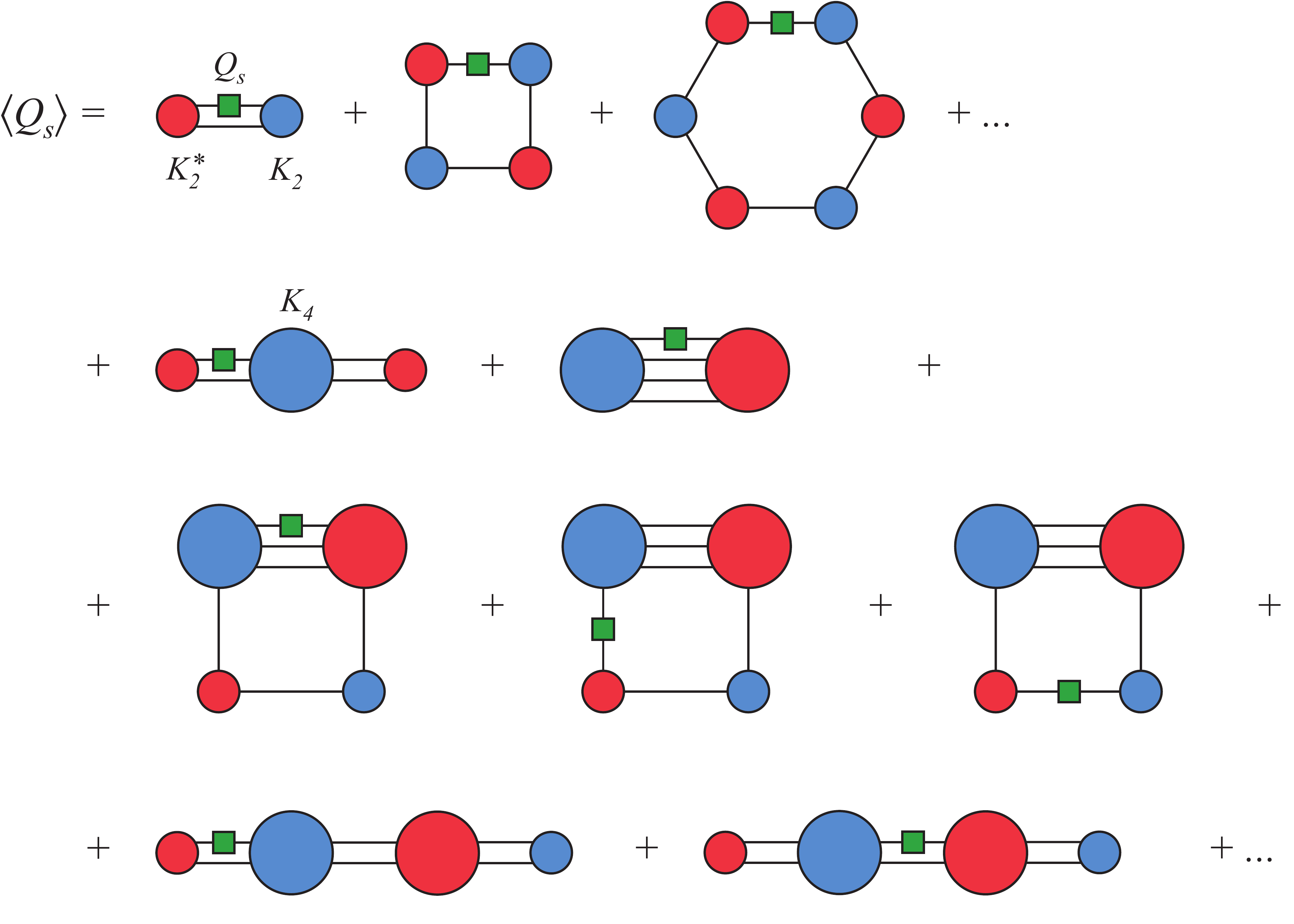}
\par\end{centering}

\protect\caption{\label{fig}Diagrammatic expansion of the expectation value of a local
charge in the state $|B\rangle$. The bra-excitations (red circles)
and the ket-excitations (blue circles) must be contracted with each
other and with the charge operator (green square) in all possible
fully connected ways. Contractions are denoted by lines, each of which
connects two circles of different \foreignlanguage{british}{colours}.
The green square can be inserted on any of these lines. Small circles
correspond to pair excitations and have two legs, larger circles correspond
to 4-particle excitations and have four legs, and so on (the full
equation contains red and blue circles with any even number of legs).}
\end{figure}

The expectation values of $Q_{s}$ in the state $|\Psi\rangle$ are
\begin{equation}
\frac{\langle\Psi|Q_{s}|\Psi\rangle}{\langle\Psi|\Psi\rangle}\equiv\langle\Psi|Q_{s}|\Psi\rangle_{conn}\:,
\end{equation}
and correspond to the sum of all possible ways to contract left and
right excitations in the expansion of (\ref{eq:psi}) with the charge
operator and with each other, in such a way that no part is disconnected
from the rest. Diagrammatically this is represented by fully connected
graphs as in Fig.\ref{fig}. In order to determine the scaling of
thermodynamic quantities with $L$ we should simply count the number
of redundant $\delta$-function factors of momentum variables in the
expectation values of the corresponding operator. As mentioned above,
the amplitudes $K_{2n}^{\Psi}$ contain a factor $\delta(\sum_{i}p(\theta_{i}))$
due to the translation invariance. Taking into account all of those
$\delta$-function factors involved in each connected graph and assuming
that the $K_{2n}^{\Psi}$ do not contain any other $\delta$-function
factors, we can easily see that each graph has exactly one left over
factor of $\delta(0)$, which is nothing but a factor equal to the
system size $L$. This means that the contribution of such graphs
is linear in $L$ i.e. extensive. If however some $K_{2n}^{\Psi}$
contains two $\delta$-function singularities in the momentum variables,
then the corresponding graphs contain one extra $\delta(0)\sim L$
factor i.e. scale more than linearly with $L$ and are not extensive.
For any single charge $Q_{s}$, extensivity may still be restored
by cancellation among the contributions scaling with a higher than
linear power of $L$. Therefore we must assume that the available
charges $Q_{s}$ are functionally independent and complete; then the
above argument means that all of the graphs must be extensive in order
for the charges to be such and this is true only if there is no amplitude
$K_{2n}^{\Psi}$ with more than one $\delta$-function. The same condition
ensures the extensivity of the generating function $\log Z=\log\langle\Psi|\Psi\rangle$
(analogous to the free energy) from which other thermodynamic quantities
can be derived.

Note that the argument requires the functional completeness of the
charges, which may require taking into consideration quasi-local charges
recently introduced in \cite{Prosen,PPSA,completeGGE}; similar extension
of the class of charges was also advocated for field theories in \cite{QFT_nonlocalcharges}.
However, as long as these additional charges are extensive for large
system sizes, such as the case e.g. for the charges recently used
in completing the GGE for the XXZ spin chain \cite{completeGGE},
the argument is left unchanged.

\subsubsection{\label{sub:Pair-structure-from} Pair structure from integrable dressing}

As already mentioned, due to the symmetry of the pre-quench Hamiltonian
under the $Z_{2}$ transformation $\phi\to-\phi$, the initial state
consists only of excitations with \emph{even} number of particles.
However in the present case it is expected to satisfy an even stronger
condition: to consist solely of \emph{pairs of particles with opposite
momenta}. An argument for this can be formulated by considering how
switching on the integrable interaction dresses the initial state.
To start with, any quench within the free case $g=0$ corresponds
to an initial state consisting of pairs. 

Now a quench to an interacting point $g\neq0$ can be considered perturbatively
and the issue now is how the state is dressed by turning on an integrable
interaction. Note that such a dressing must map the non-interacting
vacuum state $|0\rangle$ to the interacting vacuum state $|\Omega\rangle$
and the free one-particle states to the asymptotic one-particle states
of the interacting theory, which is rather nontrivial. However, there
is at least one case in which we know the result of such a dressing:
integrable boundary states. Considering only the parity invariant
case, let us start from the free field theory with Robin boundary
condition 
\begin{equation}
\mathcal{L}=\frac{1}{2}\left(\partial\phi\right)^{2}-\frac{m^{2}}{2}\phi^{2}\qquad\partial_{x}\phi|_{x=0}=-\lambda\phi|_{x=0}
\end{equation}
for which the boundary state has the form
\begin{equation}
|R\rangle=\mathcal{N}\exp\left[\frac{1}{2}\int_{-\infty}^{+\infty}\frac{\, d\theta}{2\pi}K_{R}(\theta)A{}^{\dagger}(\theta)A^{\dagger}(-\theta)\right]|0\rangle\label{RobinB}
\end{equation}
with 
\begin{equation}
K_{R}(\theta)=R_{R}\left(\frac{i\pi}{2}-\theta\right)=\frac{\cosh\theta-\lambda/m}{\cosh\theta+\lambda/m}
\end{equation}
where 
\begin{equation}
R_{R}(\theta)=\frac{\sinh\theta-i\lambda/m}{\sinh\theta+i\lambda/m}
\end{equation}
is the Robin reflection factor. When switching to Sinh-Gordon theory
(with an integrable boundary condition), $K_{R}$ gets dressed up
into \cite{corrigan_taormina} 
\begin{equation}
|S\rangle=\mathcal{N}\exp\left[\frac{1}{2}\int_{-\infty}^{+\infty}\,\frac{d\theta}{2\pi}K_{S}(\theta)A{}^{\dagger}(\theta)A^{\dagger}(-\theta)\right]|0\rangle\label{sinhG_parity}
\end{equation}
with
\begin{equation}
K_{S}(\theta)=-\frac{\cosh\theta-\cos\pi E/2}{\cosh\theta+\cos\pi E/2}K_{D}(\theta)\label{eq:sinhG_reflection_factor}
\end{equation}
where $E$ parametrizes the boundary interaction, and can be considered
as a dressed version of $\lambda$. 

We see that in the above case turning on an integrable interaction
dresses the state so that the pair structure is preserved. It is plausible
that a perturbative proof can be given, similar in spirit to the mechanism
of how particle number changing amplitudes cancel at each order of
perturbation theory in the Sinh-Gordon model \cite{dorey_exactS}.
We expect the pair structure of the dressed state to be the consequence
of the pair structure of the starting state and the integrability
of the dressing interaction, and to hold in general. In any case,
by analogy we expect the initial state to have this pair structure
for any quench from some mass and zero interaction to any other mass
and any interaction. 

We remark that an approximate pair structure is expected to hold for
small quenches irrespective of integrability. A quench is considered
small when the post-quench energy density is small compared to the
natural scale $m^{2}$ where $m$ is the mass gap in the post-quench
system. In this case the density of particles created after a quench
is small, and generally%
\footnote{There exist some exceptions to this scenario, for example a quench
in a $\phi^{4}$ coupling when perturbatively the leading process
is the creation of a quartet. For quenches in the mass parameter,
however, one expects pair creation to be the dominant elementary process. %
} their creation is dominated by pairs separated by a distance larger
than the correlation length $\xi=m^{-1}$. One expects that the pair
creation amplitude $K_{2}(\theta)$ is small, and furthermore, the
amplitude for creating any higher number of particles is well-approximated
by the product of amplitudes corresponding to independent pairs, leading
to an initial state (\ref{eq:psi}) having the form
\begin{equation}
|B\rangle=\exp\left(\int_{0}^{\infty}K(\theta)Z^{\dagger}(-\theta)Z^{\dagger}(\theta)d\theta\right)|0\rangle\:,\label{eq:squeezed}
\end{equation}
where we shortened $K_{2}$ to $K$.

\subsubsection{Exponentiation from extensivity}

In the cumulant expansion of a state with such pair structure, all
functions $K_{2n}^{\Psi}$ would contain $n$ $\delta$-functions
in order to account for the pairing of momenta. According to the above,
such a state would not satisfy the extensivity requirement unless
all $K_{2n}^{\Psi}$ with $n>1$ vanish. We therefore conclude that
the pair structure and the extensivity requirement constrain the initial
state to be of the squeezed coherent form. We note that the above
argument is analogous to the one used to show asymptotic exponentiation
in \cite{Mussardo13}, although the actual statement and the detailed
reasoning are different. Our argument is an application of the formalism
developed in \cite{BS15_prep}, extended to the integrable case.

In this special class of initial states, the one test particle equation
(\ref{eq:hierarchy_explct-1}) becomes 
\begin{align}
 & f_{1,0}^{\phi}(\theta|)\left(E_{0}(p(\theta))-E(\theta)\right)\nonumber \\
 & +2\sum_{r=1}^{\infty}\frac{1}{(r-1)!}\int\prod_{k=1}^{r-1}\frac{d\beta_{k}}{2\pi}\,\\
 & \quad\times\Biggl[f_{0,2r-1}^{\phi}(|-\theta,\beta_{r-1},-\beta_{r-1},\dots,\beta_{1},-\beta_{1})\left(E_{0}(p(\theta))+E(\theta)+2\sum_{k=1}^{r-1}E(\beta_{k})\right)K(\beta_{1})\dots K(\beta_{r-1})K(\theta)\nonumber \\
 & \quad+\frac{1}{2r}\int\frac{d\beta_{r}}{2\pi}\, f_{1,2r}^{\phi}(\theta|\beta_{r},-\beta_{r},\dots,\beta_{1},-\beta_{1})\left(E_{0}(p(\theta))-E(\theta)+2\sum_{k=1}^{r}E(\beta_{k})\right)K(\beta_{1})\dots K(\beta_{r})\Biggr]=0\:,\label{eq:oneparticle_equation_explicit}
\end{align}
as we can see by substituting 
\begin{equation}
K_{2n}(\theta_{1},...,\theta_{2n})=\frac{(2n)!}{n!}\,\prod_{s=1}^{n}2\pi\delta(\theta_{2s-1}+\theta_{2s})K(\theta_{2s})\:.
\end{equation}
into (\ref{eq:hierarchy_explct-1}). Note that based on the Zamolodchikov-Faddeev
algebra (\ref{eq:ZF}) the pair amplitude $K$ satisfies the relation
$K(\theta)=K(-\theta)S(2\theta)$.

\subsection{An Ansatz for the solution of the hierarchy}

Let us now consider the initial state on the basis of our heuristic
argument for the pair structure. We attributed the preservation of
the pair structure to the fact that when switching on the coupling,
the state gets dressed by an integrable interaction. In the free case,
the state can be expanded as 
\begin{equation}
|B\rangle_{g=0}=\mathcal{N}\left\{ |0\rangle-\frac{1}{2}\int_{-\infty}^{+\infty}\frac{\, dk}{2\pi}K^{free}(k)|k,-k\rangle_{0}+\dots\right\} \:,\label{eq:kfree}
\end{equation}
where index $0$ of the two-particle state refers to $g=0$, and $|0\rangle$
is the free boson vacuum. In the interacting case the initial state
takes the form
\begin{equation}
|B\rangle=\mathcal{N}\left\{ |\Omega\rangle+\frac{1}{2}\int_{-\infty}^{+\infty}\frac{\, d\theta}{2\pi}K(\theta)|\theta,-\theta\rangle+\dots\right\} \label{eq:kinteracting}
\end{equation}
with $|\Omega\rangle$ denoting the vacuum state of the interacting
theory. We remark that the relative sign between (\ref{eq:kfree})
and (\ref{eq:kinteracting}) is consistent with our earlier conventions
(\ref{sv}) and (\ref{exp}). Note that the interaction gives a non-trivial
renormalization of the vacuum state and also of the particles. 

Motivated by the dressing argument of Subsection \ref{sub:Pair-structure-from},
let us look for the pair amplitude $K$ in the form
\begin{equation}
K(\theta)=K^{free}(k)D(k)\:,
\end{equation}
where $k=m\sinh\theta$ and the dressing factor $D(k)$ only depends
on the parameters of the post-quench Hamiltonian, i.e. it is independent
of the pre-quench mass $m_{0}$. Now consider the limit for the pre-quench
mass $m_{0}\rightarrow\infty$. In this case the free amplitude tends
to $1$, while the amplitude $K$ is expected to become identical
to (\ref{eq:KD}) for the integrable Dirichlet boundary state, which
means that $D(k)=K_{D}(\theta)$. This leads to the proposal of the
following solution%
\footnote{The transition from integration over $k$ to integration over $\theta$
involves a Jacobi determinant factor, but it is eventually included
in the way $K_{D}(\theta)$ is given in eqn. (\ref{eq:KD}).%
} 
\begin{equation}
K(\theta)=K^{free}(k)K_{D}(\theta)\:,
\end{equation}
which gives 
\begin{equation}
|B\rangle=\mathcal{N}\exp\left[\frac{1}{2}\int_{-\infty}^{+\infty}\frac{\, d\theta}{2\pi}K(\theta)Z{}^{\dagger}(\theta)Z^{\dagger}(-\theta)\right]|\Omega\rangle\label{Ansatz_state}
\end{equation}
for the initial state, where 
\begin{eqnarray}
K(\theta) & = & \frac{E_{0}(p)-E(\theta)}{E_{0}(p)+E(\theta)}K_{D}(\theta)\:,\label{eq:Ansatz}\\
 &  & p=m\sinh\theta\quad,\quad E(\theta)=m\cosh\theta\quad,\quad E_{0}(p)=\sqrt{p^{2}+m_{0}^{2}}\:.\nonumber 
\end{eqnarray}

This is exactly the Ansatz proposed in our previous work \cite{STM}
on a slightly different basis, and it was demonstrated to solve the
one-particle equation of the hierarchy to a good approximation; note
that it has no free parameters to play with. In fact our arguments
suggest that its functional form is exact; however, our reasoning
does not preclude a coupling constant dependent renormalization of
the parameter $m_{0}/m$ on which the $K^{free}$ factor effectively
depends, similarly to $\lambda$ in the case of the integrable reflection
factor (\ref{eq:sinhG_reflection_factor}). However, in view of the
excellent agreement of our Ansatz with the numerical solution discussed
in the next Section, such a renormalization seems unlikely.

\section{Numerical solution of the hierarchy \label{sec:Numerical-solution-of}}

Following the arguments of the previous section we can now assume
that the initial state has the form

\begin{eqnarray}
|B\rangle & = & \exp\left(\int_{0}^{\infty}d\theta\; K(\theta)Z^{\dagger}(-\theta)Z^{\dagger}(\theta)\right)|\Omega\rangle\nonumber \\
 & = & \sum_{n=0}^{\infty}\frac{1}{n!}\prod_{j=1}^{n}\int\, d\theta_{j}K(\theta_{j})|-\theta_{1},\theta_{1},...,-\theta_{i},\theta_{i}\rangle\:.\label{expansion}
\end{eqnarray}
As we will see below, for a numerical solution for the function $K$
one needs to consider analytic continuation of the equations (\ref{eq:oneparticle_equation_explicit})
to complex rapidities. These can be derived by deforming the integration
contours and taking into account the residues of kinematical poles
given by (\ref{axioms}). An alternative way to obtain them is provided
by the finite volume formalism briefly reviewed in Appendix \ref{sec:Finite-volume-regularization}.
To be certain that the equations are correctly computed, we performed
the finite volume derivation and then cross-checked the result against
(\ref{eq:oneparticle_equation_explicit}).

\subsection{Keeping the vacuum and two-particle terms \label{sub:Keeping-the-vacuum}}

\begin{figure}
\begin{centering}
\includegraphics[width=0.5\textwidth]{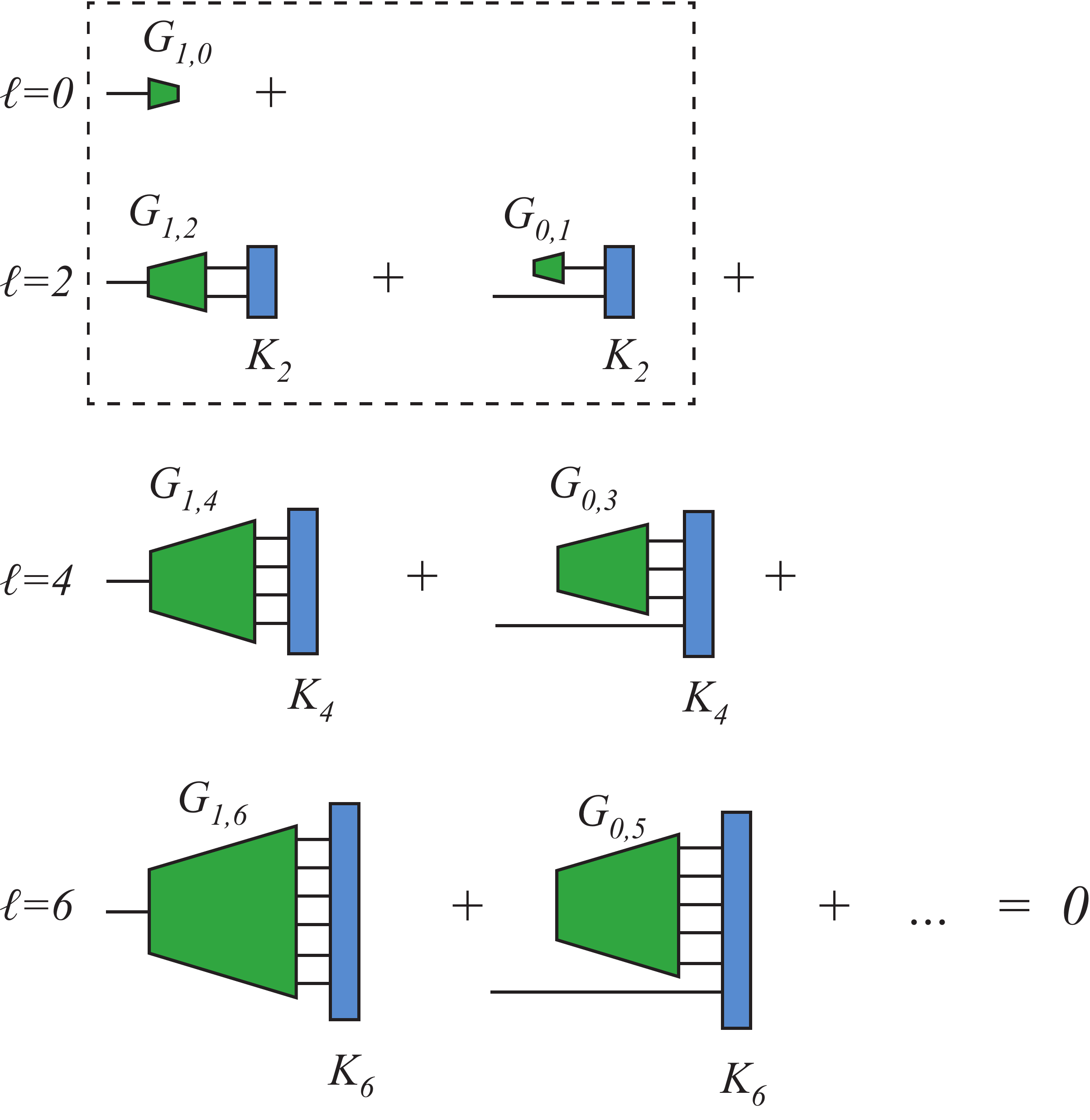}
\par\end{centering}

\protect\caption{\label{fig:Diagrammatic-expansion-of-1pt}Diagrammatic expansion of
the one test particle equation. The dashed frame encloses those diagrams
in which the bra and ket states differ by only one particle. These
three terms are the ones kept in the truncated equation of Section
\ref{sub:Keeping-the-vacuum}.}
\end{figure}

Keeping the first few terms in the expansion (\ref{expansion}) and
applying a one-particle test state $\langle\theta|$, the integral
equation 

\begin{equation}
\begin{aligned}0= & F_{1}^{\phi}+\frac{1}{2}F_{1}^{\phi}K_{D}(\theta)(1+S(-2\theta))\\
 & +\frac{1}{2}\int_{-\infty+i\varepsilon}^{+\infty+i\varepsilon}\frac{\, d\theta'}{2\pi}F_{3}^{\phi}(\theta+i\pi,-\theta',\theta')K_{D}(\theta')\\
 & +\frac{1}{4}\int_{-\infty}^{+\infty}\frac{\, d\theta'}{2\pi}(S(-2\theta)K_{D}(\theta)+S(\theta-\theta')S(\theta+\theta')K_{D}(\theta))F_{3}^{\phi}(-\theta,-\theta',\theta')K_{D}(\theta')\\
 & +\frac{1}{8}\int_{-\infty+i\varepsilon}^{+\infty+i\varepsilon}\frac{\, d\theta_{1}'}{2\pi}\int_{-\infty+i\varepsilon}^{+\infty+i\varepsilon}\frac{\, d\theta_{2}'}{2\pi}F_{5}^{\phi}(\theta+i\pi,-\theta_{1}',\theta_{1}',-\theta_{2}',\theta_{2}')K_{D}(\theta_{1}')K_{D}(\theta_{2}')+...
\end{aligned}
\label{onept_Dirichlet_upper}
\end{equation}
can be written for the Dirichlet case $m_{0}=\infty$, when the operator
equation (\ref{eq:hierarchy_opform}) simplifies to $\hat{\phi}(p)|D\rangle=0$
\cite{STM}. For the iteration procedure we need to generalize (\ref{onept_Dirichlet_upper})
to complex test rapidities. As long as $\text{Im }\theta<\varepsilon$,
(\ref{onept_Dirichlet_upper}) turns out to be valid, but for $\text{Im }\theta>\varepsilon$, 

\begin{equation}
\begin{aligned}0= & F_{1}^{\phi}+F_{1}^{\phi}K_{D}(\theta)\\
 & +\frac{1}{2}\int_{-\infty+i\varepsilon}^{+\infty+i\varepsilon}\frac{\, d\theta'}{2\pi}F_{3}^{\phi}(\theta+i\pi,-\theta',\theta')K_{D}(\theta')\\
 & +\frac{1}{2}\int_{-\infty}^{+\infty}\frac{\, d\theta'}{2\pi}S(\theta-\theta')S(\theta+\theta')K_{D}(\theta)F_{3}^{\phi}(-\theta,-\theta',\theta')K_{D}(\theta')\\
 & +\frac{1}{8}\int_{-\infty+i\varepsilon}^{+\infty+i\varepsilon}\frac{\, d\theta_{1}'}{2\pi}\int_{-\infty+i\varepsilon}^{+\infty+i\varepsilon}\frac{\, d\theta_{2}'}{2\pi}F_{5}^{\phi}(\theta+i\pi,-\theta_{1}',\theta_{1}',-\theta_{2}',\theta_{2}')K_{D}(\theta_{1}')K_{D}(\theta_{2}')+...
\end{aligned}
\label{onept_Dirichlet_lower}
\end{equation}
has to be used. This can be derived from (\ref{onept_Dirichlet_upper})
by analytical continuation using the analytical properties of the
form factors. In the case of finite mass quenches 

\begin{equation}
\begin{aligned}0= & [E_{0}(p)-E(\theta)]F_{1}^{\phi}+\frac{1}{2}[E_{0}(p)+E(\theta)]F_{1}^{\phi}K(\theta)(1+S(-2\theta))\\
 & +\frac{1}{2}\int_{-\infty+i\varepsilon}^{+\infty+i\varepsilon}\frac{\, d\theta'}{2\pi}[E_{0}(p)-E(\theta)+2E(\theta')]F_{3}^{\phi}(\theta+i\pi,-\theta',\theta')K(\theta')\\
 & +\frac{1}{4}\int_{-\infty}^{+\infty}\frac{\, d\theta'}{2\pi}[E_{0}(p)+E(\theta)+2E(\theta')](S(-2\theta)K(\theta)+S(\theta-\theta')S(\theta+\theta')K(\theta))\\
 & \hspace{1cm}\times F_{3}^{\phi}(-\theta,-\theta',\theta')K(\theta')\\
 & +\frac{1}{8}\int_{-\infty+i\varepsilon}^{+\infty+i\varepsilon}\hspace{-0.195cm}\frac{\, d\theta_{1}'}{2\pi}\int_{-\infty+i\varepsilon}^{+\infty+i\varepsilon}\hspace{-0.195cm}\frac{\, d\theta_{2}'}{2\pi}[E_{0}(p)-E(\theta)+2E(\theta_{1}')+2E(\theta_{2}')]\\
 & \hspace{1cm}\times F_{5}^{\phi}(\theta+i\pi,-\theta_{1}',\theta_{1}',-\theta_{2}',\theta_{2}')K(\theta_{1}')K(\theta_{2}')\\
 & +...\:,
\end{aligned}
\label{onept_upper}
\end{equation}
holds for $\text{Im }\theta<\varepsilon$ and 
\begin{equation}
\begin{aligned}0= & [E_{0}(p)-E(\theta)]F_{1}^{\phi}+[E_{0}(p)+E(\theta)]F_{1}^{\phi}K(\theta)\\
 & +\frac{1}{2}\int_{-\infty+i\varepsilon}^{+\infty+i\varepsilon}\frac{\, d\theta'}{2\pi}[E_{0}(p)-E(\theta)+2E(\theta')]F_{3}^{\phi}(\theta+i\pi,-\theta',\theta')K(\theta')\\
 & +\frac{1}{2}\int_{-\infty}^{+\infty}\frac{\, d\theta'}{2\pi}[E_{0}(p)+E(\theta)+2E(\theta')]S(\theta-\theta')S(\theta+\theta')K(\theta)F_{3}^{\phi}(-\theta,-\theta',\theta')K(\theta')\\
 & +\frac{1}{8}\int_{-\infty+i\varepsilon}^{+\infty+i\varepsilon}\hspace{-0.195cm}\frac{\, d\theta_{1}'}{2\pi}\int_{-\infty+i\varepsilon}^{+\infty+i\varepsilon}\hspace{-0.195cm}\frac{\, d\theta_{2}'}{2\pi}[E_{0}(p)-E(\theta)+2E(\theta_{1}')+2E(\theta_{2}')]\\
 & \hspace{1cm}\times F_{5}^{\phi}(\theta+i\pi,-\theta_{1}',\theta_{1}',-\theta_{2}',\theta_{2}')K(\theta_{1}')K(\theta_{2}')+...\:.
\end{aligned}
\label{onept_lower}
\end{equation}
for $\text{Im }\theta>\varepsilon$ \cite{STM}. The structure of
the integral equation is shown in Fig \ref{fig:Diagrammatic-expansion-of-1pt}.

Numerical calculations immediately show that the vacuum and two-particle
terms (those that are zeroth and first order in $K$) in eqns. (\ref{onept_upper})-(\ref{onept_Dirichlet_lower})
are much larger than the rest, which is expected from the considerations
of Subsection \ref{sub:Uniqueness-of-the} as they correspond to $\Delta n=\pm1$.
As a consequence, it makes sense to construct an iterative method
based on the truncated version of (\ref{onept_upper})-(\ref{onept_Dirichlet_lower})
that includes only the first three terms, while the rest are omitted:

\begin{equation}
\begin{aligned}0\approx & F_{1}^{\phi}+\frac{1}{2}F_{1}^{\phi}K_{D}(\theta)(1+S(-2\theta))\\
 & +\frac{1}{2}\int_{-\infty+i\varepsilon}^{+\infty+i\varepsilon}\frac{\, d\theta'}{2\pi}F_{3}^{\phi}(\theta+i\pi,-\theta',\theta')K_{D}(\theta')\quad(\mbox{if Im }\theta<\varepsilon)\:,
\end{aligned}
\label{ugly1}
\end{equation}

\begin{equation}
\begin{aligned}0\approx & F_{1}^{\phi}+F_{1}^{\phi}K_{D}(\theta)\\
 & +\frac{1}{2}\int_{-\infty+i\varepsilon}^{+\infty+i\varepsilon}\frac{\, d\theta'}{2\pi}F_{3}^{\phi}(\theta+i\pi,-\theta',\theta')K_{D}(\theta')\quad(\mbox{if Im }\theta>\varepsilon)\:,
\end{aligned}
\label{ugly2}
\end{equation}
for the Dirichlet, and 

\begin{equation}
\begin{aligned}0\approx & F_{1}^{\phi}\frac{E_{0}(p(\theta))-E(\theta)}{E_{0}(p(\theta))+E(\theta)}+\frac{1}{2}F_{1}^{\phi}K(\theta)(1+S(-2\theta))\\
 & +\frac{1}{2}\int_{-\infty+i\varepsilon}^{+\infty+i\varepsilon}\frac{\, d\theta'}{2\pi}\frac{E_{0}(p(\theta))-E(\theta)+2E(\theta')}{E_{0}(p(\theta))+E(\theta)}F_{3}^{\phi}(\theta+i\pi,-\theta',\theta')K(\theta')\quad(\mbox{if Im }\theta<\varepsilon)\:,
\end{aligned}
\label{ugly3}
\end{equation}

\begin{equation}
\begin{aligned}0\approx & F_{1}^{\phi}\frac{E_{0}(p(\theta))-E(\theta)}{E_{0}(p(\theta))+E(\theta)}+F_{1}^{\phi}K(\theta)\\
 & +\frac{1}{2}\int_{-\infty+i\varepsilon}^{+\infty+i\varepsilon}\frac{\, d\theta'}{2\pi}\frac{E_{0}(p(\theta))-E(\theta)+2E(\theta')}{E_{0}(p(\theta))+E(\theta)}F_{3}^{\phi}(\theta+i\pi,-\theta',\theta')K(\theta')\quad(\mbox{if Im }\theta>\varepsilon)\:,
\end{aligned}
\label{ugly4}
\end{equation}
for the finite mass quench.

\subsubsection{Dirichlet problem}

\begin{figure}[H]
\begin{centering}
\includegraphics[width=7.5cm]{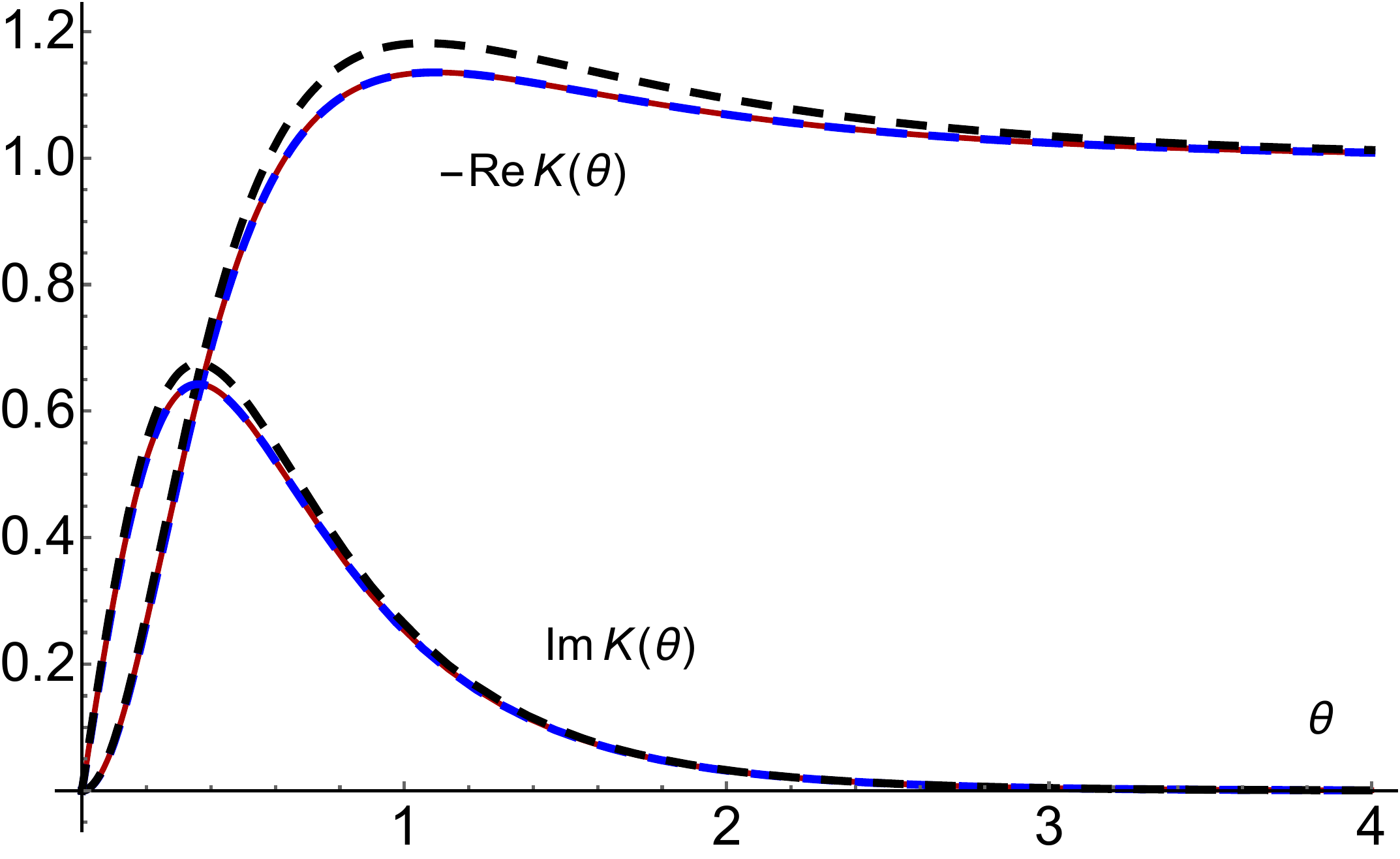} \hspace{0.5cm} \includegraphics[width=7.5cm]{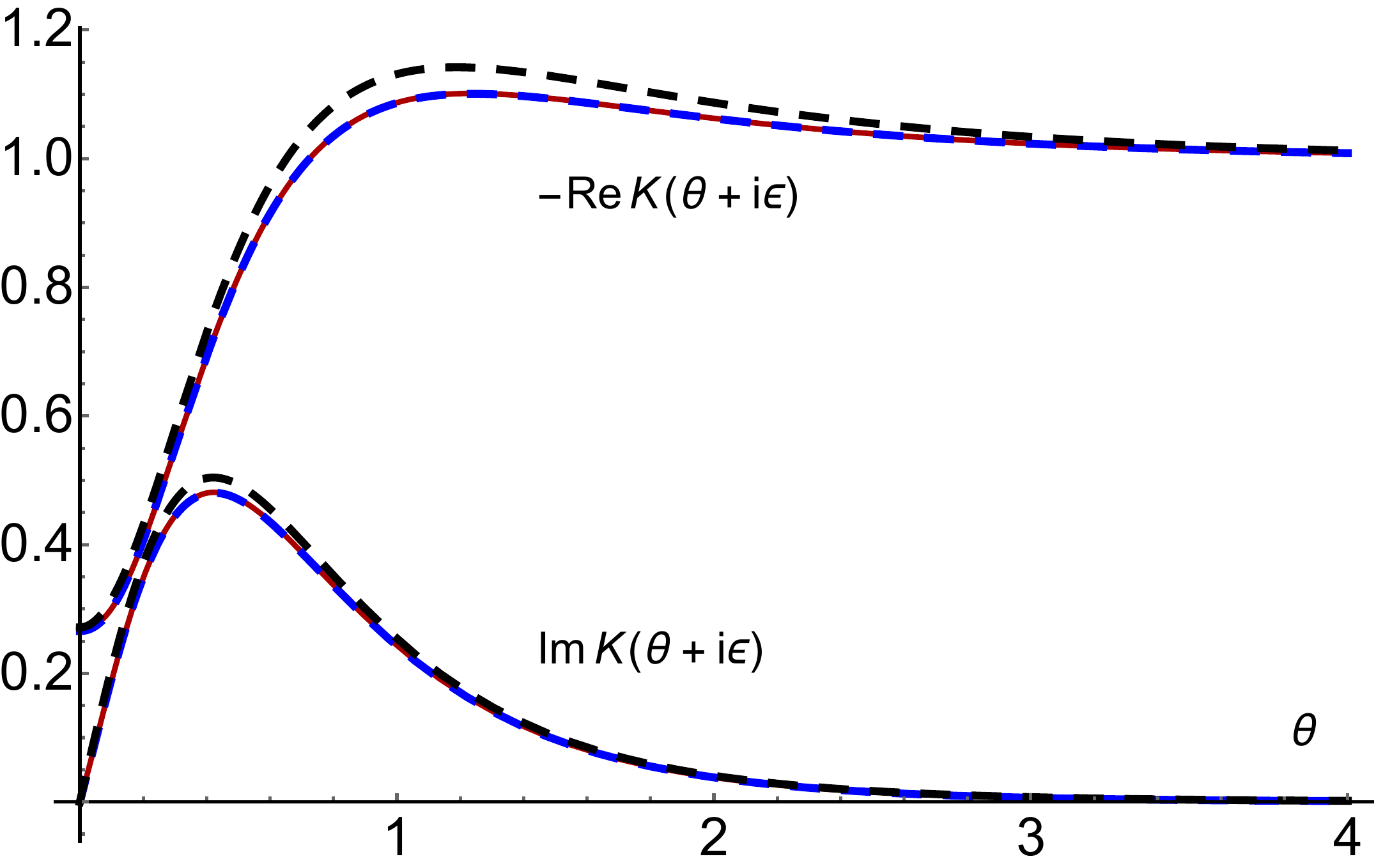}
\par\end{centering}

\protect\caption{\label{DirichletIter3term} \textit{6th (continuous red) and 7th (blue
dashed) iterative functions for the Dirichlet problem together with
the Dirichlet solution $K_{D}(\theta)$ (black dashed line) for real
and shifted rapidities. $\varepsilon_{1}=0.05,\varepsilon=\varepsilon_{2}=0.1,B=0.6$.
The input of the first iteration was the Dirichlet solution }(\ref{eq:KD})\textit{
itself.}}
\end{figure}

Discussing first the Dirichlet problem, our aim is to numerically
calculate the $K_{D}(\theta$) function for real test rapidities,
however, the equations contain this function for complex rapidities
$\theta'+i\varepsilon$ as well. Consequently, to obtain a closed
iterative scheme, we also have to plug in complex test rapidities
with imaginary part larger than the shift of the contour to the equations,
that is, in each iteration we have to calculate two iterative functions.
Instead of calculating the iterative functions at real and shifted
rapidities, for practical reasons, both of them are calculated at
shifted rapidities $0<\varepsilon_{1}<\varepsilon_{2}$ . At first,
$K(\theta+i\varepsilon_{1})^{(k+1)}$ denoted by $K(\theta)_{\varepsilon_{1}}^{(k+1)}$
is calculated based on (\ref{ugly1}) as the integration contour is
shifted with $\varepsilon_{2}$, and then $K(\theta)_{\varepsilon_{2}}^{(k+1)}$
is calculated based on (\ref{ugly2}) as the integration contour is
now shifted with $\varepsilon_{1}$. The equations of this iterative
scheme derived from (\ref{ugly1}) and (\ref{ugly2}) read

\begin{equation}
\begin{aligned}K(\theta)_{\varepsilon_{1}}^{(k+1)} & =-\frac{1}{2}\frac{1}{1+S(-2(\theta+i\varepsilon_{1}))}\left(2+\left.\frac{1}{F_{1}^{\phi}}\int_{-\infty}^{+\infty}\frac{\, d\theta'}{2\pi}F_{3}^{\phi}(\theta+i(\pi+\varepsilon_{1}),-\theta'-i\varepsilon_{2},\theta'+i\varepsilon_{2})K(\theta')_{\varepsilon_{2}}^{(k)}\right)\right.\\
 & +\frac{1}{2}K(\theta)_{\varepsilon_{1}}^{(k)}\:,\\
K(\theta)_{\varepsilon_{2}}^{(k+1)} & =-\frac{1}{2}\left(1+\left.\frac{1}{2F_{1}^{\phi}}\int_{-\infty}^{+\infty}\frac{\, d\theta'}{2\pi}F_{3}^{\phi}(\theta+i(\pi+\varepsilon_{2}),-\theta'-i\varepsilon_{1},\theta'+i\varepsilon_{1})K(\theta')_{\varepsilon_{1}}^{(k)}\right)+\frac{1}{2}K(\theta)_{\varepsilon_{2}}^{(k)}\:.\right.
\end{aligned}
\label{iterDir1}
\end{equation}
As can be seen from (\ref{iterDir1}), averaging with the previous
iterative function is also performed in the scheme.The solution along
the real axis can be obtained by the following equation

\begin{equation}
\begin{aligned}K(\theta) & =\frac{-1}{1+S(-2\theta)}\left(2+\frac{1}{F_{1}^{\phi}}\int_{-\infty}^{+\infty}\frac{\, d\theta'}{2\pi}F_{3}^{\phi}(\theta+i\pi,-\theta'-i\varepsilon_{2},\theta'+i\varepsilon_{2})K(\theta')_{\varepsilon_{2}}^{(k)}\right)\:.\end{aligned}
\label{realDir1}
\end{equation}
The above iteration scheme will be denoted by S2D. The iterations
under this scheme rapidly converge, either the functions with shifted
rapidities in their arguments or the real rapidity solutions obtained
with (\ref{realDir1}) are concerned. This is shown by calculating
the function difference of the consecutive iterative functions with
the integral

\begin{equation}
\int d\theta\left|\mbox{Re}[K(\theta)^{(k+1)}-K(\theta)^{(k)}]\right|\:,\label{eq:FunctionDifference}
\end{equation}
for the real and complex rapidity solutions as well. We took the real
part of the functions, as the imaginary parts were much smaller and
slowly varying. For a growing number of iterations all the three function
differences decreased monotonously and rapidly, and we stopped the
iterations when the largest of the three function differences crossed
a threshold value from above. In Fig. (\ref{DirichletIter3term})
the Dirichlet solution can be seen together with the 6th and 7th iterative
functions as the iteration was stopped after the 7th run, the largest
function difference being 0.0033.

The iterative solution is always very close to the expected exact
result (\ref{eq:KD}) over all the fundamental range $0\leq B\leq1$
of the coupling strength. The small deviation between the numerical
and the exact solution can be attributed to the truncation of the
infinite integral equation to the vacuum and two-particle terms. This
is supported by the results presented in Subsection \ref{sub:Adding-the-},
where it is shown that the iterative solution of the integral equation
is much closer to the exact result $K_{D}$ when higher particle number
terms are taken into account.

\subsubsection{Finite mass quench}

\begin{figure}[H]
\begin{centering}
\includegraphics[width=7.5cm]{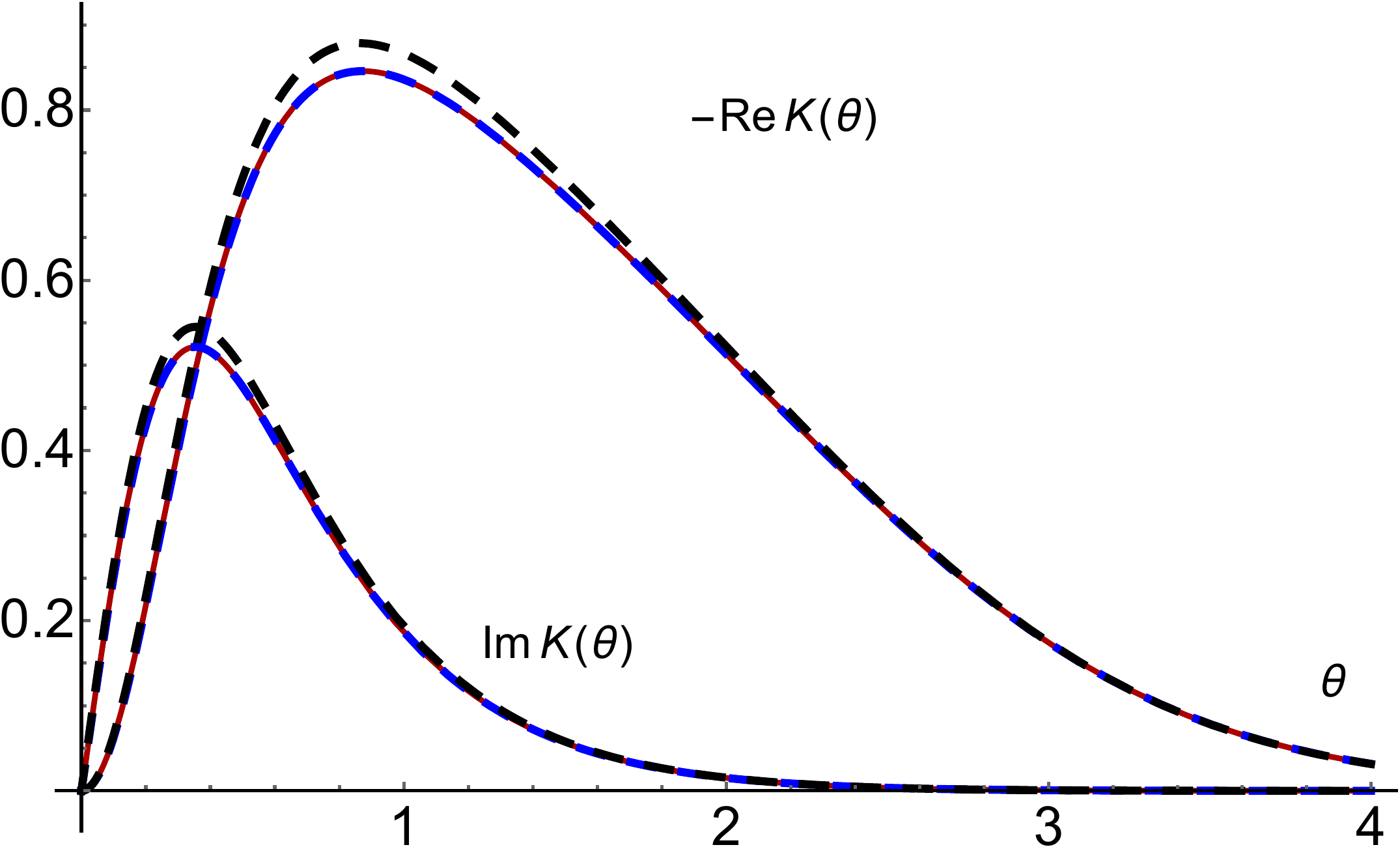} \hspace{0.5cm} \includegraphics[width=7.5cm]{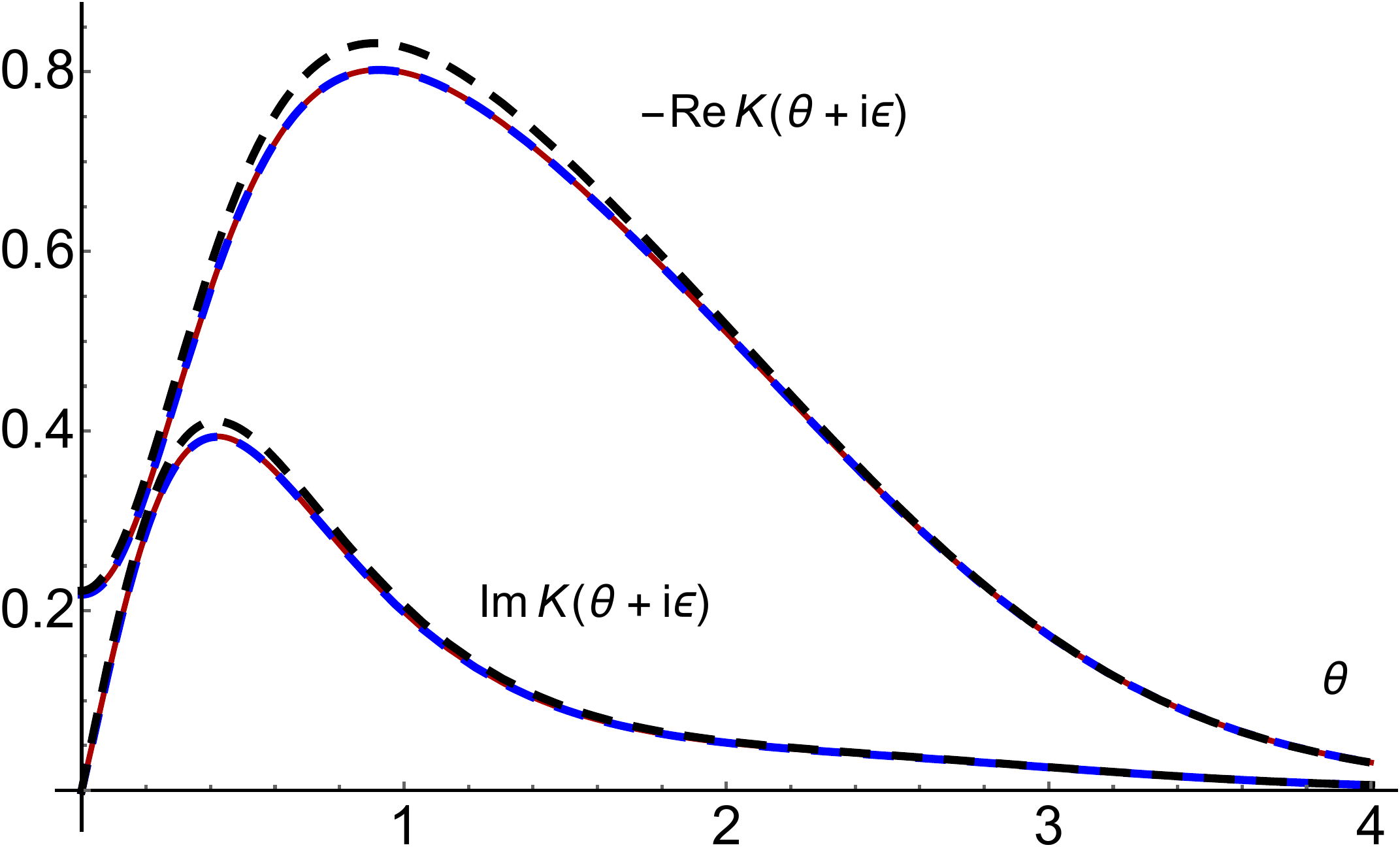}
\par\end{centering}

\protect\caption{\label{FinMassIter3term} \textit{6th (continuous red) and 7th (blue
dashed) iterations for the finite mass quench problem together with
the }\emph{proposed solution (\ref{eq:Ansatz})}\textit{ (black dashed
line) for real and shifted rapidities. $\varepsilon_{1}=0.05,\varepsilon=\varepsilon_{2}=0.1,B=0.6,m=1,m_{0}=10$.
The input of the first iteration was }\emph{(\ref{eq:Ansatz})}\textit{
itself.}}
\end{figure}

For the finite mass problem (\ref{iterDir1}) is modified as

\begin{equation}
\begin{aligned}\hspace{-1.5cm}K(\theta)_{\varepsilon_{1}}^{(k+1)} & =-\frac{1}{2}\frac{1}{1+S(-2(\theta+i\varepsilon_{1}))}\left(\vphantom{\frac{1}{F_{1}}\int_{-\infty}^{+\infty}\frac{\, d\theta'}{2\pi}F_{3}(\theta+i(\pi+\varepsilon),-\theta'-i\varepsilon,\theta'+i\varepsilon)K(\theta')_{\varepsilon_{1}}^{(k)}}2\frac{E_{0}(p(\theta+i\varepsilon_{1}))-E(\theta+i\varepsilon_{1})}{E_{0}(p(\theta+i\varepsilon_{1}))+E(\theta+i\varepsilon_{1})}+\right.\\
\\
 & +\left.\frac{1}{F_{1}^{\phi}}\int_{-\infty}^{+\infty}\frac{\, d\theta'}{2\pi}\frac{E_{0}(p(\theta+i\varepsilon_{1}))-E(\theta+i\varepsilon_{1})+2E(\theta'+i\varepsilon_{2})}{E_{0}(p(\theta+i\varepsilon_{1}))+E(\theta+i\varepsilon_{1})}\right.\\
 & \left.F_{3}^{\phi}(\theta+i(\pi+\varepsilon_{1}),-\theta'-i\varepsilon_{2},\theta'+i\varepsilon_{2})K(\theta')_{\varepsilon_{2}}^{(k)}\vphantom{\frac{1}{F_{1}}\int_{-\infty}^{+\infty}\frac{\, d\theta'}{2\pi}\frac{E_{0}(\theta+i\varepsilon_{1})-E(\theta+i\varepsilon_{1})+E(\theta'+i\varepsilon_{2})}{E_{0}(\theta+i\varepsilon_{1})+E(\theta+i\varepsilon_{1})}F_{3}(\theta+i(\pi+\varepsilon_{1}),-\theta'-i\varepsilon_{2},\theta'+i\varepsilon_{2})K(\theta')_{\varepsilon_{2}}^{(k)}}\right)+\frac{1}{2}K(\theta)_{\varepsilon_{1}}^{(k)}\:,\\
\\
\\
\hspace{-1.5cm}K(\theta)_{\varepsilon_{2}}^{(k+1)} & =-\frac{1}{2}\left(\vphantom{\frac{1}{2F_{1}}\int_{-\infty}^{+\infty}\frac{\, d\theta'}{2\pi}F_{3}(\theta+i(\pi+\varepsilon),-\theta'-i2\varepsilon,\theta'+i2\varepsilon)K(\theta')_{\varepsilon_{2}}^{(k)}}1\frac{E_{0}(p(\theta+i\varepsilon_{2}))-E(\theta+i\varepsilon_{2})}{E_{0}(p(\theta+i\varepsilon_{2}))+E(\theta+i\varepsilon_{2})}+\right.\\
\\
 & +\left.\frac{1}{2F_{1}^{\phi}}\int_{-\infty}^{+\infty}\frac{\, d\theta'}{2\pi}\frac{E_{0}(p(\theta+i\varepsilon_{2}))-E(\theta+i\varepsilon_{2})+2E(\theta'+i\varepsilon_{1})}{E_{0}(p(\theta+i\varepsilon_{2}))+E(\theta+i\varepsilon_{2})}\right.\\
 & \left.F_{3}^{\phi}(\theta+i(\pi+\varepsilon_{2}),-\theta'-i\varepsilon_{1},\theta'+i\varepsilon_{1})K(\theta')_{\varepsilon_{1}}^{(k)}\vphantom{\frac{1}{F_{1}}\int_{-\infty}^{+\infty}\frac{\, d\theta'}{2\pi}\frac{E_{0}(\theta+i\varepsilon_{1})-E(\theta+i\varepsilon_{1})+E(\theta'+i\varepsilon_{2})}{E_{0}(\theta+i\varepsilon_{1})+E(\theta+i\varepsilon_{1})}F_{3}(\theta+i(\pi+\varepsilon_{1}),-\theta'-i\varepsilon_{2},\theta'+i\varepsilon_{2})K(\theta')_{\varepsilon_{2}}^{(k)}}\right)+\frac{1}{2}K(\theta)_{\varepsilon_{2}}^{(k)}\:,
\end{aligned}
\label{iterMass1}
\end{equation}
yielding scheme S2F. The solution along the real axis is obtained
by

\begin{equation}
\begin{aligned}K(\theta) & =\frac{-1}{1+S(-2\theta)}\left(\vphantom{\frac{1}{F_{1}}\int_{-\infty}^{+\infty}\frac{\, d\theta'}{2\pi}F_{3}(\theta+i(\pi+\varepsilon_{1}),-\theta'-i2\varepsilon,\theta'+i2\varepsilon)K(\theta')_{\varepsilon}^{(k)}}2\frac{E_{0}(p(\theta))-E(\theta)}{E_{0}(p(\theta))+E(\theta)}+\right.\\
 & +\left.\frac{1}{F_{1}^{\phi}}\int_{-\infty}^{+\infty}\frac{\, d\theta'}{2\pi}\frac{E_{0}(\theta)-E(\theta)+2E(\theta'+i\varepsilon_{2})}{E_{0}(\theta)+E(\theta)}F_{3}^{\phi}(\theta+i\pi,-\theta'-i\varepsilon_{2},\theta'+i\varepsilon_{2})K(\theta')_{\varepsilon_{2}}^{(k)}\right)\:.
\end{aligned}
\label{realMass1}
\end{equation}
For the finite mass quench problem fast convergence is witnessed again,
furthermore, the iterative solution appears to be close to the proposed
solution (\ref{eq:Ansatz}) (Fig. \ref{FinMassIter3term}). Just like
in the Dirichlet case, these observations hold for a large regime
of the coupling strength $B$ and the quench parameter $m_{0}$. Note
that the deviation between the iterative solution and the Ansatz (\ref{eq:Ansatz})
is of the same magnitude as in the Dirichlet case, therefore it can
safely be attributed to the truncation of the form factor series,
which will be strongly confirmed after taking into account the four-particle
contribution in Subsection (\ref{sub:Adding-the-}).

\subsection{Adding the $O(K^{2})$ terms \label{sub:Adding-the-}}

To construct an iteration scheme including the four-particle contributions
of (\ref{onept_upper})-(\ref{onept_Dirichlet_lower}), we can simply
add them (\ref{ugly1})-(\ref{ugly4}). For a closed iteration, however,
we now need also an iterative function that is defined for real rapidities.
Although it is possible to construct a scheme working with only two
iterative functions, the one presented below uses three of them: for
$K(\theta)^{(k+1)}$ and $K(\theta)_{\varepsilon_{1}}^{(k+1)}$ the
integration contour is shifted with $\varepsilon_{2}$, and for $K(\theta)_{\varepsilon_{2}}^{(k+1)}$
the contour is shifted with $\varepsilon_{1}$. Unlike in the equation
for $K(\theta)_{\varepsilon_{1}}^{(k+1)}$, where the denominator
is $1+S(-2(\theta+i\varepsilon_{1})),$ in the equation for $K(\theta)^{(k+1)}$,
the denominator is $1+\frac{1}{2F_{1}^{\phi}}\int_{-\infty}^{+\infty}\frac{\, d\theta'}{2\pi}\left(S(-2\theta)+S(\theta-\theta')S(\theta+\theta')\right)F_{3}^{\phi}(-\theta,-\theta',\theta')K(\theta')^{(k)}$
instead. The equations for this scheme S4D read
\begin{equation}
\begin{aligned}K(\theta)_{\varepsilon_{1}}^{(k+1)} & =-\frac{1}{2}\frac{1}{1+S(-2(\theta+i\varepsilon_{1}))}\left(\vphantom{\frac{1}{F_{1}}\int_{-\infty}^{+\infty}\frac{\, d\theta'}{2\pi}F_{3}(\theta+i(\pi+\varepsilon),-\theta'-i2\varepsilon,\theta'+i2\varepsilon)K(\theta')_{\varepsilon}^{(k)}}2+\left.\frac{1}{F_{1}^{\phi}}\int_{-\infty}^{+\infty}\frac{\, d\theta'}{2\pi}F_{3}^{\phi}(\theta+i(\pi+\varepsilon_{1}),-\theta'-i\varepsilon_{2},\theta'+i\varepsilon_{2})K(\theta')_{\varepsilon_{2}}^{(k)}\right.\right.\\
\  & +\left.\frac{1}{2F_{1}^{\phi}}\int_{-\infty}^{+\infty}\frac{\, d\theta'}{2\pi}\left(S(-2(\theta+i\varepsilon_{1}))K(\theta)_{\varepsilon_{1}}^{(k)}+S(\theta+i\varepsilon_{1}-\theta')S(\theta+i\varepsilon_{1}+\theta')K(\theta)_{\varepsilon_{1}}^{(k)}\right)\right.\\
 & \left.F_{3}^{\phi}(-\theta-i\varepsilon_{1},-\theta',\theta')K(\theta')^{(k)}\right.\\
\  & +\left.\frac{1}{4F_{1}^{\phi}}\int_{-\infty}^{+\infty}\hspace{-0.195cm}\frac{\, d\theta_{1}'}{2\pi}\int_{-\infty}^{+\infty}\hspace{-0.195cm}\frac{\, d\theta_{2}'}{2\pi}F_{5}^{\phi}(\theta+i(\pi+\varepsilon_{1}),-\theta_{1}'-i\varepsilon_{2},\theta_{1}'+i\varepsilon_{2},-\theta_{2}'-i\varepsilon_{2},\theta_{2}'+i\varepsilon_{2})\right.\\
 & \left.K(\theta_{1}')_{\varepsilon_{2}}^{(k)}K(\theta_{2}')_{\varepsilon_{2}}^{(k)}\vphantom{\frac{1}{F_{1}}\int_{-\infty}^{+\infty}\frac{\, d\theta'}{2\pi}F_{3}(\theta+i(\pi+\varepsilon),-\theta'-i2\varepsilon,\theta'+i2\varepsilon)K(\theta')_{\varepsilon}^{(k)}}\right)+\frac{1}{2}K(\theta)_{\varepsilon_{1}}^{(k+1)}\:,
\end{aligned}
\end{equation}

\begin{equation}
\begin{aligned}K(\theta)_{\varepsilon_{2}}^{(k+1)} & =-\frac{1}{2}\left(\vphantom{\frac{1}{2F_{1}}\int_{-\infty}^{+\infty}\frac{\, d\theta'}{2\pi}F_{3}(\theta+i(\pi+\varepsilon),-\theta'-i2\varepsilon,\theta'+i2\varepsilon)K(\theta')_{\varepsilon}^{(k)}}1+\frac{1}{2F_{1}^{\phi}}\int_{-\infty}^{+\infty}\frac{\, d\theta'}{2\pi}F_{3}^{\phi}(\theta+i(\pi+\varepsilon_{2}),-\theta'-i\varepsilon_{1},\theta'+i\varepsilon_{1})K(\theta')_{\varepsilon_{1}}^{(k)}\right.\\
\  & +\left.\frac{1}{2F_{1}^{\phi}}\int_{-\infty}^{+\infty}\frac{\, d\theta'}{2\pi}\left(S(\theta+i\varepsilon_{2}-\theta')S(\theta+i\varepsilon_{2}+\theta')K(\theta)_{\varepsilon_{1}}^{(k)}\right)\right.\\
 & \left.F_{3}^{\phi}(-\theta-i\varepsilon_{2},-\theta',\theta')K(\theta')^{(k)}\right.\\
\  & +\left.\frac{1}{8F_{1}^{\phi}}\int_{-\infty}^{+\infty}\hspace{-0.195cm}\frac{\, d\theta_{1}'}{2\pi}\int_{-\infty}^{+\infty}\hspace{-0.195cm}\frac{\, d\theta_{2}'}{2\pi}F_{5}^{\phi}(\theta+i(\pi+\varepsilon_{2}),-\theta_{1}'-i\varepsilon_{1},\theta_{1}'+i\varepsilon_{1},-\theta_{2}'-i\varepsilon_{1},\theta_{2}'+i\varepsilon_{1})\right.\\
 & \left.K(\theta_{1}')_{\varepsilon_{1}}^{(k)}K(\theta_{2}')_{\varepsilon_{1}}^{(k)}\vphantom{\frac{1}{F_{1}}\int_{-\infty}^{+\infty}\frac{\, d\theta'}{2\pi}F_{3}(\theta+i(\pi+\varepsilon),-\theta'-i2\varepsilon,\theta'+i2\varepsilon)K(\theta')_{\varepsilon}^{(k)}}\right)+\frac{1}{2}K(\theta)_{\varepsilon_{2}}^{(k+1)}\:,
\end{aligned}
\label{iterDir2-2}
\end{equation}

\begin{equation}
\begin{aligned}K(\theta)^{(k+1)} & =-\frac{1}{2}\left[1+\frac{1}{2F_{1}^{\phi}}\int_{-\infty}^{+\infty}\frac{\, d\theta'}{2\pi}\left(S(-2\theta)+S(\theta-\theta')S(\theta+\theta')\right)\left.F_{3}^{\phi}(-\theta,-\theta',\theta')K(\theta')^{(k)}\vphantom{\frac{1}{2F_{1}}\int_{-\infty}^{+\infty}\frac{\, d\theta'}{2\pi}\left(S(-2\theta)K(\theta)^{(k)}+S(\theta-\theta')S(\theta+\theta')K(\theta)^{(k)}\right)F_{3}(-\theta,-\theta',\theta')K(\theta')^{(k)}}\right]^{-1}\right.\\
 & \left(\vphantom{\frac{1}{F_{1}}\int_{-\infty}^{+\infty}\frac{\, d\theta'}{2\pi}F_{3}(\theta+i(\pi+\varepsilon),-\theta'-i2\varepsilon,\theta'+i2\varepsilon)K(\theta')_{\varepsilon}^{(k)}}2+S(-2\theta)K(\theta)^{(k)}+\left.\frac{1}{F_{1}^{\phi}}\int_{-\infty}^{+\infty}\frac{\, d\theta'}{2\pi}F_{3}^{\phi}(\theta+i\pi,-\theta'-i\varepsilon_{2},\theta'+i\varepsilon_{2})K(\theta')_{\varepsilon_{2}}^{(k)}\right.\right.\\
\  & +\left.\frac{1}{4F_{1}^{\phi}}\int_{-\infty}^{+\infty}\hspace{-0.195cm}\frac{\, d\theta_{1}'}{2\pi}\int_{-\infty}^{+\infty}\hspace{-0.195cm}\frac{\, d\theta_{2}'}{2\pi}F_{5}^{\phi}(\theta+i\pi,-\theta_{1}'-i\varepsilon_{2},\theta_{1}'+i\varepsilon_{2},-\theta_{2}'-i\varepsilon_{2},\theta_{2}'+i\varepsilon_{2})\right.\\
 & \left.K(\theta_{1}')_{\varepsilon_{2}}^{(k)}K(\theta_{2}')_{\varepsilon_{2}}^{(k)}\vphantom{\frac{1}{F_{1}}\int_{-\infty}^{+\infty}\frac{\, d\theta'}{2\pi}F_{3}(\theta+i(\pi+\varepsilon),-\theta'-i2\varepsilon,\theta'+i2\varepsilon)K(\theta')_{\varepsilon}^{(k)}}\right)+\frac{1}{2}K(\theta)^{(k)}\:.
\end{aligned}
\label{iterDir2}
\end{equation}
We omit the equations of the iteration scheme for the finite mass
case, as the corresponding finite mass scheme S4F is easily obtained
from S4D by plugging the extra $\frac{E_{0}-E}{E_{0}+E}$ type factors.
Similarly to schemes S2D and S2F, averaging with the previous iterative
functions is present in each iterative step. 

This scheme was chosen from other possibilities by observing that
it always performed better than all other schemes we tried. Unlike
S2D and S2F and just like other schemes including the four-particle
terms, S4D and S4F iterations are unstable. To overcome this issue
we stopped the iteration, when the difference between two consecutive
iterative functions is the smallest. The function difference was measured
by the (\ref{eq:FunctionDifference}) as, similarly to the previous
schemes, the difference between the imaginary parts tends to be much
smaller and slowly varying.

\begin{figure}[H]
\begin{centering}
\includegraphics[width=7.5cm]{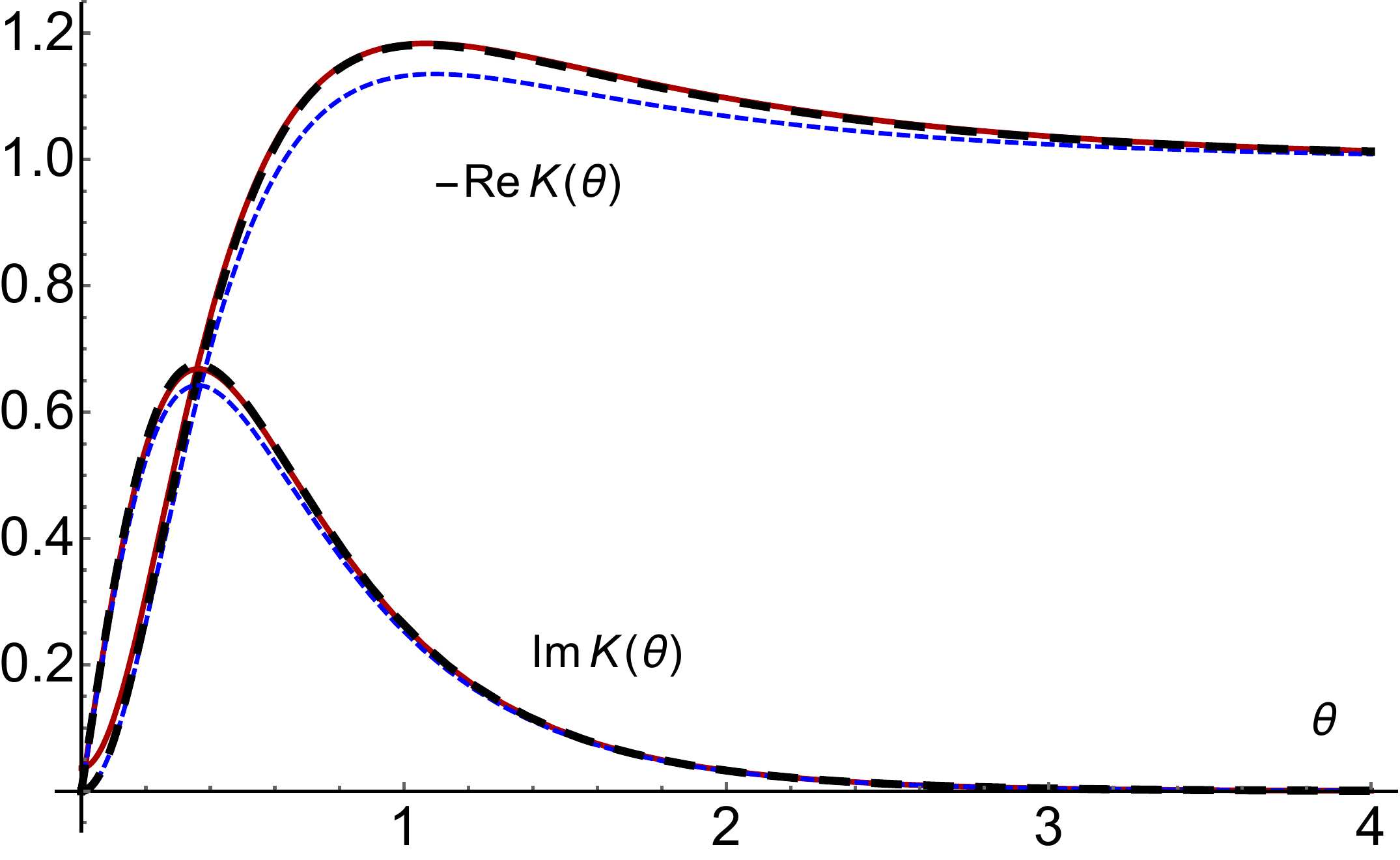} \hspace{0.5cm} \includegraphics[width=7.5cm]{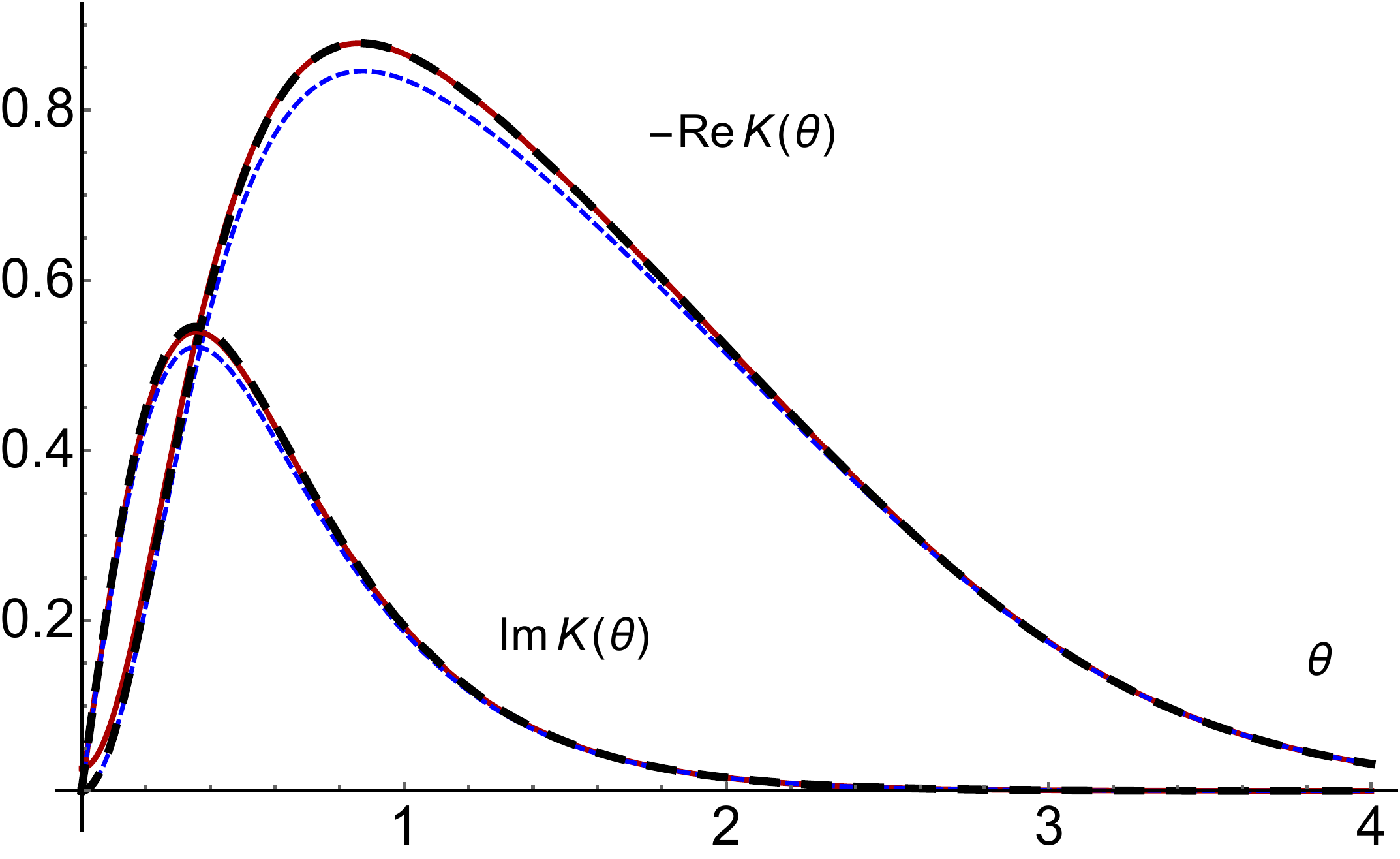}
\par\end{centering}

\protect\caption{\label{DirFinMassIter5term}\textit{ Left: 7th (blue dashed) iterative
functions obtained by S2D and the optimal iterative function (red
continuous) obtained by S4D for the Dirichlet problem together with
the Dirichlet solution $K_{D}(\theta)$ (black dashed line) for real
rapidities.$\varepsilon_{1}=0.05,\varepsilon=\varepsilon_{2}=0.1,B=0.6$.
The input of the first S2D iteration was the Dirichlet solution itself,
whereas the S4D input was the 7th S2D iterative function. Right: 7th
(blue dashed) iterative functions obtained by S2F and the optimal
iterative function (red continuous) obtained by S4F for the finite
mass quench problem together with the proposed solution }\emph{(\ref{eq:Ansatz})}\textit{
(black dashed line) for real rapidities. $\varepsilon_{1}=0.05,\varepsilon=\varepsilon_{2}=0.1,B=0.6,m=1,m_{0}=10$.
The input of the first S2F iteration was }\emph{(\ref{eq:Ansatz})}\textit{
itself, wheres the S4F input was the 7th S2F iterative function.}}
\end{figure}

As demonstrated in Fig. \ref{DirFinMassIter5term} the solutions of
S4D and S4F schemes are indeed remarkably close to the exact Dirichlet
solution and the Ansatz, respectively. In fact, the solution curves
now lie on top of the exact Dirichlet amplitude and the Ansatz, respectively.
This pattern again holds for all values of the coupling strength $B$
and a large range of the quench parameter $m_{0}$.

\subsection{The three-particle condition}

For a numerical verification of the exponential form (\ref{sv}) of
the initial state, it is necessary to consider higher members of the
hierarchy. This rapidly becomes impractical due to the computational
cost of the integrals involving very high particle form factors. In
addition, constructing the form of the equation with the integration
contours arranged conveniently for numerical evaluation is also quite
tedious. 

For this reason, we restrict ourselves here to the case of three-particle
test states. The three-particle condition is a sum 
\begin{equation}
0=T_{0}+T_{2}+T_{4}+T_{6}+\dots\label{eq:3ptcond}
\end{equation}
with $T_{n}$ denoting the $n$-particle contribution from initial
state (\ref{sv}). Since the iterative solution of the one-particle
condition is very well described by the Ansatz (\ref{eq:Ansatz}),
the latter can be used to perform the evaluations. We have computed
the first three terms explicitly with the result given in (\ref{eq:NotCrossedoutBigFinMass-1}).
For the contribution $T_{6}$, deriving the integral form proved to
be so tedious that we decided to estimate its contribution directly
evaluating the corresponding term in the finite volume form (\ref{eq:finvol_hierarchy})
for a large value of the volume ($mL=250$). 

To verify that (\ref{eq:3ptcond}) holds\emph{,} we computed the sum
for several values of the Sinh-Gordon coupling in the fundamental
range $0\leq B\leq1$, for different values of the test rapidities
at each point, and for several different quenches, parametrized by
the mass ratio $m_{0}/m$. 

To make certain that the integral form (\ref{eq:NotCrossedoutBigFinMass-1})
of the condition was derived correctly, we supplemented the calculation
of $T_{2}$ and $T_{4}$ by a direct evaluation of the finite volume
sum for $mL=250$. The finite volume form was always found to agree
with the integral form within the numerical precision of the former.
Note that for practical evaluation the finite volume sum must be truncated
to states below some upper energy cutoff. For $T_{2}$ and $T_{4}$
it was possible to keep this truncation high enough to achieve better
than 3 digits accuracy; however, for $T_{6}$ this was not always
the case, as we discuss below. 

A sample of the resulting data is collected in Appendix \ref{sec:Three-particle-condition}.
As a benchmark, we always quote the Dirichlet case $m_{0}/m=\infty$,
which can be obtained by omitting the energy factors in square brackets
from each term. For the Dirichlet case, the equation must hold exactly
when terms $T_{n}$ are included for all $n$, since then (\ref{sv})
corresponds to a boundary state known exactly from reflection factor
bootstrap \cite{GhoshalZamo,Ghoshal}. We can see that the terms $T_{2}$
and $T_{4}$ (corresponding to $\Delta n=\pm1$) are always dominant,
and typically cancel each other within a few percent. To go further,
it is necessary to include $T_{0}$ and $T_{6}$, corresponding to
$\Delta n=\pm2$. In most cases, this improves the cancellation to
better than a percent. There are two exceptions to this, however.
First, when the sum $T_{0}+T_{2}+T_{4}$ is relatively small, to verify
the improvement $T_{6}$ would need to be evaluated to a very high
precision which is not possible using the finite volume summation.
Second, when some of the test rapidities are relatively large, the
cut-off necessary to evaluate the finite volume sum for $T_{6}$ prevents
the evaluation of all the dominant contributions, as these come from
regions which cannot be explored within reasonable computer time.
However, in all these cases we also see the same deviations for the
$m_{0}/m=\infty$ case, for which we known the full equation should
hold exactly. Taken altogether, these facts show that the deviations
can be explained by the approximations made during numerical evaluation.
On the other hand, we have data for a much larger number of couplings
and rapidities than shown in Appendix \ref{sec:Three-particle-condition},
and all of them fall in the same pattern. 

Summing up, the numerical evaluation of the three-particle member
of the hierarchy strongly confirms the exponential form of the initial
state.

\section{Conclusions \label{sec:Conclusions}}

The present paper continues a program started in \cite{STM}, which
is aimed at constructing the initial state of quantum quenches in
integrable quantum field theories in terms of the post-quench eigenstates.
Such a construction amounts to the determination of the overlaps of
the initial state with the eigenstates of the post-quench Hamiltonian,
which is an essential input for computing the temporal evolution and
the steady state expectation values of physical observables.

The particular class of quenches considered in this paper was within
the Sinh-Gordon model, starting from the ground state of mass $m_{0}$
with zero coupling, to a post-quench system with a mass $m$ and a
nonzero value of $g$. The important characteristics of this quench
is that the initial state can be specified via an operator condition.
This allows the derivation of an infinite hierarchy of integral equations
based on the form factor bootstrap for the overlap functions $K_{2n}$.
Each integral equation can be written as a form factor expansion,
consisting of an infinite number of terms.

We then examined the hierarchy, and presented a number of arguments
concerning the nature of its solutions. Some of these arguments are
likely valid for more general models and initial states, described
by similar integral hierarchies. The results we expect to be generally
valid are the following:
\begin{enumerate}
\item For a free field, the unique solution of the hierarchy is the usual
squeezed state obtained from the Bogoliubov transformation, and perturbation
theory considerations imply the existence and uniqueness of its solution
for the interacting case. 
\item The terms in the hierarchy are ordered in magnitude by the difference
between the bra and ket state particle numbers $\Delta n$; the larger
$\Delta n$, the smaller is the corresponding term. This means that
each equation in the hierarchy can be well approximated by a finite
truncation, and that higher equations corresponding to test states
with more particles probe overlap amplitudes with states containing
more particles. For the elementary field $\phi$ considered in this
paper it is the$\Delta n=\pm1$ terms that dominate, and a rather
good approximation can be obtained by truncating to only two terms.
\item The iterative solution method that was developed on the basis of the
aforementioned truncatability property.
\item The result that exponentiation follows for states that are built out
of zero-momentum particle pairs, which follows from the extensivity
of local conserved charges. 
\end{enumerate}
In addition, for the particular class of quenches we presented a heuristic
argument, dubbed ``integrable dressing'', that the state only contains
pair states. This argument is heuristic in many respects. First, it
supposes the construction of a mapping of the free theory eigenstates
to the interacting eigenstates; there are general results such as
Haag's theorem \cite{haag} showing that such a mapping does not exist
in mathematically strict sense. This can be avoided by introducing
an ultraviolet cut-off and removing it after renormalization, but
then it must be shown that the resulting relations really confirm
the argument. An equivalent approach is to consider fast but smooth
quantum quenches \cite{DGM} in which the time interval during which
the quench is performed tends to zero but kept larger than the inverse
energy cut-off. Either way, this is definitely out of the scope of
the present paper; however, completing the argument or finding an
alternative is definitely worthwhile to consider, especially in view
of extending the result to other integrable quenches. 

The heuristic nature of the pairing argument makes it very important
to supplement the analytical considerations by numerical ones. First,
the iterative solution confirms to high precision that the proposed
factorized Ansatz (\ref{eq:Ansatz}) indeed solves the one-particle
test state condition, which is the lowest member of the hierarchy.
Second, an independent test of the squeezed state form, which includes
the pair structure and the exponentiation, was provided by checking
the next member of the hierarchy, i.e. the three-particle condition.
Note that since the Ansatz contains no free parameters, there is no
way to adjust it: it either fails or passes. In the end, the Ansatz
passed all the tests we could pose; while these are limited by computing
power, they still impose very stringent constraints. This shows that
the Ansatz is at least a very good approximation to the solution,
and raises the possibility that it is eventually exact; in this regard
it could be of interest to work out the integrable dressing argument
(cf. Subsection \ref{sub:Pair-structure-from}) in more detail for
the quench situation.

Our results show that an exponential cut-off regularisation of the
state of the form (\ref{eq:extrapol_length}) cannot provide a consistent
description of the initial state (or its asymptotic behaviour at large
times) in the limit of large initial mass $m_{0}$. This is because
the original amplitude $K(\theta)$ of (\ref{eq:Ansatz}) decays algebraically
as a function of momentum $p$ for large momenta (it behaves as $1/p^{2}$)
and therefore physical observables with different dependence on momentum
scales would exhibit different scaling behaviour in the Ansatz and
in the exponentially cut-off state as $m_{0}\to\infty$. In other
words, fixing the correspondence between $\tau_{0}$ and $m_{0}$
for some test observable would give inconsistent results for other
observables. A demonstration of this discrepancy was presented in
\cite{STM}. The same argument would hold for any other choice of
regularisation, different from the exponential cut-off of (\ref{eq:extrapol_length}),
unless it decays exactly as the Ansatz itself.

Interesting open issues include applying the present approach to quenches
in other field theoretical models, such as sine-Gordon theory or the
O(3) sigma model, that are more relevant for applications in condensed
matter. 

Another challenge is to provide a less heuristic argument for the
pair structure, or to prove the exponentiation of the initial state
directly from the hierarchy equations. One possible approach to tackle
this problem is to first focus on the Dirichlet case, for which both
the exponential form and the amplitude $K_{D}$ are fully determined,
in a completely different way, from the functional constraints imposed
by the boundary bootstrap \cite{GhoshalZamo}. Since the pole structure
and asymptotic behaviour of $K_{D}$ are linked to the above constraints,
which are sufficient to determine it, one expects that they must also
be sufficient (along with the properties of the form factors) in order
to verify analytically that it satisfies the hierarchy equations.
On the other hand, exponentiation may also be explained by a reduction
of the $N$-particle equation to the one-particle one in a way analogous
to the free case, though definitely more elaborate.

It would also be worthwhile to find more efficient numerical methods
to solve the hierarchy determining the overlap amplitudes.

\paragraph*{Acknowledgments }

The authors thank G. Mussardo and J. Cardy for fruitful discussions.
SS acknowledges financial support by the ERC under Starting Grant
279391 EDEQS, while HD and GT were supported by the Momentum grant
LP2012-50 of the Hungarian Academy of Sciences. This work was also
partially supported by the CNR-HAS bilateral grant SNK-84/2013.

\appendix

\section{Finite volume regularization of the integral equation hierarchy \label{sec:Finite-volume-regularization}}

Here we briefly summarize the finite volume regularization for the
hierarchy and show that after an appropriate redefinition of integration
contours the resulting equations are equivalent to those obtained
from the form factor representation (\ref{eq:genff_expansion}) of
the field.

\subsection{The finite volume regularization}

In a finite volume representation the initial state takes the form
\cite{Kormos:2010}
\begin{equation}
|B\rangle_{L}=|\Omega\rangle_{L}+\sum_{I'>0}N_{2}(\theta')_{L}K(\theta')|-\theta',\theta'\rangle_{L}+\sum_{I_{1}'>I_{2}'>0}N_{4}(\theta_{1}',\theta_{2}')_{L}K(\theta_{1}')K(\theta_{2}')|-\theta_{1}',\theta_{1}',-\theta_{2}',\theta_{2}'\rangle_{L}+\dots\:,
\end{equation}
where the summations run over all positive integer quantum numbers
for the particles, and the rapidities are related to the quantum numbers
by the Bethe-Yang equations:
\begin{equation}
\bar{Q}^{(2)}(\theta')=mL\sinh\theta'-i\log S(2\theta')=2\pi I
\end{equation}
for the two-particle, and 
\begin{eqnarray}
\bar{Q}_{1}^{(4)}(\theta_{1}',\theta_{2}') & = & mL\sinh\theta_{1}'-i\log S(2\theta_{1}')-i\log S(\theta_{1}'-\theta_{2}')-i\log S(\theta_{1}'+\theta_{2}')=2\pi I_{1}\nonumber \\
\bar{Q}_{2}^{(4)}(\theta_{1}',\theta_{2}') & = & mL\sinh\theta_{2}'-i\log S(2\theta_{2}')-i\log S(\theta_{2}'-\theta_{1}')-i\log S(\theta_{2}'+\theta_{1}')=2\pi I_{2}
\end{eqnarray}
for the four-particle term. These are quantization conditions for
states that are constrained to have a pair structure, and similar
equations can be written for higher particle numbers. 

We can also define the unconstrained multi-particle states
\[
|\theta_{1},\dots,\theta_{n}\rangle_{L}\:,
\]
which (up to corrections that vanish exponentially with the volume)
satisfy the Bethe-Yang quantization equations 
\begin{equation}
Q_{j}^{(n)}(\theta_{1},\dots,\theta_{n})=mL\sinh\theta_{j}-i\sum_{k\neq j}\log S(\theta_{j}-\theta_{k})=2\pi I_{j}\qquad j=1,\dots,n\:.\label{eq:generalBY}
\end{equation}
The density of unconstrained states in rapidity space is given by
the Jacobi determinant \cite{Pozsgay:2007kn}
\begin{equation}
\rho_{n}(\theta_{1},\dots,\theta_{n})_{L}=\det\left\{ \frac{\partial Q_{j}^{(n)}}{\partial\theta_{k}}\right\} _{j,k=1,\dots,n}\:,
\end{equation}
while those of the paired states read \cite{Kormos:2010}
\begin{equation}
\bar{\rho}_{2n}(\theta_{1}',\dots,\theta_{n}')_{L}=\det\left\{ \frac{\partial\bar{Q}_{j}^{(2n)}}{\partial\theta_{k}'}\right\} _{j,k=1,\dots,n}\:.
\end{equation}
In these notations, the finite volume normalization factors are \cite{Kormos:2010}
\begin{equation}
N_{2n}(\theta_{1}',\dots,\theta_{n}')_{L}=\frac{\sqrt{\rho_{2n}(-\theta_{1}',\theta_{1}',\dots-\theta_{n}',\theta_{n}')_{L}}}{\bar{\rho}_{2n}(\theta_{1}',\dots,\theta_{n}')_{L}}\:.
\end{equation}
Let us introduce the short-hand notation
\begin{equation}
\mathcal{O}(p)=E_{0}(p)\hat{\phi}(p)+\left[\hat{\phi}(p),H\right]\:,
\end{equation}
so that we can write the operator form of our hierarchy (\ref{eq:hierarchy_opform})
as 
\begin{equation}
\mathcal{O}(p)|B\rangle=0\:.
\end{equation}
In finite volume, we can write the following form for the $n$-particle
equation in the hierarchy
\begin{equation}
\sum_{k=0}^{\infty}\frac{1}{k!}\sum_{I_{1}'}\dots\sum_{I_{k}'}N_{2k}(\theta_{1}',\dots,\theta_{k}')_{L}K(\theta_{1}')\dots K(\theta_{k}')\langle\theta_{1},\dots,\theta_{n}|\mathcal{O}(p)|-\theta_{1}',\theta_{1}',\dots,-\theta_{k}',\theta_{k}'\rangle_{L}=0\:,\label{eq:finvol_hierarchy}
\end{equation}
where $\theta_{1},\dots,\theta_{n}$ must also be a state satisfying
the quantization relations with some arbitrarily chosen quantum numbers.
For the finite volume form factors of $\mathcal{O}$ we can use the
relation \cite{Pozsgay:2007kn,Pozsgay:2007gx}
\begin{equation}
\langle\theta_{1},\dots,\theta_{m}|\mathcal{O}(p)|\theta_{1}',\dots,\theta_{n}'\rangle_{L}=\frac{F_{m+n}^{\mathcal{O}(p)}(\theta_{m}+i\pi,\dots,\theta_{1}+i\pi,\theta_{1}',\dots,\theta_{n}')}{\sqrt{\rho_{m}(\theta_{1},\dots,\theta_{m})\rho(\theta_{1}',\dots,\theta_{n}')}}\label{eq:genff_finvol}
\end{equation}
valid up to exponential corrections in the volume $L$, where 
\begin{equation}
F_{n}^{\mathcal{O}(p)}(\theta_{1},\dots,\theta_{n})=\langle\Omega\mid\mathcal{O}(p)\mid\theta_{1},\theta_{2},\ldots,\theta_{n}\rangle_{in}
\end{equation}
are the infinite volume multi-particle form factors of $\mathcal{O}(p)$.
They can be expressed using the form factors of the elementary field
$\phi(x)$ as
\begin{eqnarray}
F_{n}^{\mathcal{O}(p)}(\theta_{1},\dots,\theta_{n}) & = & F_{n}^{\phi}(\theta_{1},\dots,\theta_{n})\left(E_{0}(p)+\sum_{i=1}^{n}m\cosh\theta_{i}\right)\delta(p+\sum_{i=1}^{n}m\sinh\theta_{i})\:,\nonumber \\
F_{n}^{\phi}(\theta_{1},\dots,\theta_{n}) & = & \langle\Omega\mid\phi(x=0)\mid\theta_{1},\theta_{2},\ldots,\theta_{n}\rangle_{in}\:.
\end{eqnarray}
Using these formulas, one can derive the equations satisfied by the
state $|B\rangle$ in the infinite volume limit by applying the methods
developed in \cite{Pozsgay:2010cr}. The trick is to rewrite the summation
over the quantum number using multi-dimensional residue integrals,
and open the contours which necessitates compensating for some additional
singularities by subtracting their residues. In the Supplementary
Material of the previous work \cite{STM} we have shown an explicit
example of this sort of calculation; therefore here we refrain from
going into more details.

For one-particle test states, the resulting equation up to (and including)
four-particle terms from the initial state can be written as (\ref{onept_upper})
valid as long as $\mbox{Im }\theta<\varepsilon$. 

For $\mbox{Im }\theta>\varepsilon$, the equation is changed by kinematic
poles of the form factors $F_{3}^{\phi}$ and $F_{5}^{\phi}$ crossing
the contours of integrations; the appropriate analytic continuation
of the equation reads (\ref{onept_lower}). In the limit $m_{0}\rightarrow\infty$,
the energy terms $E(\theta)$ can be dropped and the equation divided
through by $E_{0}(p)$, leading to the simplified form (\ref{onept_Dirichlet_upper})
for $\mbox{Im }\theta<\varepsilon$, and (\ref{onept_Dirichlet_lower})
for $\mbox{Im }\theta>\varepsilon$. These express the condition
\begin{equation}
\phi(x)|D\rangle=0
\end{equation}
satisfied by the Dirichlet state.

For three-particle test states, the result is 

\begin{equation}
\begin{aligned}0= & T_{0}+T_{2}+T_{4}+\dots\end{aligned}
\label{eq:NotCrossedoutBigFinMass-1}
\end{equation}
with $T_{n}$ denoting the $n$-particle contribution from the state
$|B\rangle$:
\begin{equation}
\begin{aligned}T_{0}= & \left[E_{0}(p)-E(\theta{}_{1})-E(\theta_{2})-E(\theta_{3})\right]F(\theta_{3}+i\pi,\theta_{2}+i\pi,\theta_{1}+i\pi)\end{aligned}
\:,
\end{equation}
\begin{equation}
\begin{aligned}T_{2}= & \frac{1}{2}\int_{-\infty}^{+\infty}\frac{\mathrm{d}\theta'}{2\pi}\Big[[E_{0}(p)-E(\theta_{1})-E(\theta_{2})-E(\theta_{3})+2E(\theta'+i\varepsilon)]\\
 & \;\times F(\theta_{3}+i\pi,\theta_{2}+i\pi,\theta_{1}+i\pi,-\theta'-i\varepsilon,\theta'+i\varepsilon)K(\theta'+i\varepsilon)\Big]\\
 & +\frac{1}{2}[E_{0}(p)+E(\theta_{1})-E(\theta_{2})-E(\theta_{3})][S(\theta_{2}-\theta_{1})S(\theta_{3}-\theta_{1})+S(-2\theta_{1})]F(\theta_{3}+i\pi,\theta_{2}+i\pi,-\theta_{1})K(\theta_{1})\\
 & +\frac{1}{2}[E_{0}(p)-E(\theta_{1})+E(\theta_{2})-E(\theta_{3})][S(\theta_{3}-\theta_{2})+S(\theta_{2}-\theta_{1})S(-2\theta_{2})]F(\theta_{3}+i\pi,\theta_{1}+i\pi,-\theta_{2})K(\theta_{2})\\
 & +\frac{1}{2}[E_{0}(p)-E(\theta_{1})-E(\theta_{2})+E(\theta_{3})][1+S(\theta_{3}-\theta_{2})S(\theta_{3}-\theta_{1})S(-2\theta_{3})]F(\theta_{2}+i\pi,\theta_{1}+i\pi,-\theta_{3})K(\theta_{3})\:,
\end{aligned}
\end{equation}
\begin{equation}
\begin{aligned}T_{4}= & \frac{1}{8}\int_{-\infty}^{+\infty}\frac{\mathrm{d}\theta'_{1}}{2\pi}\int_{-\infty}^{+\infty}\frac{\mathrm{d}\theta'_{2}}{2\pi}\Big[[E_{0}(p)-E(\theta_{1})-E(\theta_{2})-E(\theta_{3})+2E(\theta_{1}'+i\varepsilon)+2E(\theta_{2}'+i\varepsilon)]\\
 & \;\times F(\theta_{3}+i\pi,\theta_{2}+i\pi,\theta_{1}+i\pi,-\theta'_{1}-i\varepsilon,\theta'_{1}+i\varepsilon,-\theta'_{2}-i\varepsilon,\theta'_{2}+i\varepsilon)K(\theta'_{1}+i\varepsilon)K(\theta'_{2}+i\varepsilon)\Big]\\
 & +\frac{1}{4}\int_{-\infty}^{+\infty}\frac{\mathrm{d}\theta'}{2\pi}[E_{0}(p)+E(\theta_{1})-E(\theta_{2})-E(\theta_{3})+2E(\theta'+i\varepsilon)]F(\theta_{3}+i\pi,\theta_{2}+i\pi,-\theta_{1},-\theta'-i\varepsilon,\theta'+i\varepsilon)\\
 & \;\times[S(\theta_{3}-\theta_{1})S(\theta_{2}-\theta_{1})S(\theta_{1}-\theta'-i\varepsilon)S(\theta_{1}+\theta'+i\varepsilon)+S(-2\theta_{1})]K(\theta'+i\varepsilon)K(\theta{}_{1})\\
 & +\frac{1}{4}\int_{-\infty}^{+\infty}\frac{\mathrm{d}\theta'}{2\pi}[E_{0}(p)-E(\theta_{1})+E(\theta_{2})-E(\theta_{3})+2E(\theta'+i\varepsilon)]F(\theta_{3}+i\pi,\theta_{1}+i\pi,-\theta_{2},-\theta'-i\varepsilon,\theta'+i\varepsilon)\\
 & \;\times[S(\theta_{3}-\theta_{2})S(\theta_{2}-\theta'-i\varepsilon)S(\theta_{2}+\theta'+i\varepsilon)+S(\theta_{2}-\theta_{1})S(-2\theta_{2})]K(\theta'+i\varepsilon)K(\theta{}_{2})\\
 & +\frac{1}{4}\int_{-\infty}^{+\infty}\frac{\mathrm{d}\theta'}{2\pi}[E_{0}(p)-E(\theta_{1})-E(\theta_{2})+E(\theta_{3})+2E(\theta'+i\varepsilon)]F(\theta_{2}+i\pi,\theta_{1}+i\pi,-\theta_{3},-\theta'-i\varepsilon,\theta'+i\varepsilon)\\
 & \;\times[S(\theta_{3}-\theta'-i\varepsilon)S(\theta_{3}+\theta'+i\varepsilon)+S(\theta_{3}-\theta_{1})S(\theta_{3}-\theta_{2})S(-2\theta_{3})]K(\theta'+i\varepsilon)K(\theta{}_{3})\\
 & +\frac{1}{4}[E_{0}(p)+E(\theta_{1})+E(\theta_{2})-E(\theta_{3})]F(\theta_{3}+i\pi,-\theta_{1},-\theta{}_{2})[S(\theta_{3}-\theta_{1})S(\theta_{2}+\theta_{1})S(\theta_{3}-\theta_{2})K(\theta_{1})K(\theta_{2})\\
 & \;+S(\theta_{3}-\theta_{1})K(\theta_{1})K(-\theta_{2})+S(\theta_{3}-\theta_{2})K(-\theta_{1})K(\theta_{2})+S(-\theta_{2}-\theta_{1})K(-\theta_{1})K(-\theta_{2})]\\
 & +\frac{1}{4}[E_{0}(p)+E(\theta_{1})-E(\theta_{2})+E(\theta_{3})]F(\theta_{2}+i\pi,-\theta_{1},-\theta{}_{3})[S(\theta_{2}-\theta_{1})S(\theta_{3}+\theta_{1})K(\theta_{1})K(\theta_{3})\\
 & \;+S(\theta_{2}-\theta_{1})S(\theta_{3}-\theta_{2})K(\theta_{1})K(-\theta_{3})+K(-\theta_{1})K(\theta_{3})+S(\theta_{3}-\theta_{2})S(-\theta_{3}-\theta_{1})K(-\theta_{1})K(-\theta_{3})]\\
 & +\frac{1}{4}[E_{0}(p)-E(\theta_{1})+E(\theta_{2})+E(\theta_{3})]F(\theta_{1}+i\pi,-\theta_{2},-\theta{}_{3})[S(\theta_{2}+\theta_{3})K(\theta_{2})K(\theta_{3})\\
 & \;+S(\theta_{3}-\theta_{1})K(\theta_{2})K(-\theta_{3})+S(\theta_{2}-\theta_{1})K(-\theta_{2})K(\theta_{3})\\
 & \;+S(-\theta_{3}-\theta_{2})S(\theta_{3}-\theta_{1})S(\theta_{2}-\theta_{1})K(-\theta_{1})K(-\theta_{3})]\:,
\end{aligned}
\end{equation}
and $p=m\sinh\theta_{1}+m\sinh\theta_{2}+m\sinh\theta_{3}$, valid
as long as $\mbox{Im }\theta_{i}<\varepsilon$. For other complex
values of the test rapidities it can be continued analytically similarly
to the one-particle equation; the condition satisfied by the Dirichlet
function $K_{D}$ can be obtained by dropping the terms containing
combinations of $E_{0}$ and $E$ in the square brackets.

\subsection{Comparing to the infinite volume formalism}

The equation hierarchy can be obtained directly from substituting
the infinite volume matrix element (\ref{eq:general_ff}) into (\ref{eq:hierarchy_explct}).
Considering the case of a one-particle test state, from (\ref{eq:oneparticle_equation_explicit})
one obtains 
\begin{equation}
\begin{aligned}0= & [E_{0}(p)-E(\theta)]F_{1}^{\phi}+[E_{0}(p)+E(\theta)]F_{1}^{\phi}K(\theta)\\
 & +\frac{1}{2}\int_{-\infty}^{+\infty}\frac{\, d\theta'}{2\pi}[E_{0}(p)-E(\theta)+2E(\theta')]F_{3}^{\phi}(\theta+i\pi+i0,-\theta'-i0,\theta'-i0)K(\theta')\\
 & +\frac{1}{2}\int_{-\infty}^{+\infty}\frac{\, d\theta'}{2\pi}[E_{0}(p)+E(\theta)+2E(\theta')]F_{3}^{\phi}(-\theta'-i0,\theta'-i0,-\theta-i0)K(\theta')K(\theta)\\
 & +\frac{1}{8}\int_{-\infty}^{+\infty}\hspace{-0.195cm}\frac{\, d\theta_{1}'}{2\pi}\int_{-\infty}^{+\infty}\hspace{-0.195cm}\frac{\, d\theta_{2}'}{2\pi}[E_{0}(p)-E(\theta)+2E(\theta_{1}')+2E(\theta_{2}')]\\
 & \hspace{1cm}\times F_{5}^{\phi}(\theta+i\pi+i0,-\theta_{1}'-i0,\theta_{1}'-i0,-\theta_{2}'-i0,\theta_{2}'-i0)K(\theta_{1}')K(\theta_{2}')+...\:.
\end{aligned}
\label{onept_aqft}
\end{equation}
In the finite volume formula it is necessary to take the form (\ref{onept_lower})
valid for $\mbox{Im}\theta>\varepsilon$ to have the same ordering
of the imaginary parts between the unprimed and primed rapidity variables
as in (\ref{onept_aqft}) above. Shifting back the contours to the
real axis, and absorbing $S$-matrix factors by reordering the rapidity
variables in the corresponding form factor gives
\[
\begin{aligned}0= & [E_{0}(p)-E(\theta)]F_{1}^{\phi}+[E_{0}(p)+E(\theta)]F_{1}^{\phi}K(\theta)\\
 & +\frac{1}{2}\int_{-\infty}^{+\infty}\frac{\, d\theta'}{2\pi}[E_{0}(p)-E(\theta)+2E(\theta')]F_{3}^{\phi}(\theta+i\pi+i\varepsilon,-\theta',\theta')K(\theta')\\
 & +\frac{1}{2}\int_{-\infty}^{+\infty}\frac{\, d\theta'}{2\pi}[E_{0}(p)+E(\theta)+2E(\theta')]K(\theta)F_{3}^{\phi}(-\theta',\theta',-\theta)K(\theta')\\
 & +\frac{1}{8}\int_{-\infty}^{+\infty}\hspace{-0.195cm}\frac{\, d\theta_{1}'}{2\pi}\int_{-\infty}^{+\infty}\hspace{-0.195cm}\frac{\, d\theta_{2}'}{2\pi}[E_{0}(p)-E(\theta)+2E(\theta_{1}')+2E(\theta_{2}')]\\
 & \hspace{1cm}\times F_{5}^{\phi}(\theta+i\pi+i\varepsilon,-\theta_{1}',\theta_{1}',-\theta_{2}',\theta_{2}')K(\theta_{1}')K(\theta_{2}')+...\:.
\end{aligned}
\]
 Due to the $+i0$ shifts in the unprimed rapidities, the $-i0$ shifts
in (\ref{onept_aqft}) can be eliminated, making the two equations
identical. 

Similar identity can be demonstrated for the three-test particle condition;
as it contains no essential novelty compared to the one-particle case,
for the sake of brevity we omit the details here.

\section{\label{sec:Three-particle-condition}Tables for the numerical verification
of the three-particle condition }

The tables in this Section contain a sample of numerical data obtained
from numerical evaluation of the three-particle condition (\ref{eq:NotCrossedoutBigFinMass-1}).
The first three terms $T_{0}$, $T_{2}$ and $T_{4}$ were evaluated
using the integral formulae. The first line labeled ``sum'' gives
the sum of these three terms, and verifies how precisely the condition
is satisfied in this truncation. For the evaluation of $T_{6}$ it
proved practical to use the finite volume sum form (\ref{eq:genff_finvol}).
The second line labeled ``sum'' gives the value of the three-particle
condition once the computed result for $T_{6}$ is included as well.

\subsection{B=0.1}

{\footnotesize{}}%
\begin{tabular}{|c|c|c|c|c|c|c|}
\hline 
\multicolumn{4}{|c|}{{\footnotesize{}$\{\theta_{1},\theta_{2},\theta_{3}\}=\{0.0607,0.1277,0.2606\}$}} & \multicolumn{3}{c|}{{\footnotesize{}$\{\theta_{1},\theta_{2},\theta_{3}\}=\{-0.0866,0.0495,0.1621\}$}}\tabularnewline
\hline 
{\footnotesize{}$m_{0}$ } & {\footnotesize{}$\infty$ } & {\footnotesize{}10 } & {\footnotesize{}2 } & {\footnotesize{}$\infty$ } & {\footnotesize{}10 } & {\footnotesize{}2 }\tabularnewline
\hline 
{\footnotesize{}$T_{0}$ } & {\footnotesize{}-0.0138-0.0067 i } & {\footnotesize{}-0.0963-0.0464 i } & {\footnotesize{}0.0137 +0.0066 i } & {\footnotesize{}-0.0175-0.0175 i } & {\footnotesize{}-0.1222-0.1222 i } & {\footnotesize{}0.0178 +0.0178 i }\tabularnewline
\hline 
{\footnotesize{}$T_{2}$ } & {\footnotesize{}-1.8826-0.908 i } & {\footnotesize{}-13.8152-6.6628 i } & {\footnotesize{}-0.6397-0.3085 i } & {\footnotesize{}26.5853 +26.5881 i } & {\footnotesize{}195.393 +195.413 i } & {\footnotesize{}8.7483 +8.7492 i }\tabularnewline
\hline 
{\footnotesize{}$T_{4}$ } & {\footnotesize{}1.9096 +0.9209 i } & {\footnotesize{}13.9958 +6.7499 i } & {\footnotesize{}0.6241 +0.301 i } & {\footnotesize{}-26.9231-26.9261 i } & {\footnotesize{}-197.666-197.688 i } & {\footnotesize{}-8.7886-8.7895 i }\tabularnewline
\hline 
{\footnotesize{}sum } & {\footnotesize{}0.0131 +0.0063 i } & {\footnotesize{}0.0843 +0.0406 i } & {\footnotesize{}-0.0019-0.0009 i } & {\footnotesize{}-0.3553-0.3556 i } & {\footnotesize{}-2.3951-2.3972 i } & {\footnotesize{}-0.0225-0.0226 i }\tabularnewline
\hline 
{\footnotesize{}$T_{6}$ } & {\footnotesize{}-0.0122-0.0059 i } & {\footnotesize{}-0.0826-0.0398 i } & {\footnotesize{}-0.0015-0.0007 i } & {\footnotesize{}0.3523 +0.3524 i } & {\footnotesize{}2.4211 +2.4213 i } & {\footnotesize{}0.0492 +0.0492 i }\tabularnewline
\hline 
{\footnotesize{}sum } & {\footnotesize{}0.0009 +0.0004 i } & {\footnotesize{}0.0017 +0.0008 i } & {\footnotesize{}-0.0034-0.0016 i } & {\footnotesize{}-0.0029-0.0032 i } & {\footnotesize{}0.026 +0.0241 i } & {\footnotesize{}0.0267 +0.0266 i }\tabularnewline
\hline 
\end{tabular}{\footnotesize{}}\\
{\footnotesize{}~}\\
{\footnotesize{}}%
\begin{tabular}{|c|c|c|c|c|c|c|}
\hline 
\multicolumn{4}{|c|}{{\footnotesize{}$\{\theta_{1},\theta_{2},\theta_{3}\}=\{-0.078,0.4879,0.2606\}$}} & \multicolumn{3}{c|}{{\footnotesize{}$\{\theta_{1},\theta_{2},\theta_{3}\}=\{0.7633,0.8322,0.9462\}$}}\tabularnewline
\hline 
{\footnotesize{}$m_{0}$ } & {\footnotesize{}$\infty$ } & {\footnotesize{}10 } & {\footnotesize{}2 } & {\footnotesize{}$\infty$ } & {\footnotesize{}10 } & {\footnotesize{}2 }\tabularnewline
\hline 
{\footnotesize{}$T_{0}$ } & {\footnotesize{}0.0415 -0.0275 i } & {\footnotesize{}0.2859 -0.1889 i } & {\footnotesize{}-0.0446+0.0295 i } & {\footnotesize{}-0.013-0.0046 i } & {\footnotesize{}-0.0813-0.0289 i } & {\footnotesize{}0.0086 +0.0031 i }\tabularnewline
\hline 
{\footnotesize{}$T_{2}$ } & {\footnotesize{}-21.9148+14.4858 i } & {\footnotesize{}-159.244+105.261 i } & {\footnotesize{}-6.7734+4.4773 i } & {\footnotesize{}-0.0402-0.0143 i } & {\footnotesize{}-0.2756-0.0978 i } & {\footnotesize{}-0.0201-0.0071 i }\tabularnewline
\hline 
{\footnotesize{}$T_{4}$ } & {\footnotesize{}22.1749 -14.6577 i } & {\footnotesize{}160.966 -106.399 i } & {\footnotesize{}6.7582 -4.4672 i } & {\footnotesize{}0.0424 +0.0151 i } & {\footnotesize{}0.2876 +0.1021 i } & {\footnotesize{}0.0109 +0.0039 i }\tabularnewline
\hline 
{\footnotesize{}sum } & {\footnotesize{}0.3016 -0.1994 i } & {\footnotesize{}2.0086 -1.3278 i } & {\footnotesize{}-0.0598+0.0395 i } & {\footnotesize{}-0.0108-0.0038 i } & {\footnotesize{}-0.0692-0.0246 i } & {\footnotesize{}-0.0006-0.0002 i }\tabularnewline
\hline 
{\footnotesize{}$T_{6}$ } & {\footnotesize{}-0.3085+0.2039 i } & {\footnotesize{}-2.0947+1.3846 i } & {\footnotesize{}-0.0402+0.0265 i } & {\footnotesize{}0.0112 +0.004 i } & {\footnotesize{}0.0695 +0.0247 i } & {\footnotesize{}0.0009 +0.0003 i }\tabularnewline
\hline 
{\footnotesize{}sum } & {\footnotesize{}-0.0068+0.0045 i } & {\footnotesize{}-0.0861+0.0568 i } & {\footnotesize{}-0.1+0.0661 i } & {\footnotesize{}0.0003 +0.0001 i } & {\footnotesize{}0.0003 +0.0001 i } & {\footnotesize{}0.0003 +0.0001 i }\tabularnewline
\hline 
\end{tabular}{\footnotesize{}}\\
{\footnotesize{}~}\\
{\footnotesize{}}%
\begin{tabular}{|c|c|c|c|c|c|c|}
\hline 
\multicolumn{4}{|c|}{{\footnotesize{}$\{\theta_{1},\theta_{2},\theta_{3}\}=\{-0.9706,1.5852,-2.13\}$}} & \multicolumn{3}{c|}{{\footnotesize{}$\{\theta_{1},\theta_{2},\theta_{3}\}=\{0.5267,1.5444,2.1303\}$}}\tabularnewline
\hline 
{\footnotesize{}$m_{0}$ } & {\footnotesize{}$\infty$ } & {\footnotesize{}10 } & {\footnotesize{}2 } & {\footnotesize{}$\infty$ } & {\footnotesize{}10 } & {\footnotesize{}2 }\tabularnewline
\hline 
{\footnotesize{}$T_{0}$ } & {\footnotesize{}0.0095 +0.0009 i } & {\footnotesize{}0.0199 +0.0018 i } & {\footnotesize{}-0.0451-0.0041 i } & {\footnotesize{}0.0411 -0.0194 i } & {\footnotesize{}0.177 -0.0838 i } & {\footnotesize{}-0.0263+0.0124 i }\tabularnewline
\hline 
{\footnotesize{}$T_{2}$ } & {\footnotesize{}-0.6149-0.0565 i } & {\footnotesize{}-2.6313-0.2417 i } & {\footnotesize{}0.0003 } & {\footnotesize{}0.0509 -0.0241 i } & {\footnotesize{}0.2783 -0.1317 i } & {\footnotesize{}0.0221 -0.0104 i }\tabularnewline
\hline 
{\footnotesize{}$T_{4}$ } & {\footnotesize{}0.6181 +0.0568 i } & {\footnotesize{}2.687 +0.2469 i } & {\footnotesize{}0.0311 +0.0029 i } & {\footnotesize{}-0.0542+0.0257 i } & {\footnotesize{}-0.2771+0.1311 i } & {\footnotesize{}-0.0052+0.0024 i }\tabularnewline
\hline 
{\footnotesize{}sum } & {\footnotesize{}0.0127 +0.0012 i } & {\footnotesize{}0.0757 +0.007 i } & {\footnotesize{}-0.0137-0.0013 i } & {\footnotesize{}0.0377 -0.0179 i } & {\footnotesize{}0.1783 -0.0844 i } & {\footnotesize{}-0.0094+0.0044 i }\tabularnewline
\hline 
{\footnotesize{}$T_{6}$ } & {\footnotesize{}0.0044 +0.0004 i } & {\footnotesize{}0.0191 +0.0018 i } & {\footnotesize{}0.0001 } & {\footnotesize{}0.0016 -0.0007 i } & {\footnotesize{}0.008 -0.0038 i } & {\footnotesize{}0.0001 }\tabularnewline
\hline 
{\footnotesize{}sum } & {\footnotesize{}0.0171 +0.0016 i } & {\footnotesize{}0.0948 +0.0087 i } & {\footnotesize{}-0.0136-0.0012 i } & {\footnotesize{}0.0393 -0.0186 i } & {\footnotesize{}0.1862 -0.0881 i } & {\footnotesize{}-0.0093+0.0044 i }\tabularnewline
\hline 
\end{tabular}{\footnotesize \par}

\subsection{B=0.5}

{\footnotesize{}}%
\begin{tabular}{|c|c|c|c|c|c|c|}
\hline 
\multicolumn{4}{|c|}{{\footnotesize{}$\{\theta_{1},\theta_{2},\theta_{3}\}=\{0.0534,0.1261,0.2692\}$}} & \multicolumn{3}{c|}{{\footnotesize{}$\{\theta_{1},\theta_{2},\theta_{3}\}=\{-0.0958,0.05,0.1708\}$}}\tabularnewline
\hline 
{\footnotesize{}$m_{0}$ } & {\footnotesize{}$\infty$ } & {\footnotesize{}10 } & {\footnotesize{}2 } & {\footnotesize{}$\infty$ } & {\footnotesize{}10 } & {\footnotesize{}2 }\tabularnewline
\hline 
{\footnotesize{}$T_{0}$ } & {\footnotesize{}-0.005+0.0072 i } & {\footnotesize{}-0.0346+0.0504 i } & {\footnotesize{}0.0049 -0.0072 i } & {\footnotesize{}-0.012+0.0132 i } & {\footnotesize{}-0.0836+0.0919 i } & {\footnotesize{}0.0122 -0.0134 i }\tabularnewline
\hline 
{\footnotesize{}$T_{2}$ } & {\footnotesize{}-0.5063+0.7382 i } & {\footnotesize{}-3.7154+5.4174 i } & {\footnotesize{}-0.1732+0.2526 i } & {\footnotesize{}10.3305 -11.353 i } & {\footnotesize{}75.9036 -83.4169 i } & {\footnotesize{}3.3881 -3.7235 i }\tabularnewline
\hline 
{\footnotesize{}$T_{4}$ } & {\footnotesize{}0.5317 -0.7753 i } & {\footnotesize{}3.8838 -5.6629 i } & {\footnotesize{}0.1681 -0.2452 i } & {\footnotesize{}-10.7851+11.8527 i } & {\footnotesize{}-78.8828+86.6909 i } & {\footnotesize{}-3.4052+3.7423 i }\tabularnewline
\hline 
{\footnotesize{}sum } & {\footnotesize{}0.0205 -0.0299 i } & {\footnotesize{}0.1338 -0.1951 i } & {\footnotesize{}-0.0001+0.0002 i } & {\footnotesize{}-0.4666+0.5128 i } & {\footnotesize{}-3.0628+3.366 i } & {\footnotesize{}-0.005+0.0055 i }\tabularnewline
\hline 
{\footnotesize{}$T_{6}$ } & {\footnotesize{}-0.0201+0.0293 i } & {\footnotesize{}-0.134+0.1954 i } & {\footnotesize{}-0.0023+0.0033 i } & {\footnotesize{}0.4811 -0.5287 i } & {\footnotesize{}3.2224 -3.5414 i } & {\footnotesize{}0.0552 -0.0606 i }\tabularnewline
\hline 
{\footnotesize{}sum } & {\footnotesize{}0.0004 -0.0006 i } & {\footnotesize{}-0.0002+0.0003 i } & {\footnotesize{}-0.0024+0.0035 i } & {\footnotesize{}0.0145 -0.0159 i } & {\footnotesize{}0.1596 -0.1754 i } & {\footnotesize{}0.0502 -0.0552 i }\tabularnewline
\hline 
\end{tabular}{\footnotesize{}}\\
{\footnotesize{}~}\\
{\footnotesize{}}%
\begin{tabular}{|c|c|c|c|c|c|c|}
\hline 
\multicolumn{4}{|c|}{{\footnotesize{}$\{\theta_{1},\theta_{2},\theta_{3}\}=\{-0.0775,-0.4968,0.1173\}$}} & \multicolumn{3}{c|}{{\footnotesize{}$\{\theta_{1},\theta_{2},\theta_{3}\}=\{0.7577,0.8313,0.9519\}$}}\tabularnewline
\hline 
{\footnotesize{}$m_{0}$ } & {\footnotesize{}$\infty$ } & {\footnotesize{}10 } & {\footnotesize{}2 } & {\footnotesize{}$\infty$ } & {\footnotesize{}10 } & {\footnotesize{}2 }\tabularnewline
\hline 
{\footnotesize{}$T_{0}$ } & {\footnotesize{}0.0579 -0.1144 i } & {\footnotesize{}0.398 -0.7868 i } & {\footnotesize{}-0.0625+0.1236 i } & {\footnotesize{}-0.0035+0.0059 i } & {\footnotesize{}-0.0221+0.0366 i } & {\footnotesize{}0.0023 -0.0039 i }\tabularnewline
\hline 
{\footnotesize{}$T_{2}$ } & {\footnotesize{}-16.5658+32.7473 i } & {\footnotesize{}-120.284+237.778 i } & {\footnotesize{}-5.0646+10.0117 i } & {\footnotesize{}-0.0086+0.0143 i } & {\footnotesize{}-0.0597+0.0991 i } & {\footnotesize{}-0.0047+0.0078 i }\tabularnewline
\hline 
{\footnotesize{}$T_{4}$ } & {\footnotesize{}17.2361 -34.0724 i } & {\footnotesize{}124.58 -246.27 i } & {\footnotesize{}5.1212 -10.1237 i } & {\footnotesize{}0.0104 -0.0173 i } & {\footnotesize{}0.0712 -0.118 i } & {\footnotesize{}0.0026 -0.0043 i }\tabularnewline
\hline 
{\footnotesize{}sum } & {\footnotesize{}0.7282 -1.4395 i } & {\footnotesize{}4.6937 -9.2786 i } & {\footnotesize{}-0.0059+0.0116 i } & {\footnotesize{}-0.0017+0.0028 i } & {\footnotesize{}-0.0106+0.0176 i } & {\footnotesize{}0.0002 -0.0003 i }\tabularnewline
\hline 
{\footnotesize{}$T_{6}$ } & {\footnotesize{}-0.7247+1.4325 i } & {\footnotesize{}-4.7758+9.4408 i } & {\footnotesize{}-0.0747+0.1476 i } & {\footnotesize{}0.0017 -0.0028 i } & {\footnotesize{}0.01 -0.0166 i } & {\footnotesize{}0.0001 -0.0002 i }\tabularnewline
\hline 
{\footnotesize{}sum } & {\footnotesize{}0.0035 -0.007 i } & {\footnotesize{}-0.0821+0.1622 i } & {\footnotesize{}-0.0806+0.1593 i } & {\footnotesize{}-0.0001+0.0001 i } & {\footnotesize{}-0.0006+0.001 i } & {\footnotesize{}0.0003 -0.0005 i }\tabularnewline
\hline 
\end{tabular}{\footnotesize{}}\\
{\footnotesize{}~}\\
{\footnotesize{}}%
\begin{tabular}{|c|c|c|c|c|c|c|}
\hline 
\multicolumn{4}{|c|}{{\footnotesize{}$\{\theta_{1},\theta_{2},\theta_{3}\}=\{-0.9692,1.5855,-2.1307\}$}} & \multicolumn{3}{c|}{{\footnotesize{}$\{\theta_{1},\theta_{2},\theta_{3}\}=\{0.5223,1.5347,2.1319\}$}}\tabularnewline
\hline 
{\footnotesize{}$m_{0}$ } & {\footnotesize{}$\infty$ } & {\footnotesize{}10 } & {\footnotesize{}2 } & {\footnotesize{}$\infty$ } & {\footnotesize{}10 } & {\footnotesize{}2 }\tabularnewline
\hline 
{\footnotesize{}$T_{0}$ } & {\footnotesize{}0.1249 +0.0499 i } & {\footnotesize{}0.2626 +0.105 i } & {\footnotesize{}-0.5949-0.238 i } & {\footnotesize{}-0.0439-0.4658 i } & {\footnotesize{}-0.1891-2.0068 i } & {\footnotesize{}0.0282 +0.2994 i }\tabularnewline
\hline 
{\footnotesize{}$T_{2}$ } & {\footnotesize{}-6.3313-2.5327 i } & {\footnotesize{}-26.9395-10.7763 i } & {\footnotesize{}0.0664 +0.0266 i } & {\footnotesize{}-0.0486-0.5161 i } & {\footnotesize{}-0.2742-2.9104 i } & {\footnotesize{}-0.0215-0.2286 i }\tabularnewline
\hline 
{\footnotesize{}$T_{4}$ } & {\footnotesize{}6.4396 +2.576 i } & {\footnotesize{}27.724 +11.0901 i } & {\footnotesize{}0.3068 +0.1227 i } & {\footnotesize{}0.0606 +0.6434 i } & {\footnotesize{}0.3182 +3.3773 i } & {\footnotesize{}0.0054 +0.057 i }\tabularnewline
\hline 
{\footnotesize{}sum } & {\footnotesize{}0.2331 +0.0932 i } & {\footnotesize{}1.0471 +0.4189 i } & {\footnotesize{}-0.2216-0.0887 i } & {\footnotesize{}-0.0319-0.3384 i } & {\footnotesize{}-0.1451-1.54 i } & {\footnotesize{}0.012 +0.1278 i }\tabularnewline
\hline 
{\footnotesize{}$T_{6}$ } & {\footnotesize{}0.0389 +0.0156 i } & {\footnotesize{}0.1561 +0.0624 i } & {\footnotesize{}0.0004 +0.0002 i } & {\footnotesize{}0.0003 +0.0029 i } & {\footnotesize{}0.0019 +0.0203 i } & {\footnotesize{}0.0001 +0.0005 i }\tabularnewline
\hline 
{\footnotesize{}sum } & {\footnotesize{}0.272 +0.1088 i } & {\footnotesize{}1.2032 +0.4813 i } & {\footnotesize{}-0.2212-0.0885 i } & {\footnotesize{}-0.0316-0.3355 i } & {\footnotesize{}-0.1432-1.5197 i } & {\footnotesize{}0.0121 +0.1283 i }\tabularnewline
\hline 
\end{tabular}{\footnotesize \par}

\subsection{B=0.9}

{\footnotesize{}}%
\begin{tabular}{|c|c|c|c|c|c|c|}
\hline 
\multicolumn{4}{|c|}{{\footnotesize{}$\{\theta_{1},\theta_{2},\theta_{3}\}=\{0.0526,0.1259,0.2702\}$}} & \multicolumn{3}{c|}{{\footnotesize{}$\{\theta_{1},\theta_{2},\theta_{3}\}=\{-0.097,0.05,0.1719\}$}}\tabularnewline
\hline 
{\footnotesize{}$m_{0}$ } & {\footnotesize{}$\infty$ } & {\footnotesize{}10 } & {\footnotesize{}2 } & {\footnotesize{}$\infty$ } & {\footnotesize{}10 } & {\footnotesize{}2 }\tabularnewline
\hline 
{\footnotesize{}$T_{0}$ } & {\footnotesize{}-0.0039+0.0084 i } & {\footnotesize{}-0.0275+0.0586 i } & {\footnotesize{}0.0039 -0.0084 i } & {\footnotesize{}-0.0098+0.0164 i } & {\footnotesize{}-0.0686+0.1144 i } & {\footnotesize{}0.01 -0.0167 i }\tabularnewline
\hline 
{\footnotesize{}$T_{2}$ } & {\footnotesize{}-0.3545+0.7563 i } & {\footnotesize{}-2.6022+5.5512 i } & {\footnotesize{}-0.1218+0.2598 i } & {\footnotesize{}7.2362 -12.0605 i } & {\footnotesize{}53.1652 -88.6102 i } & {\footnotesize{}2.3711 -3.9519 i }\tabularnewline
\hline 
{\footnotesize{}$T_{4}$ } & {\footnotesize{}0.3772 -0.8046 i } & {\footnotesize{}2.7507 -5.868 i } & {\footnotesize{}0.1176 -0.2509 i } & {\footnotesize{}-7.6368+12.7282 i } & {\footnotesize{}-55.7598+92.9346 i } & {\footnotesize{}-2.3777+3.963 i }\tabularnewline
\hline 
{\footnotesize{}sum } & {\footnotesize{}0.0187 -0.0399 i } & {\footnotesize{}0.121 -0.2582 i } & {\footnotesize{}-0.0002+0.0005 i } & {\footnotesize{}-0.4105+0.6841 i } & {\footnotesize{}-2.6632+4.4387 i } & {\footnotesize{}0.0034 -0.0057 i }\tabularnewline
\hline 
{\footnotesize{}$T_{6}$ } & {\footnotesize{}-0.0179+0.0381 i } & {\footnotesize{}-0.1182+0.2523 i } & {\footnotesize{}-0.0019+0.0041 i } & {\footnotesize{}0.4182 -0.697 i } & {\footnotesize{}2.7766 -4.6278 i } & {\footnotesize{}0.045 -0.0751 i }\tabularnewline
\hline 
{\footnotesize{}sum } & {\footnotesize{}0.0009 -0.0018 i } & {\footnotesize{}0.0028 -0.006 i } & {\footnotesize{}-0.0021+0.0046 i } & {\footnotesize{}0.0077 -0.0128 i } & {\footnotesize{}0.1134 -0.189 i } & {\footnotesize{}0.0484 -0.0807 i }\tabularnewline
\hline 
\end{tabular}{\footnotesize{}}\\
{\footnotesize{}~}\\
{\footnotesize{}}%
\begin{tabular}{|c|c|c|c|c|c|c|}
\hline 
\multicolumn{4}{|c|}{{\footnotesize{}$\{\theta_{1},\theta_{2},\theta_{3}\}=\{-0.0771,-0.4988,0.1191\}$}} & \multicolumn{3}{c|}{{\footnotesize{}$\{\theta_{1},\theta_{2},\theta_{3}\}=\{0.7571,0.8312,0.9526\}$}}\tabularnewline
\hline 
{\footnotesize{}$m_{0}$ } & {\footnotesize{}$\infty$ } & {\footnotesize{}10 } & {\footnotesize{}2 } & {\footnotesize{}$\infty$ } & {\footnotesize{}10 } & {\footnotesize{}2 }\tabularnewline
\hline 
{\footnotesize{}$T_{0}$ } & {\footnotesize{}0.0582 -0.1498 i } & {\footnotesize{}0.4 -1.0295 i } & {\footnotesize{}-0.0629+0.1619 i } & {\footnotesize{}-0.0027+0.0066 i } & {\footnotesize{}-0.0171+0.0412 i } & {\footnotesize{}0.0018 -0.0044 i }\tabularnewline
\hline 
{\footnotesize{}$T_{2}$ } & {\footnotesize{}-13.8111+35.5416 i } & {\footnotesize{}-100.262+258.014 i } & {\footnotesize{}-4.2068+10.8259 i } & {\footnotesize{}-0.006+0.0145 i } & {\footnotesize{}-0.042+0.1009 i } & {\footnotesize{}-0.0034+0.0083 i }\tabularnewline
\hline 
{\footnotesize{}$T_{4}$ } & {\footnotesize{}14.5091 -37.3378 i } & {\footnotesize{}104.671 -269.361 i } & {\footnotesize{}4.2464 -10.9278 i } & {\footnotesize{}0.0077 -0.0186 i } & {\footnotesize{}0.0527 -0.1268 i } & {\footnotesize{}0.0019 -0.0045 i }\tabularnewline
\hline 
{\footnotesize{}sum } & {\footnotesize{}0.7562 -1.9459 i } & {\footnotesize{}4.8094 -12.3765 i } & {\footnotesize{}-0.0233+0.06 i } & {\footnotesize{}-0.001+0.0025 i } & {\footnotesize{}-0.0064+0.0153 i } & {\footnotesize{}0.0003 -0.0006 i }\tabularnewline
\hline 
{\footnotesize{}$T_{6}$ } & {\footnotesize{}-0.7363+1.8948 i } & {\footnotesize{}-4.804+12.3627 i } & {\footnotesize{}-0.0705+0.1815 i } & {\footnotesize{}0.0008 -0.0019 i } & {\footnotesize{}0.0048 -0.0116 i } & {\footnotesize{}0.0001 -0.0001 i }\tabularnewline
\hline 
{\footnotesize{}sum } & {\footnotesize{}0.0198 -0.0511 i } & {\footnotesize{}0.0054 -0.0138 i } & {\footnotesize{}-0.0938+0.2414 i } & {\footnotesize{}-0.0002+0.0006 i } & {\footnotesize{}-0.0015+0.0037 i } & {\footnotesize{}0.0003 -0.0007 i }\tabularnewline
\hline 
\end{tabular}{\footnotesize{}}\\
{\footnotesize{}~}\\
{\footnotesize{}}%
\begin{tabular}{|c|c|c|c|c|c|c|}
\hline 
\multicolumn{4}{|c|}{{\footnotesize{}$\{\theta_{1},\theta_{2},\theta_{3}\}=\{-0.9687,1.5857,-2.131\}$}} & \multicolumn{3}{c|}{{\footnotesize{}$\{\theta_{1},\theta_{2},\theta_{3}\}=\{0.5205,1.5437,2.1324\}$}}\tabularnewline
\hline 
{\footnotesize{}$m_{0}$ } & {\footnotesize{}$\infty$ } & {\footnotesize{}10 } & {\footnotesize{}2 } & {\footnotesize{}$\infty$ } & {\footnotesize{}10 } & {\footnotesize{}2 }\tabularnewline
\hline 
{\footnotesize{}$T_{0}$ } & {\footnotesize{}0.2713 +0.1465 i } & {\footnotesize{}0.5702 +0.3079 i } & {\footnotesize{}-1.2928-0.6982 i } & {\footnotesize{}-0.4238-0.7565 i } & {\footnotesize{}-1.8254-3.2585 i } & {\footnotesize{}0.2728 +0.487 i }\tabularnewline
\hline 
{\footnotesize{}$T_{2}$ } & {\footnotesize{}-12.4191-6.7073 i } & {\footnotesize{}-52.7054-28.465 i } & {\footnotesize{}0.1829 +0.0988 i } & {\footnotesize{}-0.4415-0.7882 i } & {\footnotesize{}-2.5287-4.5139 i } & {\footnotesize{}-0.1971-0.3518 i }\tabularnewline
\hline 
{\footnotesize{}$T_{4}$ } & {\footnotesize{}12.6765 +6.8463 i } & {\footnotesize{}54.3523 +29.3545 i } & {\footnotesize{}0.5922 +0.3198 i } & {\footnotesize{}0.5858 +1.0456 i } & {\footnotesize{}3.0962 +5.5269 i } & {\footnotesize{}0.0496 +0.0885 i }\tabularnewline
\hline 
{\footnotesize{}sum } & {\footnotesize{}0.5287 +0.2855 i } & {\footnotesize{}2.2171 +1.1974 i } & {\footnotesize{}-0.5177-0.2796 i } & {\footnotesize{}-0.2796-0.4991 i } & {\footnotesize{}-1.2579-2.2455 i } & {\footnotesize{}0.1253 +0.2237 i }\tabularnewline
\hline 
{\footnotesize{}$T_{6}$ } & {\footnotesize{}-0.013-0.007 i } & {\footnotesize{}-0.0768-0.0415 i } & {\footnotesize{}-0.0012-0.0007 i } & {\footnotesize{}0.0174 +0.0311 i } & {\footnotesize{}0.0901 +0.1609 i } & {\footnotesize{}0.0009 +0.0015 i }\tabularnewline
\hline 
{\footnotesize{}sum } & {\footnotesize{}0.5157 +0.2785 i } & {\footnotesize{}2.1403 +1.1559 i } & {\footnotesize{}-0.5189-0.2802 i } & {\footnotesize{}-0.2621-0.4679 i } & {\footnotesize{}-1.1678-2.0846 i } & {\footnotesize{}0.1262 +0.2252 i }\tabularnewline
\hline 
\end{tabular}
\end{document}